\documentclass[onecolumn,conference,moreauthors, a4paper]{IEEEtran}
\usepackage{filecontents}
\usepackage{cite}
\usepackage{amssymb,amsmath,epsfig,latexsym,graphicx,bm}
\usepackage{tabu}
\usepackage{tabularx}
\usepackage{graphicx,bm,makecell,array,optidef}
\usepackage[justification=centering]{caption}
\usepackage{subcaption}
\usepackage{float}
\usepackage{tabularx}
\usepackage{amsmath,amssymb,amsfonts}
\usepackage{algorithmic}
\usepackage{amsmath}
\usepackage[left = 1 in, top=1 in,right=1 in,bottom=1 in,nohead,nofoot]{geometry}
\usepackage{textcomp}
\usepackage{hyperref}
\usepackage{multirow}
\usepackage{array}
\usepackage{makecell}
\usepackage{ragged2e}
\usepackage{epstopdf}
\usepackage{array}
\usepackage{longtable}
\usepackage{lipsum}
\usepackage{ragged2e}
\usepackage[T1]{fontenc}
\usepackage{lmodern}

\title{A Contemporary Survey on Free Space Optical Communication: Potential, Technical Challenges, Recent Advances and Research Direction}
\newcommand{\orcid}[1]{\href{https://orcid.org/#1}{\textcolor[HTML]{A6CE39}}}

\author{\IEEEauthorblockN{Abu Jahid, Mohammed H. Alsharif and Trevor J. Hall \\
\centering }}
\providecommand{\keywords}[1]{\textbf{\textit{Keywords---}} #1}
\begin{document}
\bstctlcite{IEEEexample:BSTcontrol}
\maketitle

\begin{abstract}

Due to the unprecedented growth of high speed multimedia services and diversified applications initiated from the massive connectivity of IoT devices, 5G and beyond 5G (B5G) cellular communication systems, the existing electromagnetic spectrum under RF ranges incapable to tackle the enormous future data rate demands. Optical wireless communication (OWC) covering an ultra-wide range of unlicensed spectrum has emerged as an extent efficient solution to mitigate conventional RF spectrum scarcity ranging from communication distances from nm to several kilometers. Free space optical (FSO) systems operating near IR (NIR) band in OWC links has received substantial attention for enormous data transmission between fixed transceivers covering few kilometers path distance due to high optical bandwidth and higher bit rate as well. FSO offers a broad range of applications in outdoor and indoor services for instance, backhaul for mobile cellular networks, local area network (LAN) and metropolitan area network (MAN) connectivity as well as extensions, wireless video surveillances, data centers, terrestrial, mobile networks, military demands, last mile solution, backup of fiber connectivity, medical usage, space communications, and so on. Despite the potential benefits of FSO technology, its widespread link reliability suffers especially in the long-range deployment due to atmospheric turbulence, cloud induced fading, some other environmental factors such as fog, aerosol, temperature variations, storms, heavy rain, cloud, pointing error, and scintillation. FSO has the potential to offloading massive traffic demands from RF networks, consequently the combined application of FSO/RF and radio over FSO (RoFSO) systems is regarded as an excellent solution to support 5G and beyond for improving the limitations of an individual system. The aggregate deployment of wireless scheme can attain significant system performance in the form of spectral efficiency, reliability, and energy efficiency of distinct networks. Over the last decade, numerous FSO research have been carried out in the context of optical wireless communication to address the prominent technical challenges focusing on channel modeling, adaptive transmission and modulation, cooperative diversity, mitigation techniques of channel induced factors, novel design of FSO transmitter and receiver architectures. This survey presents the overview of several key technologies and implications of state-of-the-art criteria in terms of spectrum reuse, classification, architecture and applications are described for understanding FSO. This paper provides principle, significance, demonstration, and recent technological development of FSO technology among different appealing optical wireless technologies. The opportunities in the near future, the potential challenges that need to be addressed to realize the successful deployment of FSO schemes are outlined.

\end{abstract}


\section{Introduction}		\label{intro}

In recent years, diverse types of multimedia applications are expanding enormously, generating a mass volume of mobile data together with high data-rate wireless connectivity. The forthcoming 5G technology offers various attractive services such as massive system capacity, huge device connectivity, high-level security, ultra-low latency, extremely low power consumption with tremendous quality of experience (QoE) \cite{ijaz2016enabling, jaber20165g, shafi20175g, chowdhury2020optical}. Notably, 5G communication contemplated with ultra-dense heterogeneous networks allowing hundred times additional wireless device connectivity as well as transmission rate compared to existing wireless networks \cite{chowdhury2018comparative}. Therefore, 5G and beyond networks required high-capacity backhaul connectivity in order to support hyper-dense fast access network, minimal power consumption and little end-to-end delays \cite{siddique2015wireless, olwal2016survey, mustafa2015separation}. It becomes challenging to handle unprecedented high volume of information for the 5G connectivity and hence, a robust technical solutions are required in order to guaranteed quality of service (QoS) for the end users. It is widely accepted that radio frequency (RF) is commonly used in wireless communications which are more limited due to the shortage of spectrum resources \cite{ghassemlooy2015emerging, obeed2018joint}. Envisioning the concept of IoT, it enable real-time communications, sensing, monitoring and resource sharing in huge smart device connectivity ranging from social, industrial and business purpose.  With the rapid growth of Internet of Things (IoT)/ Internet of Everything (IoE) technologies, a massive number of physical smart devices are connected to the networks is accelerating exponentially \cite{palattella2016internet, hassan2017feasibility, schulz2017latency}. Consequently, IoT devices generates a huge volume of data. To meet the ever-increasing demand of 5G networks and serving the huge demands of IoT paradigm, it is predicted that currently available electromagnetic frequency band is insufficient. Meanwhile, this RF frequency band suffers a small spectrum range, also have limitations related to regulation based spectrum use and intense level of interference from the surrounded RF access points. In many cases, RF sub-bands are entirely allocated to mobile cellular operators, TV broadcasting and end-to-end microwave links. From the indication of RF based wireless network drawbacks, researchers are looking alternative approach millimeter and nanometer waves for wireless communication. As a consequence, academia and industries are currently interested in license free optical spectrum covering 1 mm-10 nm as an emerging alternative of RF for future ultra-density and ultra-capacity networks in perspective of optical wireless communication (OWC) \cite{koonen2017indoor, cruz2018optical, boucouvalas2015standards}. The term OWC is referred as the wireless connectivity using optical spectrum. OWC technologies offers a number of remarkable features for example broad spectrum, ultra-high data rate, very low latency, low cost and lower power consumption, addressing the massive demand requirements of 5G/B5G communications. 

In contrary to RF-enabled networks, OWC technology offers some remarkable advantages like as high data rate transmission capacity ranging from nanometers to several kilometers both for indoor and outdoor applications. Moreover, OWC technology has the potential to provide some outstanding communication features such as reliable security, electromagnetic interference free transmission, and high system efficiency due to using broad optical spectrum \cite{arnon2012advanced, grobe2013high}. Reference \cite{tsonev2015towards} demonstrates that OWC technology can attain 100 Gb/s at standard indoor illuminations and capable to provide energy efficient communication. The most important feature of OWC technologies it do not require a comprehensive infrastructure and thereby reducing the installation cost maintaining all the green agenda for high speed communication \cite{karunatilaka2015led, zhang2017space, khreishah2018hybrid, arienzo2019green}. Visible light communication (VLC) and light fidelity (Li-Fi) techniques under OWC technology use the existing illumination structure to realize wireless data transfer \cite{haas2015lifi, ghassemlooy2016optical, wang2016dimmable}. OWC provides the greater level of data security since the light waves does not penetrate the enclosed walls. Typically, visible light (VL), infrared (IR) or ultraviolet (UV) spectra are used as propagation media. Most promising wireless systems in OWC such as VLC, LiFi, optical camera communication (OCC), and free space optical (FSO) communication are being developed based on those three optical bands. However, VLC, LiFi, OCC, and FSO technologies have some similarities and disparity by means of communication protocol, propagation media, architecture and applications. On the other hand, the aggregate option of OWC and RF refereed as a hybrid RF/OWC systems may establish an effective solution for tremendous upcoming user demands. 

Despite the remarkable advantages of OWC technologies, some factors for instance limited coverage area, dependence on line-of-sight (LOS), interference incurred by multiple light sources, outdoor atmosphere, and low transmit power  degrade the system performance. Moreover, the performance OWC suffers from external disturbances and limited their transmit power as well as signal to interference plus noise ratio (SINR). Hence, it is a challenging issue to exploit successful OWC deployment addressing these limitations. The use of RF spectrum across existing wireless applications has almost been exhausted. However, RF-based communication is highly sensitive to interference but provides better performance under non-LOS (NLOS) conditions. This key feature of RF systems capable to overcome some weakness of OWC schemes particularly for the end users. In order to ensure desired QoS, the conjunction of RF and optical wireless based networks in the context of wireless communication play a significant role through integrating a diverse frequency bands. The heterogeneous (endeavoring of two or more network architectures operates simultaneously) operation of access network technologies cooperate each other for traffic offloading, thereby mitigate the QoS constraints. Due to the complementary characteristics of RF and OWC technologies, researchers have proposed some hybrid RF/OWC approaches to support IoT/IoE, 5G/B5G communications systems \cite{guo2015parallel, pan2017secure, kafafy2018novel, liang2017multi, jin2014resource, buyukccorak2017bayesian, kashef2017transmit, sharma2018eh, rahaim2017beam, li2016mobility, wang2017learning, nguyen2019vlc, mufutau2020demonstration, tonini2019cost, esmail20195g}. The co-deployment of RF/optical wireless hybrid system solves the last mile problem where end users can be benefited from RF coverage. The hybrid approach incorporates two or more different related technologies (e.g., RF/FSO, WiFi/LiFi/, VLC/FSO, LiFi,/OCC, acoustic/optical for underwater communication) that can enhance the system performance in terms of throughput, bit error rate (BER), reliability, and energy efficiency providing the combined benefits of both technologies. Hybrid networks can be deployed in many applications for instance seamless movement, load balancing, high speed wireless connectivity in remote areas and link performance improvement \cite{aldalbahi2017visible, basnayaka2017design, feng2016applying, baig2018high, wang2015dynamic, rakia2017cross, zhou2017distributed, hasan2019real, hasan2018fuzzy, tsai2019256}. This survey preliminary discuss the comparative overview of OWC technologies competitive to FSO technique. Thereafter, an extensive analysis of FSO technology illustrating the potential applications including recently developed intriguing features, wide range of deployment challenges and possible mitigation methods, reliability analysis, and research directions.


\begin{table}[]
\centering
\caption{Nomenclature} 
\label{Table1}
{\tabulinesep=1mm
\begin{tabular}{|l|l|l|l|} \hline
5G    & $5^{th}$ Generation           & OOC  & Optical Orthogonal Codes	 \\ \hline
AMC  & Adaptive Modulation and Coding &OFDMA & Orthogonal Frequency-Division Multiple Access \\ \hline
AP   & Access Point   &  OOK & On-Off Keying \\ \hline
AWGN & Additive White Gaussian Noise & OPPM & Optical Pulse Position Modulation \\\hline
B5G  & Beyond 5G      & OSTBC  & Orthogonal Space-Time Block Code \\ \hline 
BER  & Bit Error Rate    & OSSK & Optical Space Shift Keying  \\\hline
BCHB & Bose–Chaudhuri–Hocquenghem Block &OWC & Optical Wireless Communication \\\hline
BICM & Bit Interleaved Coded Modulation & P2P&Point-to-Point \\\hline
BMST & Block Markov Superposition Transmission &P2mP & Point-to-multi Point\\\hline 
BPSK  & Binary Phase Shift Keying  & EE & Energy Efficiency \\ \hline
CAP & Carrierless Amplitude and Phase & PAPR &Peak-to-Average Power Ratio\\\hline
CBFSK    & Coherent Frequency Shift Keying     & PAM & Pulse Amplitude Modulation  \\ \hline
CBPSK& Coherent BPSK &PAT &  Pointing Acquisition Tracking \\ \hline 
CN  & Core Network & PD & Photodetector \\  \hline 
CSI & Channel State Information  &  SM  & Spatial Modulation  \\\hline  
CSK & Color Shift Keying &PDF & Probability Density Function \\\hline
CSMA   & Carrier Sense Multiple Access       &PEP &Pairwise Error Probability  \\ \hline
D2D   & Device-to-Device           & PLC  & Power-Line Communication \\ \hline
DCO  & DC-biased Optical &PLNC &  Physical Layer Network Coding \\\hline
DBPSK    & Differential BPSK           & PM & Pulse Modulation \\ \hline 
DPIWM & Digital Pulse Interval and Width Modulation & PPM & Pulse Position Modulation\\ \hline
DPPM & Differential Pulse Position Modulation & PWM & Pulse Width Modulation\\ \hline 
E2E   & End-to-End          & QAM  & Quadrature Amplitude Modulation \\ \hline
EO    & Electro Optic           &  QPSK & Quadrature Phase Shift Keying \\ \hline 
FEC & Forward Error Correction & QoE & Quality of Experience   \\ \hline
FTTH    & Fiber to the Home           & QoS  & Quality of Service \\ \hline
GPS    & Global Positioning System           &  RADAR & Radio Detection And Ranging \\ \hline
HetNets    & Heterogeneous Networks           & RC & Repetition Coding \\ \hline
ICI    & Inter-cell Interference          & RCPC  &  Rate Compatible Punctured Convolutional \\ \hline
IMC & Inverted Manchester Code  &RF &  Radio Frequency \\ \hline
IM/DD   & Intensity Modulation/Direct Detection         & RoFSO & Radio Over FSO \\ \hline
IoE   & Internet of Everything           & RS & Reed-Solomon \\ \hline
IoT     & Internet of Things           & RSS & Received Signal Strength \\ \hline
IR    & Infrared Radiation           & SDN & Software Defined Networking \\ \hline
ISI  & Inter Symbol Interference   & SE &  Spectral Efficiency  \\ \hline
LD    & Laser Diode           &SER  & Symbol Error Rate \\ \hline
LDPC & Low Density Parity Check  & SIM& Subcarrier Intensity Modulation \\ \hline
LED    & Light Emitting Diode           & SINR  & Signal-to-Interference-Plus-Noise Ratio \\ \hline  
LiDAR    & Light Detection And Ranging           & ST & Space-Time \\ \hline
LiFi    & Light Fidelity          & STTC &  Space-Time Trellis Code \\ \hline
LOS    & Line-of-Sight           & UAV & Unmanned Area Vehicle \\ \hline
LT & Lubry Transform   & UE  & User Equipment  \\ \hline
M2M   & Machine-to-Machine          & UV & Ultraviolet \\ \hline
MAI   & Multi Access Interference          & UWC   & Underwater Communication \\ \hline
MBS     & Macrocellular Base Station          & UWOC & Underwater Wireless Optical Communication \\ \hline
mmWave     & Millimeter-Wave           & V2X & Vehicle-to-Vehicle, Vehicle-to-Infrastructure \\ \hline
MIMO    & Multiple-Input and Multiple-Output           & VANET & Vehicular Ad-hoc Network \\ \hline
MPPM & Multi-pulse Pulse Position Modulation &VCSEL  & Vertical-Cavity Surface Emitting Laser  \\ \hline
MZM   & Mach Zehnder Modulator           & VHO & Vertical Handover \\ \hline
NIR    & Near Infrared            & VLC  & Visible Light Communication \\ \hline
NLOS     & Non-Line-of-Sight           & WDM & Wavelength Division Multiplexing \\ \hline
NOMA   &Non –Orthogonal Multiple Access          & WiFi  & Wireless Fidelity \\ \hline
OCC     & Optical Camera Communication           & WLAN & Wireless Local Area Network \\ \hline
OCDMA   & Optical Code Division Multiple Access           & WSN & Wireless Sensor Networks \\ \hline
\end{tabular}}
\end{table}

\section{Overview of OWC}			\label{overview_OWC}

The wide range of electromagnetic (EM) frequency spectrum allows to extend the use of frequency band for optical wireless communication purpose to tackle the congested spectrum of RF. An enormous portion of upper band EM spectrum over microwave range is available for OWC for next generation HetNets as an alternative solution to RF. Each optical frequency band of near IR (NIR), VL and UV has their unique features and characteristics. Based on the related optical bands properties, different OWC technologies are being developed. For example, infrared spectrum can be used for applications where illumination is not required. On the other hand, UV frequency band support high speed data rate for LOS and NLOS communication over short and distances. The OWC technologies offers an excellent features of high data rate communication link that can used in a broad range of applications ranging from ultra-short to long distance communications. Another important key features include unregulated high optical bandwidth, low-energy consumption, greater level of data security, low deployment cost, immune of interference from RF based networks, low bit error rate (BER) and ease of integration with existing infrastructure. In contrast, the OWC performance degraded due to blocking obstacles along the communication link including several other limiting factors. For ease of reference, notations used in this paper are summarized in Table \ref{Table1}.

\subsection{OWC Technologies}			\label{OWC_tech}

Optical carrier transmission in the form of infrared (IR), visible light (VL), or ultra-violet (UV) spectrum have distinctive properties enabling different key communications \cite{ghassemlooy2015emerging, boucouvalas2015standards, maier2019towards}. Terrestrial free space optical communication systems (FSO) are operated at IR and VL frequency band. Whereas UV spectrum supports high speed non line-of-sight (NLOS) and LOS optical communications \cite{ghassemlooy2015emerging}. Several OWC technologies such as VLC, light fidelity (LiFi), optical camera communication (OCC), light detection and ranging (LiDAR) and free space optical (FSO) communication are developed to meet the widespread wireless demand. LiFi is a relatively new technology complementing WiFi that provides high data rate transmission in the speed of light illumination aiming the same purposes to be served \cite{haas2015lifi, haas2020introduction}. Light emitting diode (LED) or laser diode (LD) with an optical diffuser can be used as transmitters and photodetector (PD) as receiver \cite{chi2015phosphorous}. However, VLC and LiFi technologies can provide extreme transmission speed for indoor applications and not effective for outdoor services due to obstacle of light. Moreover, they cannot cover long range of communication distance. On the other hand, LEDs used as transmitter and a camera or image sensor used as receiver in OCC sub-system under OWC \cite{goto2016new, zheng2017accuracy}. OCC achieves a high SNR performance cancel out associated interference even under the outdoor environments. Besides, it exhibits stable performance for longer link coverage whether the communication take place in indoor or outdoor \cite{zheng2017accuracy}. LiDAR \cite{qin2011toward, jin2019receiver, zhang2019tracking} is a remote sensing 3D laser scanning method of a target with high-resolution maps. A LiDAR device comprises of a LD, a scanner and a specialized global positioning system (GPS) receiver. It capture information about a landscape and measure things using high resolution sensors for analyzing conditions and characteristics. Recently, free space optical (FSO) communication has emerged as a promising solution for next generation lightwave networking regardless indoor or outdoor environments, despite the system is severely suffered by the channel impairments such as atmospheric turbulence and pointing error effect between the optical transmitter and receiver \cite{yang2014free, malik2015free, li2019hybrid, sharma2019effect}.

\subsection{OWC Technologies using IR/VL/UV Spectrum}		\label{spectra}

According to the transmission spectrum allocation \cite{khalighi2014survey}, OWC can be classified into three categories, namely infrared, visible light and ultraviolet. Near IR (NIR) band is widely operated for point-to-point FSO communication including ultra-short (750 nm to 1450 nm), short (1400 nm to 3000 nm), medium (3000 nm to 8000 nm), long (8000 nm to 15000 mm), and ultra-long distances (15000 nm to 1 mm). However, ultra-short IR and short IR optical spectrum are used for long distance FSO communications. Whereas, medium IR and long IR spectrum can be used in military applications and thermal imaging purpose respectively. Besides, IR also commonly used in LiFi and OCC for short and medium transmission distances; covers long distances for LiDAR technologies. The operating wavelength of NIR is 760nm-1mm and particularly suitable for LOS communication where no illumination is needed. However, it’s not always safe for human and limited application for NLOS communication. Secondly, visible light (VL) is commonly used for VLC, LiFi and OCC covering short and medium distances. In some limited cases, VL spectrum (operating wavelength of 360-760nm) is also used for FSO and LiDAR over various broadcasting range \cite{chowdhury2018comparative}. The benefit of VL spectrum are: safe for human and can be exploited for illumination and communication purpose simultaneously. Likewise NIR, VL is perfect for LOS communication and exhibit poor data rate for NLOS transmission through reflection of light. Finally, a considerable amount of interest in UV communication found for providing high speed NLOS/LOS communication links. UV spectrum (operating wavelength of 10-400nm) is pertinent for short and medium ranges in LiFi, ultra-short, short, medium, long and ultra-long coverage distance in FSO systems \cite{chowdhury2020optical}.

\subsection{Brief Description of OWC Technologies}


Each of the OWC technologies (VLC, LiFi, OCC, LiDAR and FSO) have a distinctive features by means of communication protocol, architecture, principle of operation, modulation techniques, channel media, transmitting scheme, receiving system, and application scenario. Each technologies have some advantages and disadvantages that are briefly discussed in this subsection. Typically, either light emitting diodes (LEDs) or laser diodes (LDs) are used as physical transmitter whereas photodetectors (PDs) or cameras are used as optical receivers. Besides, infrared (IR), visible light (VL) or ultra-violet (UV) spectra are used as communication media. A substantial comparison among different OWC technologies and RF systems is demonstrated in Table \ref{Table:2}.
\vspace{2mm}

\subsubsection{Light Fidelity (LiFi)} \cite{haas2015lifi}, \cite{alshaer2018bidirectional, chen2015downlink, soni2016looming, haas2018lifi, dimitrov2015principles, lu201656, wu2019load, haas2020introduction, abdallah2020optical, wu2019mobility, soltani2019bidirectional, demirkol2019powering, bian201915, soltani2018modeling, islam2019hybrid, abumarshoud2019dynamic}

The operational principle of LiFi is almost similar to WiFi and can be classified as an nm-wave communication. LiFi technology provides very high-speed data rate using light illumination. Likewise VLC, LEDs or LDs is used as optical transmitters and PDs as optical receivers in LiFi that support high speed wireless connectivity, localization and illumination as well. LiFi is a bidirectional communication scheme whereas VLC can be either unidirectional or bidirectional. LiFi supports point-to-point, point-to-multipoint, multipoint-to-multipoint communications and seamless user mobility like RF based wireless systems. Therefore, LiFi is a mature wireless networking system enabling multiuser communications within existing HetNets. LiFi uses either IR, VL or UV spectra in the backward path whereas VL light spectrum is used for the forward path communication. 
A micro LED driven optical source has achieved more than 3 Gbps \cite{tsonev20143} and a vertical cavity surface emitting laser (VCSEL) based light source can attain 56 Gbps speed \cite{lu201656}. The concept of attocells LiFi can shrink cell size through offloading huge amount of data from highly congested RF systems. The purpose of illumination and communication are performed simultaneously in an indoor environment. On the other hand, only communication is accomplished for applications like controlling traffic signals, back lamps of vehicles and both communications and illumination purpose are performed for applications like controlling street lamp and vehicle forward lamp under an outdoor environments. In summary, the key features of LiFi over WiFi include stringent security, high speed bit rate, cost effectiveness, and availability of light spectrum. In contrast, the performance of LiFi is greatly suffered by sunlight and other outdoor ambient light sources. Moreover, the coverage holes is key limitations of indoor LiFi networks. However, LiFi technology is not suitable for long-haul communication and not effective in outdoor environments. 

\vspace{1mm}

\subsubsection{Visible Light Communication (VLC)}  \cite{grobe2013high, tsonev2015towards, karunatilaka2015led}, \cite{wang2016dimmable, guo2015parallel, pan2017secure, kafafy2018novel, liang2017multi, jin2014resource, buyukccorak2017bayesian, kashef2017transmit}, \cite{rahaim2017beam}, \cite{nguyen2019vlc}, \cite{zheng2017accuracy}, \cite{soni2016looming}, \cite{kishore2020spectrally, gismalla2020improve, zheng2020ofdm, dahri2020design, yin2018tractable, yang2019power, rodoplu2020characterization, liu2020power}

VLC uses LEDs and LDs as transmitters and PDs as optical receivers and visible light as the communication media. VLC provides point-to-point, point-to-multipoint communications either in unidirectional or bidirectional ways where mobility support is not mandatory. VLC technology can support 100 Gbps data rate \cite{tsonev2015towards} and offers tenfold thousand times higher bandwidth capacity compared to RF technologies \cite{karunatilaka2015led}. However, the application scenarios are almost similar to LiFi for example office, residential areas, roadsides, aircrafts, cars, etc. In addition, VL has many applications in indoor and outdoor environments for instance, V2X communications, smart lighting, hospitals, aviation, data offloading, acoustic communications, smart display boards, location aware services, LANs, wireless backhaul connectivity etc. Due to the fast switching capability of energy-efficient LED devices, VLC technology can be used for illumination, localization and communication simultaneously. A transmission rate of 10 Gbps has already been achieved using LED and 100 Gbps is attained using LDs at standard indoor illumination \cite{tsonev2015towards}. Likewise LiFi, VLC is not suitable for outdoor applications as well as long distance communications. It also creates coverage holes in indoor environments but it is harmless to human.

\subsubsection{Optical Camera Communications (OCC)}  \cite{ghassemlooy2016optical}, \cite{nguyen2019vlc}, \cite{hasan2019real, hasan2018fuzzy}, \cite{he2020high, matus2020experimentally, akram2019design, zhang2020push, lain2019experimental, saeed2019optical, hasan2019simultaneous, yang2018spatial}

OCC uses LEDs as transmitters, camera as the receiving module and IR and VL spectra as communication media. OCC technology can be implemented in the existing hardware infrastructure. The data signal transmitted from the multiple transmitters can be easily captured using a camera image sensor. OCC system can support interference free transmission in outdoor environments through separating the pixels of sunlight and other ambient sources \cite{chowdhury2018comparative}. The achievable data rate using OCC is 45 Mbps with zero BER and 54 Mbps with a BER of $10^{-5}$ were demonstrated in \cite{goto2016new}. However, VLC system can be considered as the OCC scheme when it uses LEDs as optical sources and camera as receivers instead of photodetectors. OCC can be easily overcome some limitations for example, short communication distance, high attenuation of signal strength, interference due to sunlight and other ambient light sources, background noises, and challenges in detection using PDs. OCC is very prominent for indoor positioning, V2V, V2I, I2V communications, drone-to-drone communications, digital signage, sending and detecting the control signals using body motion, non-interference communication, and good SNR quality. In addition, OCC enable data sharing among mobile users, and perform motion detection and recognition. In spite the benefit of stable performance of OCC in outdoor environments, the attainable bit rate is comparatively small. Blocking of optical signals and light flickering due to low frame rates are another example of OCC shortcomings.  

\subsubsection{Light Detection And Ranging (LiDAR)}  \cite{qin2011toward, jin2019receiver, zhang2019tracking}, \cite{kwon2017detection, abuella2019vildar, ma201966db, wu2020moving, hua2020detection, popko2020geometric, widarsson2020high}

LiDAR is optical remote sensing technology that uses near infrared (NIR) and VL spectra to find the image objects or other information of a target. The principle of LiDAR is similar to radio detection and ranging (RADAR) used for localization and object detection. The key difference is LiDAR uses light wave and RADAR uses microwave signal for energy transmission to hits an object. The reflected signal is back to the receiver at the source location and then the distance, size, other characteristics are determined analyzing the reflected signal. Laser beams are used to target dust, aerosols, non-metallic objects, rain, clouds, etc. LiDAR technology can map the physical features with the help of very high resolution laser beams and measure the properties of scattered light for 3D mapping. LiDAR system comprises of laser sources, photodetectors, scanner and position and navigation sub-systems all are used to track the position and movement of objects. The application of LiDAR includes transportation, precise location, space surveys, architectural exploration, meteorology, military operations, robotics, forestry, etc.

\subsubsection{Free Space Optical (FSO) Communication}   \cite{rahaim2017beam}, \cite{tonini2019cost, esmail20195g}, \cite{yang2014free, malik2015free, li2019hybrid, sharma2019effect}, \cite{islam2019effect, nazrul2019effect, bashir2020adaptive, bashir2019free, sharma2014high}, \cite{ahmed2020performance, kumar2020performance, li2018cooperative, poulton2019long, huang2019game, paul2019alleviation, wang2020influence, jaiswal2018free, yi2019performance, yeh2019integration}

FSO is a subset of OWC technology, is typically used NIR spectrum as channel medium where the attenuation levels are much lower. FSO can also be operated by using VL and UV spectra and do not require illumination. Narrow beam of LD light sources are used to establish high speed communication links instead of LEDs whereas PDs are used as optical receiver. Due to the coherent nature of laser technology, the FSO signal can travel long distance enabling point-to-point communication with excellent data rate. Optical power amplifiers can be used to increase the signal intensity of the modulated laser output and then the laser light beam is collected and refocused before transmission. However, different modulation schemes and channel coding are used to deliver a high optical power over a broad range of temperature. Electronic circuits are needed to convert the PD output into voltage, low pass filter (LPF) eliminates the noise levels and demodulator performs necessary process of sending the original data to the destination. The application of FSO system include video surveillance, campus connectivity, backhauling for cellular networks, LAN-to-LAN connectivity, inter-chip connectivity, underwater communication, space communication, fiber backup, etc. Despite the multiple advantages of FSO systems over a wide range of applications, the performance of FSO link suffers link reliability and high sensitivity to some limiting factors for example, outdoor weather conditions (e.g., heavy rain, fog, smoke, storms, deep clouds, snow, and scintillation), atmospheric turbulences and physical obstructions. This paper comprehensively explore the features, scope, performances, applications, system design, hybrid framework, research challenges, open issues, and future directions in the subsequent sections. Table \ref{Table:3} presents the key comparison among different RF techniques. 


\begin{table}
\centering
\caption{Comparison of Different Wireless Communication Technologies \cite{chowdhury2020optical, ghassemlooy2015emerging, tsonev2015towards, sharma2014high, saha2015survey, luo2015experimental, rajagopal2014hybrid, chen2017performance, el2017effect, chang2014100}} \label{Table:2}
{\tabulinesep=1mm
\begin{tabu}{|l|c|c|c|c|c|}
 	\hline
\textbf{Factors} & \textbf{LiFi} & \textbf{VLC} & \textbf{OCC} & \textbf{FSO} & \textbf{RF}\\ \hline
 	Light Source & LED/LD & LED/LD & LED & LD & Antenna\\\hline
 	Receiver & PD & PD & Camera & LD & Antenna\\\hline	
 	Modulation & OOK, OFDM, & OOK, OFDM, & OOK, PAM, & OOK, OFDM, & ASK, PSK, QPSK,  \\
 		&         CDMA   &   CDMA  & CDMA  &  QAM, SIK   &   QAM, OFDM  \\ \hline
 	Licensed &  No & No & No & No & Yes (except WiFi)\\
 	Spectrum & & & & &  \\\hline	
 	Link coverage & 10m & 20m & 200m & $>$1000km & $>$100km(microwave link)\\\hline	
 	Environmental  & yes & yes & No & yes & yes\\
 	effect & & & & & \\\hline	
 	Bit rate & 100 Gbps (LD) & 100 Gbps (LD) & 54 Mbps & 40 Gbps & 6 Gbps\\\hline	
 	Spectrum & IR/VL/UV & VL & VL & IR/VL/UV & Radio frequency\\\hline	
 	Path loss \&  & Medium(LOS) \cite{ghassemlooy2015emerging} & Medium(LOS)  & Low \& zero\cite{ghassemlooy2016optical} & High \& low\cite{el2017effect}  & High \& very high\\
 	interference & \& low & \& low\cite{pathak2015visible} & & &\\\hline	
 	Communication & Bidirectional & Unidirectional/  & Unidirectional & Bidirectional & Bidirectional\\
 	topology & & Bidirectional & & &   \\\hline	
 	Security & high & high & high & high & low\\\hline	
 	Key & Poor performance  & Poor performance  & Poor data rate  & Performance severely   & High\\
 	limitations &for NLOS  & for NLOS  & \& short range  & degraded by  & interference \\
 	& communication & communication & communication & environmental issues & \\\hline	
\end{tabu}}
\end{table}

\begin{table}
\centering
\caption{Key Features of Different RF Technologies}\label{Table:3}
{\tabulinesep=1mm
\begin{tabu}{|c|c|c|c|c|c|}
 	\hline
 	Technology & WiFi & mmWave/Microwave & UWC & Heterogeneous networks & Bluetooth\\\hline
 	 & Free spectrum,  & High bandwidth \& & Wide range of & Mobility support, & Unlicensed spectrum,\\
 	 & low installation, & high data rate, & coverage, support & high level of QoS & support LOS/NLOS \\
 	Merits & mobility support, & Reduced antenna size & LOS/NLOS & support LOS/NLOS & communications \\
 	& enable NLOS & at higher  & communications & communications & \\ 
 	& communications  & frequencies	& & & \\\hline
 	 & Low-level & Suffer electromagnetic  & Poor bit rate, & Limited & Limited coverage,\\
 	Demerits & security \& high & and other & high propagation & spectrum & high interference, \\
 	& interference & interference & delay & & low security \\ \hline	
\end{tabu}}
\end{table}

\paragraph*{WiFi} \cite{intel, roy2018optimal, shen2019reliable, poularakis2019joint, meng2019revealing}
technology enables wireless area networking using radio frequency under a particular range based on IEEE 802.11 standard. WiFi provides very high data rate communication up to 8 Gbps with mobility support under both LOS and NLOS environments. The prime benefits is unlicensed free spectrum and lower deployment cost. Conversely, the lower level security issue and interference effect limited its performance. 

\paragraph*{Microwave link networks} \cite{li2016multi, feng2018photonic, wang2018performance}
supports point-to-point communication with a beam of radio signals to transmit information in microwave frequency range. A microwave communication is a promising alternative to optical fiber link with high data rate particularly for remote areas. It can covers over 100 km link distance and bit rate of 12.6 Gbps is demonstrated in \cite{li2016multi}. The performance of microwave communication is suffered by electromagnetic interference, fog, rain and other atmospheric turbulence. 

\paragraph*{Cellular networks}  \cite{chowdhury2018interference, liang2019joint, shnaiwer2019opportunistic, yu2019network}
comprised of macrocell and small cells called heterogeneous networks (HetNets) are widely deployed to provide high bit rate to users. A macrocell base station (BS) provide a wide range of coverage and user mobility and supports a large number of user equipments (UEs). On the other hand, small cells network (e.g., microcell, picocell, femtocell) extend the cellular connectivity at the cell edge and filled up the coverage holes especially where the access is unavailable. However, macrocell deployment is expensive and offers low data rate whereas limited RF spectrum is allocated for small cell networks. 

\paragraph*{Bluetooth} \cite{collotta2018bluetooth, hussain2017secure, rondon2019understanding, elhence2020electromagnetic}
is a RF-based wireless technology provides data rate up to 2 Mbps using short wavelength radio signal in the 2.4 GHz unlicensed frequency band. Bluetooth devices consume low energy and support LOS/NLOS communications in a single hop topology. However, the performance bluetooth communication suffers of low level security, short operational coverage and high interference effect. 

\paragraph*{Underwater communication (UWC)} \cite{kaushal2016underwater, zeng2016survey, jouhari2019underwater, sun2020review}
is an acoustic and RF based technology to transfer data under water up to 20 km long range link distance and support both LOS/NLOS communications. However, the performance of UWC is limited by low data rate, large propagation delay and high sensitivity to the conditions of water.

\subsection{Related Works}

A number of curiosity driven research works are conducting on OWC based networks across worldwide in order to addressing the aforementioned issues. The use of optical frequency spectrum complementary to RF based wireless system provides some advantages for instance, low spectrum availability, secure transmission, and low power consumption. Researchers across worldwide conducted essential literature reviews \cite{karunatilaka2015led, sarigiannidis2014architectures, shaddad2014survey} in several aspects of OWC field. Reference \cite{chowdhury2018comparative} presents an overview of OWC focusing on operational principle of VLC, LiFi, LiDAR, FSO technologies highlighting the advantages and range of application areas. Reference \cite{karunatilaka2015led} reported a survey emphasizing on various modulation schemes, dimming functions, filtering, equalization and beamforming for indoor VLC applications. Author in \cite{sarigiannidis2014architectures} presents the research direction towards optical-wireless convergence pointing out the issues of the conjunction of two different wireless and optical networks under dynamic bandwidth allocation. The fundamental access technologies and related progress advancements are comprehensively discussed of wireless access and optical broadband technologies \cite{shaddad2014survey}. Sevincer \textit{et al.} \cite{sevincer2013lightnets} discuss the key challenges of jointly performing the lighting and networking functions in the area of integrated FSO and VLC technologies.

Reference \cite{islam2019effect} demonstrated BER evaluation considering multi access interference (MAI) and crosstalk at the SIK receiver to assess SINR accompanied by BER varying system parameters such as code length, number of simultaneous user. Results revealed that mutlti- wavelength OCDMA (MW-OCDMA) attains superior BER performance oven single wavelength OCDMA FSO scheme. However, the power penalty (PP) is more significant for higher values of jitter standard deviation and simultaneous users at a given BER of 10$^{-9}$. Note that PP can be considerably reduced with the increment of code length and number of wavelength for a given number of concurrent users. An analytical model of multi-wavelength OCDMA wavelength division multiplexing (MW-OCDMA-WDM) FSO communication systems is conducted to evaluate BER functionality and link capacity in the presence of aforementioned limiting factors for free space applications \cite{islam2016ber}. Authors decompose the evaluation of BER into two categories; Q-ary Optical PPM (Q-OPPM) with intensity modulation–direct detection (IM/DD) receiver considering the impact of refractive index variation of the optical channel link without encoding.  Secondly, an extensive assessment of optical direct sequence (DS) CDMA encode and shift inverse keying (SIK) dual optical decoder IM/DD based FSO communication is further investigated taking into considerations of strong and weak atmospheric turbulence. Furthermore, the OCDMA FSO system performance in terms of signal to interference plus noise ratio (SINR) as well as BER is carried out including the combined effect of distinctive atmospheric turbulences, multi-access interference (MAI), cross talk, and pointing error varying link length, transmission rate, cloud thickness, channel parameters, code length, and number of simultaneous users. In addition, numerous transmitter and receiver diversity like SIMO, MISO and MIMO configurations over the turbulent channel with OOK direct detection based Rake receivers are also contemplated in order to evaluate power penalty sufferings, receiver sensitivity and capacity improvement for a given BER of 10$^{-9}$. To the end, simulation results validate the effectiveness of MW-OCDM-WDM hybrid scheme analytical model with a promising candidate for future terrestrial wireless optical communication network. Reference \cite{khalighi2014survey} describes the theoretical limiting factors of FSO channels, algorithmic-level of system design approach integrating diverse channel coding, adaptive modulation schemes, spatial diversity and hybrid RF/FSO systems. Table \ref{Table:4} summarizes the key contributions of numerous research works conducted on different wireless optical schemes in the context of OWC.

\begin{table}
\centering
\caption{Related Works of OWC Technologies}\label{Table:4}
{\tabulinesep=1.2mm
\begin{tabular} {|c|p{0.8\textwidth}|c|c|} 
 	\hline
 	OWC  &  \centering Major Contribution & References \\
 	 Systems & & \\\cline{1-3}
 	 & Application of LiFi components to hybrid LiFi/WiFi networks and clarify how LiFi takes VLC to realize & \cite{haas2015lifi} \\
 	 &  fully wireless network systems &  \\\cline{2-3}
 	 & Introduce dynamically resource allocation scheme, management and configuration in LiFi attocell access network &  \cite{alshaer2018bidirectional}\\\cline{2-3}
 	 & Presented the downlink performance of optical attocell networks in terms of SINR, outage probability, and &  \cite{chen2015downlink} \\
 	 &   achievable throughput with OFDM technique &  \\\cline{2-3}
 	LiFi & The key advantages of LiFi over WiFi is presented by means of data rate and flexibility operation & \cite{soni2016looming} \\\cline{2-3}
 	 & Demonstrated LiFi based 8 Gbps transmission rate in the 5G communication & \cite{haas2018lifi} \\\cline{2-3}
 	 & Described the underlying principles of high performance indoor OWC focusing on throughput maximization, &  \\
 	 &  energy efficiency, hardware complexity integrating OFDM-MIMO transmission and non-linearity model & \cite{dimitrov2015principles} \\\cline{2-3}
 	 & Experimented of 56 Gbps bit rate based on four level pulse amplitude modulation (PAM) LiFi in outdoor environment & \cite{lu201656}  \\\cline{1-3}
 	 & Investigated 100 Gb/s transmission capabilities using LDs at indoor level considering illumination constraints & \cite{tsonev2015towards} \\\cline{2-3}
 	 & Analyzed the accuracy of VLC based indoor positioning systems based  on three methods namely, & \cite{luo2017indoor} \\
 	 &  mathematical, sensor-assisted and optimization &  \\ \cline{2-3}
 	 & Reviewed the system design, physical layer properties of VLC channel, adaptive modulation formats, medium  &  \\
 	 &  access control, characteristics of transceiver, various visible light sensing, robustness to noise, direction of & \cite{pathak2015visible}\\
 	 &    increasing communication range, enhancement of spectral efficiency, concept of  heterogeneous VLC for &  \cite{khan2017visible} \\
 	 &  short range communication, parallel VLC framework, and potential applications in indoors and outdoors, &  \cite{cuailean2017current}\\
 	 &     deployment of challenges & \\\cline{2-3}
 	VLC & Examined resource and power allocation mechanisms, access point (AP)  coordination techniques, non- &  \\
 	 &  orthogonal multiple access (NOMA) VLC networks, security concerns, fairness, energy efficiency, fairness, &\cite{obeed2019optimizing} \\
 	 &   secrecy rate, channel capacity evaluation, performance metrics, simultaneous information transmission  and &  \\
 	 &  energy harvesting in VLC networks in the context of 5G communication & \\\cline{2-3}
 	 & Discussed optimal resource allocation, fairness, spectral efficiency, achievable throughput and interference management in a multi-user VLC system employing OFDMA technique & \cite{hammouda2018resource, bawazir2018multiple}  \\\cline{2-3}
 	 & Pointed out the point-to-point ergodic channel capacity and SNR with respect to geometrical parameters for indoor VLC applications & \cite{xu2017ergodic} \\\cline{2-3}
 	 & Proposed a combined optical spectral efficient mobile fronthaul integrating low cost LED in the perspectives of spatial densified LTE networks to enable multi-tier coordination & \cite{lu2017efficient} \\\cline{2-3}
 	 & Proposed a priority based dynamic channel reservation model focusing on call blocking probability and channel utilization considering real-time call arrival rates provisioning quality of service & \cite{chowdhury2014dynamic} \\\cline{1-3}
 	 & Presented 55 Mb/s VLC signal transmission using optical communication image (OCI) sensor and optical OFDM with achievable BER <$10^{-5}$ & \cite{goto2016new} \\\cline{2-3}
 	 & Examine the channel model according to link distance between transmitter and receiver for vehicle to infrastructure varying the luminance of the central pixel projected from optical source   & \cite{takai2014optical, yamazato2015vehicle} \\\cline{2-3}
 	  & A downlink mobile camera based short-distance VLC is designed to examined the attainable data rate incorporating camera rotation compensation for multiuser access  & \cite{cahyadi2016mobile} \\\cline{2-3}
 	OCC & A vehicle localization scheme in an outdoor environment based on hybrid optical camera communication and photogrammetry technique is proposed & \cite{hossan2018new} \\\cline{2-3}
 	 & Introduced the smartphones localization technique in an indoor environments by calculating the coordinate of receiver and distance from LED transmitters & \cite{hossan2018novel} \\\cline{2-3}
 	 & Demonstrated head mounted display enabled human bond communication method of bidirectional communication among multiple users for visual applications in OCC & \cite{hossan2019human} \\\cline{1-3}
 	 & Introduce a new approach of LiDAR waveform decomposition and a new detection scheme for occluded pedestrian based on human Doppler distribution & \cite{qin2011toward, kwon2017detection} \\\cline{2-3}
 	 LiDAR & Proposed a novel speed estimation system sensing visible light variation of the vehicle’s headlamps under different road scenarios & \cite{abuella2019vildar} \\\cline{1-3}
 	 & Presented BER evaluation of a multi-wavelength OCDMA FSO system with shift inverse keying (SIK) balanced direct detection receiver and optical encoder taking into consideration of pointing error & \cite{islam2019effect} \\\cline{2-3}
 	 & Multi access interference (MAI) and crosstalk at the SIK receiver are assumed to assess SINR accompanied by BER varying system parameters such as code length, number of simultaneous user considering atmospheric turbulence & \cite{nazrul2019effect} \\\cline{2-3}
 	FSO & Analyzed adaptive acquisition system that perform better for the low probability of detection based on shotgun scanning approach in FSO & \cite{bashir2020adaptive} \\\cline{2-3}
 	 & A detector array based receiver is illustrated in deep space for FSO in order to minimize pointing loss and improvement of probability error under low SNR condition & \cite{bashir2019free} \\
 	 & Examined the feasibility of vertical backhaul/fronthaul framework where the traffic transport between the access and core networks via point-to-point FSO links under different weather conditions     &  \cite{alzenad2018fso}     \\\cline{1-3}
\end{tabular}}
\end{table}

\section{Contributions}

In this paper, we present a brief overview and extensive comparison of different optical wireless technologies over RF based systems. The comprehensive analysis of different OWC technologies is not the prime objective of this study. This article substantially covers the area of free space optical communication (FSO) pertaining different issues including the effective solution for the successful deployment of FSO systems. The application scenarios, principle of operations, transmitters, receivers, channel characterization, modulation techniques, network architecture of FSO and hybrid FSO systems highlighting the key advantages and disadvantages are clearly presented. For better illustration of FSO schemes, a comparative surveys among other OWC technologies are demonstrated by means of communication distance, data rate, and modulation techniques. In addition, different important hybrid schemes for instance RF/FSO, mmWave/FSO, acoustic/FSO, OCC/FSO and the research trends are clearly manifested in this paper. For better illustration, Table \ref{Table:5} presents the related surveys/reviews of FSO with a concise description of their major contributions. The key contributions of this survey paper can be summarized as follows.

\begin{itemize}

\item Brief overview of optical wireless technologies in perspective of different systems (e.g., 5G/B5G, IoT/IoE, underwater communications, etc.)

\item The detailed scope of FSO and hybrid FSO technologies, related works and the recent research trends are surveyed.    

\item FSO transceivers, channel modeling, QoS provisioning, radio over FSO (RoFSO) scheme, MIMO FSO systems, multi-user communications, FSO transmission in physical and TCP layer, link budget, reliability, high capacity backhaul networking, and next generation FSO systems are discussed.

\item The challenging issues encountered and the potential mitigation techniques as well as future research directions for FSO systems are extensively pointed out.

\end{itemize}

\begin{table}
\centering
\caption{Comparison of Related Surveys/Reviews on FSO Communications}\label{Table:5}
{\tabulinesep=1.3mm
\begin{tabular}{|m{12mm}|m{5mm}|m{35mm}|m{85mm}|}
 	\hline
 	Reference & Year & Journals/Book Chapters & Major Contribution \\\hline	
 	\cite{sevincer2013lightnets} & 2013 & IEEE Communications Surveys $\&$ Tutorials & Explore the VLC and FSO technologies and survey the potential of combined implications in a single field of study. Present different scenarios of limiting factors on system performance and functionality of light sources. \\\cline{1-4}
 	\cite{alkholidi2014free} & 2014  & Contemporary Issues in Wireless Communications & Review the applications of FSO, optical link budget, and empirical study of BER performance in a particular city considering few constraints  \\\cline{1-4}
 	\cite{khalighi2014survey} & 2014 & IEEE Communications Surveys $\&$ Tutorials & Survey FSO link modeling, modulation formats, spatial diversity, cooperative transmission and hybrid RF/FSO systems \\\cline{1-4}
 	\cite{son2017survey} & 2017 & Digital Communications and Networks & Demonstrate the characteristics of different classified subnetwork FSO systems and several design factors including channel modeling, network topology design, eye safety issue \\\cline{1-4}
 	\cite{kaushal2016optical} & 2017 & IEEE Communications Surveys $\&$ Tutorials & Survey numerous challenges encountered in FSO uplink/downlink space communications and inter-satellite transmission \\\cline{1-4}
 	\cite{mansour2017new} & 2017 & Optics and Lasers in Engineering & Overview the potential challenges in FSO communications and solutions including cooperative relay networks and aspects of three different channel coding  \\\cline{1-4}
 	\cite{chowdhury2018comparative} & 2018 & IEEE Access & Survey different wireless optical access technologies, and provide details of these technologies explicitly  \\\cline{1-4}
 	\cite{kaymak2018survey} & 2018 & IEEE Communications Surveys $\&$ Tutorials  & Review on pointing, acquisition, tracking, and (PAT) techniques used in FSO systems under line of sight (LOS) and non-LOS scenarios  \\\cline{1-4}
 	\cite{farooq2018survey} & 2018 & Optical and Wireless Technologies & Addressing the factors related to channel impairments in FSO applications \\\cline{1-4}
 	\cite{raj2019historical} & 2019 & IET Communications & Review the FSO channel limitations, temporal and spatial challenges, and possible mitigation techniques \\\cline{1-4}
 	\cite{hamza2018classification} & 2019 & IEEE Communications Surveys $\&$ Tutorials & Survey multi-level FSO network classifications compared to existing ones and different FSO applications including heterogeneous FSO networks \\\cline{1-4}
 	\cite{kumar2020survey} & 2020 & Intelligent Communication, Control and Devices  & Overview of challenges and possible mitigation techniques in FSO communications briefly \\\hline
	\multicolumn{3}{|c|} { This Paper} &    {Comprehensive survey dedicated to FSO wireless systems in comparison with other optical wireless technologies, discusses various issues related to wire range of FSO challenges, mitigation methods, reliability and link design aspects, extensive applications scenario incorporating next-generation functionalities} \\\cline{1-4}
\end{tabular}}
\end{table}	

The rest of the paper is organized as follows. Section \ref{4} presents the fundamentals of FSO communications, merits and demerits, wide range of application scenarios and the discussion on FSO transceiver. The reliability of including link budget is demonstrated in Section \ref{5}. The potential of mitigation techniques to address the technical challenges are extensively discussed in Section \ref{6}. Subsequent sections describes some attractive features of FSO technology pointing out on radio over FSO, mutiuser FSO, MIMO FSO, and the prospect of future generation FSO technology.  Furthermore, the summary of this survey and the research challenges, future research directions and lessons learned are presented in Section \ref{12} and \ref{13} respectively. Finally, Section \ref{14} concludes the review paper.

\section{Principle of free-space optical (FSO) communication}	\label{4}

Free space optical communication (FSO) refers to point-to-point wireless optical transmission through the unguided propagation media using infrared (IR) and visible light bands. In other words, FSO is a LOS technology that uses eye-safe laser beams that provide optical data communication wirelessly in the free space medium. FSO receivers are comprised of telescopic lenses that able to collect light stream and transmit digital information to the destination in the speed of Gbps range. The availability of broad optical spectrum extends the opportunity to tackle a massive volume of data capacity. FSO is a promising alternative to radio relay link as the light travels faster through air than glass. Generally, outdoor FSO links operate in the VL and UV bands, near IR and VL bands are used in indoor and underwater FSO communication \cite{hamza2018classification}. FSO operating at extremely high frequencies (i.e. short wavelength) make the photodetectors immune to multipath signal fluctuations whereas RF networks are highly susceptible to multipath fading. Multiple digital signals such as internet data, voice, image/video, computer files, etc. are first converted to light signals using optical transmitter, then modulated by a suitable scheme, traveling multiplexed signal through the FSO channel in optical domain, received the incoming signal using optical photodetector and finally transmitted demultiplexed signals towards the destination after necessary electronic switching.   

\subsection{Advantages and Disadvantages of FSO}

The transmission window for NIR wavelength ranging from 700 nm to 1600 nm, whereas, the wavelength for RF based system lies between 30 mm to 3 m. Therefore, optical wavelength is thousand times smaller than RF wavelength; this makes the operation bandwidth for FSO system in order of THz offers extremely high bandwidth than that of a conventional RF carrier. Typically, the allowable bandwidth for RF based communications can be around 20\% of the radio carrier frequency. On the other hand, even if the bandwidth is considered 1\% of optical carrier frequency ($10^{16}$ Hz), the allowable bandwidth is 100 THz implying enormous information carrying capacity over mmWave and microwave RF communication systems.

FSO systems are typically use very narrow spectrum laser beams as carrier signals that provides high speed data communication between two fixed nodes over distances up to few kilometers, inherent security (i.e., light is confined within a certain defined zone), large reuse factor and immunity to electromagnetic interference with other communication network or electronic equipment. Unlike radio waves, optical beam cannot penetrate walls, free from inter-cell interference and do not require frequency coordination. Hence, the same optical beam can reused for different purposes. Due to the broad optical carrier bandwidth, FSO scheme enables extreme level of data rates in compared to RF counterparts. Moreover, the coherent and low divergence nature of the optical beam allows a number of parallel FSO links installation at a given site. Experimental outcomes of OWC systems \cite{zhang20131} in terms of bit rate are tightly competing with generic fiber optic system. Some other few advantages of optical spectrum over the other frequency band, large unregulated bandwidth, very high SNR, ease of conjunction into the existing arrangement. Notably, FSO scheme is cheaper than conventional fiber optic link since the frequency band occupied by FSO technology over 300 GHz is unlicensed worldwide and hence requires no license fees. Optical wireless transmission like FSO is immune from side lobes and this invisible light does not cause any health hazards. Easy and quick deployment, lower installation and maintenance cost compared to the buried fiber grid connectivity makes the FSO systems more attractive. Unlike RF signals, optical components consume less power and exhibit ultra-low latency. 


The beam divergence is directly proportional to wavelength and follow inverse relationship to the aperture diameter. As a result, the optical beam is much narrower compared to the RF carrier which leads to increased intensity of the received signal for a given laser transmit power. Besides, the antenna size for FSO communications is incredibly lower because of lower optical wavelength to attain the same gain where the antenna gain varies inversely proportional to the square of the wavelength. Therefore, the power requirement, antenna diameter and mass is considerably lower than RF wireless systems. In addition to these benefits, the directivity and gain of FSO antennas are also remarkably higher due to smaller wavelength. However, longer optical wavelength wireless link is less affected by pointing induced fading \cite{kaushal2016optical}. Therefore, the choice of wavelength is an important design parameter by means of link performance as well as detector sensitivity. Furthermore, eye safe issue, economic factor and the availability of transceiver’s components greatly affect the selection of wavelength in FSO design. In a nutshell, the selection of operating wavelength depends on the trade-off between pointing errors and receiver sensitivity. 
Unlike RF signal, the optical beam cannot be detected by spectrum analyzers. Moreover, FSO transceiver are easy to deploy, lightweight, compact, easily expandable.

Despite the potential advantages of FSO communication, FSO networking is not yet mature as their RF counterpart because of some limitations. The performance of FSO links are intensively deteriorates the link reliability due to atmospheric turbulence effects such as heavy rain, fog, snow, cloud and pointing stability in wind. Some other limiting factors including beam dispersion, background light radiation (e.g., sunlight) and shadowing effect degrades the system performance pertaining to bit error rate (BER), SINR, ergodic capacity, outage probability. Uneven distribution of temperature on the earth, velocity variation of wind and atmospheric pressure due to irregular flow of wind causes atmospheric turbulence. Such atmospheric circulation affect the optical beam propagation along the FSO link both in tempo-spatial domain. Atmospheric turbulence causes phase shifts of the transmitted optical beams resulting in signal distortion is referred to optical aberrations. Scintillation is an intensity distortion of propagating light pulses caused by the atmospheric turbulence \cite{alimi2017toward}. Laser beam passing through scintillation experience irradiance fluctuation even in short distance communication. Moisture, aerosols, temperature and pressure variations generate refractive index deviations in air density known as eddies. However, the refractive index change of air causes wave front phase perturbation of propagating optical light. As a consequence, some other wave fronts having different phase causes an interference and amplitude variation. Therefore, the received optical signal fluctuating randomly and causes a distorting wave front of the isophase surface. Note that the optical laser beam will bends when the dimension of turbulence eddies are bigger than the beam radius. In addition, constructive and destructive interference due to arrival time variations of multiple wave fronts resulting temporal fluctuations of the light intensity at the receiver end. 

Pointing error/jitter is another noticeable parameter of FSO systems that can be defined as the digression between the expected antenna position and its actual position. In other words, mechanical misalignment of FSO transceivers and mechanical vibration causes an error of tracking systems. Owing to the strong nature of directional beam of FSO transmitter, the problem involves in pointing, acquisition and tracking (PAT) technique because each optical transceiver must be precisely aligned simultaneously at the point of communication. Any unwanted objects in LOS links break the communication between transmitter and receiver. Clouds is another important limiting factor which is collection of water droplets (i.e. excess vapor condenses of tiny particles) and crystals suspended in air. Clouds imposes the attenuation, beam scattering and absorption during the laser beam travelling through the medium. On the other hand, non-selective scattering due to rainfall and heavy fog is one of the major impairment of FSO link reliability. A larger size of rain droplets and a high dense fog causes reflection and refraction of light beam \cite{jahid2015bit}.

\subsection{FSO Classifications}

With the augmentation of wireless devices and diverse type of multimedia services impose a massive demand on RF spectrum where most sub-bands are licensed. As a consequence, RF band is outstripping supply progressively which is also fundamentally restricted in capacity and cost; the paradigm of shifting the upper portions of the electromagnetic spectrum for wireless communications has become the paramount concern. Over the past few decades, the applications of FSO is remained confined to military short range applications and inter-satellite links in space applications \cite{rabinovich2010free}. With the introduction of FSO communications products and wireless optical terrestrial links offers a variety range of services to cope with ever-growing demands of higher data rates \cite{komine2004fundamental, grubor2008broadband, le20101, elgala2011indoor, borah2012review}. Development of efficient wireless technology such as FSO is essential for heterogeneous building communication networks and can be deployed in different applications ranging from outdoor inter-building connections to satellite links. 

According to the geographical location and link distance, FSO technology can be employed in four different scenarios: atmosphere, indoor, space and underwater. However, atmospheric environment can be considered as outdoor terrestrial FSO communication, FSO home networking include the indoor communications and space scenario is classified under FSO space communication. Terrestrial FSO networks establish long-distance point-to-point outdoor optical communications following LOS mechanism over atmospheric turbulent channels. This FSO communication has offer a promising potential for wireless telecommunications especially for broadband internet access. Terrestrial FSO is emerging solution for last-mile access. Although fibre connectivity has been widely used for fibre to the home (FTTH) service, but still there are many end users are not connected to FFTH service due to geographical limitations. This technology enables high bandwidth connection for remote users over a long communication distance. In other words, terrestrial FSO technology establish a bridge for geographically separated premises, for instance, inter-building or community-to-communality communications without incorporating optical fibre cable. Most importantly, the integration of terrestrial FSO with wireless RF networks minimize the capacity and scalability problems with greater level of QoS. Note that the spectral efficiency and fairness of RF wireless networks severely attenuated when the hop length and hop count is large.
 
On the other hand, FSO home networking (FSO-HN) are commonly used for indoor broadband wireless communications either in house, building, or office. FSO-HN are used in a short-range communication ranging from few meters to tens of meter. The hardware is normally portable, inexpensive, lightweight and facilitates eye safety issue. Multiple small cell terminals connected via short-range optical wireless links are installed in the building. Unlike radio signal, optical light cannot penetrate walls, hence, each small cell is confined to a particular room and all the terminals are connected to high-speed backbone infrastructure. The optical beams can be reused as each terminal is free from mutual interference from the neighboring cells. Compared to non-LOS link, LOS FSO technology achieve high throughput and extremely low BER due to better beam steering mechanism and power budget. A diffused light source is commonly used in indoor place to disperse light beams in non-LOS links. Taking the full advantage of multipath propagations caused by the different objects such as ceiling, furniture, walls, and floor, diffused links (or non-LOS links) exhibit more robust performance \cite{son2017survey}. However, a diffused link support lower bit rate in comparison with LOS links and there exist a trade-off between link reliability and network capacity.  

FSO space links can be further categorized into inter-orbital links, inter-satellite links and deep space links. FSO technology can be used as a global space backbone for satellite communications with high quality of data services. With the augmentation of FSO technique in downlink space communications (satellite-to-earth), the end users can avail the wide range of benefits regardless the receiver is in moving (e.g. airplane, cruise shift, moving vehicles, etc.) or stationary. The limitation of cellular networks coverage in the remote areas can be overcome thanks to broadband FSO-based satellite communications. Recently, acoustic communications in underwater environments is thoroughly investigated aiming to ensure reliability considering all the limiting factors.

Based on the link coverage, the classification of FSO can be summarized as:

\begin{itemize}

\item \textit{Ultra-short distance}: integrated chip-to-chip communication in nm distance of multi-chip packages \cite{kachris2012survey, taubenblatt2011optical}

\item \textit{Short-distance}: wireless personal area networks and underwater wireless communications \cite{gabriel2013monte, kaushal2016underwater, saeed2019underwater}

\item \textit{Medium-distance}: indoor visible light communication technology for wireless local area networks (WLANs) \cite{matheus2019visible, rehman2019visible}, vehicle-to-vehicle (V2V) and V2X infrastructures \cite{sichitiu2008inter, lee2012evaluation}

\item \textit{Long-distance}: inter-building communications \cite{malik2015free, sharma2019effect, khalighi2014survey}

\item \textit{Ultra-long distance}: satellite links (e.g., inter-satellite, satellite-ground, satellite-airplane) \cite{chan2003optical, kaymak2018survey, ai2019physical, gong2019network}

\end{itemize}

\subsection{Applications of FSO}

\begin{figure}
  \centering
  \includegraphics[width=0.9\columnwidth, height = 120mm]{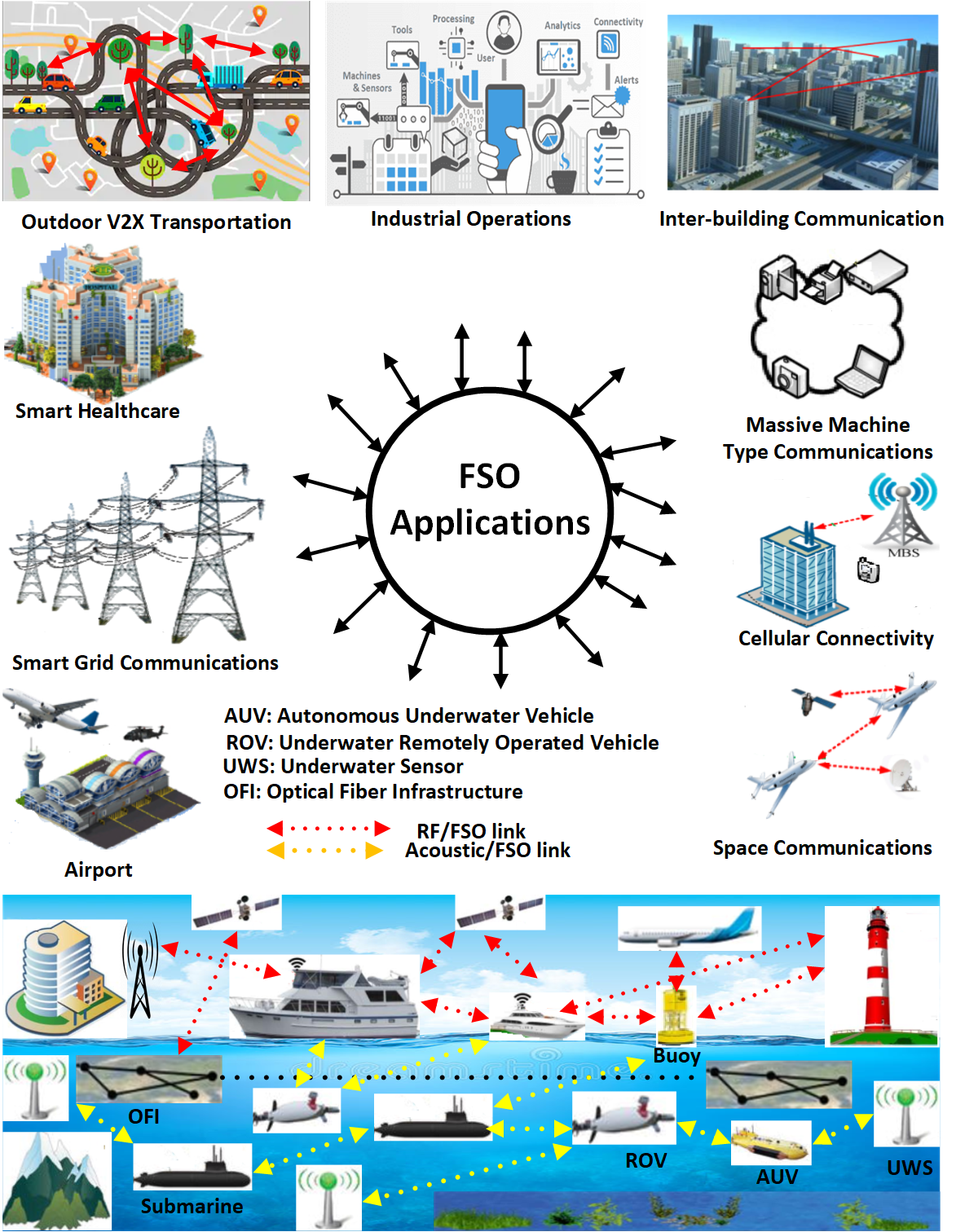}\\
  \caption{A few applications of FSO systems.}\label{Fig:1}
\end{figure}

FSO systems have primarily attracted as it provides efficient solution of the bottlenecks of last mile problem filled up the gap between optical fiber infrastructure and the destination users. To fully utilize the existing setup, telecom operators are investing considerable investments to fiber backbone and expansion of FSO links as well accompanied by the tremendous growth at the network boundary where end subscribers easily get access with high speed systems. FSO schemes can be promising alternative in cases where buried optical fiber connectivity is costly and/or infeasible. A few applications of FSO systems are illustrated in Fig. \ref{Fig:1}. Some attractive applications of FSO systems are briefly described in the following.

\begin{itemize}

\item \textit{Inter-building (e.g., Campus/Enterprise/Residential) connectivity}: 

Buildings in different infrastructures (i.e., school campus, office, corporations) are experiencing heterogeneous traffic demand (e.g., voice, internet data, fax, and multimedia services) overwhelmingly. FSO technology can setup a path among multiple buildings in certain areas supporting extreme level of data speed in a cheaper way without installing dedicated fiber links. FSO can be used as high speed data interconnection bridge among geographically separated buildings (i.e. enterprise connectivity) for example, ship-to-ship, and community to community communications.  

\item \textit{Video surveillance}:
  
Wireless video monitoring are widely employed in military applications, commercial and public safety which provides high throughput than conventional ones. FSO systems exhibits a good choice of alternative to allow ultra-quality of video transmission. 

\item	\textit{Backhaul for cellular networks}: 

A communication among cellular base stations (i.e., base transceiver station, controller section and mobile switching center) involves wireline connections by means of T1/E1 lines and microwave communications for wireless link. The deployment of FSO technology offers a backup path for extended bandwidth-intensive high speed cellular services. As a consequence, FSO can be deployed for SONET rings connectivity and core infrastructure in metropolitan area. 

\item \textit{Broadcasting and disaster monitoring:} 

Nowadays live events broadcasting such as sports, high definition (HD) services, television crime and ceremonies report, covering critical news in remote zones, signals from cameras are needed to be sent a central node. FSO system is capable of satisfying the high quality of transmission (throughput demand) among different nodes to end users. On the other hand, temporary FSO links can be readily installed in disaster situations to monitor terrorist attack, emergency situations and natural disasters. Besides, FSO links can be readily deployed in the area where no wireline connections are not feasible.

\item \textit{Security:} 

Cyber security and computational security under the realization of quantum computers world attaining deep attention for example, transfer secure electronic money transfer. Quantum cryptography provides an unconditional security which are generally connected with fiber optic infrastructure. Unwanted users cannot hack the information, since FSO signal cannot penetrate an obstacle. 

\item  \textit{Last mile solution:} 

Nowadays, a majority of the buildings are connected via coaxial cable and fiber optic cable, there are still many end-users do not have access to the optical fiber infrastructure (FTTH). FSO communication can be considered as a paramount solution for remote end users (e.g., users in rural areas) with high speed connection of Gbps range over a large link distance. FSO can provide instant service to fiber optic users during laid off their existing fiber connectivity. Moreover, broadband internet services can be readily provided in isolated area via FSO link where it is very hard to achieve conventional access technologies.  

\item \textit{IoT and 5G:}

IoT extends the widespread internet connectivity beyond the conventional multimedia devices enabling to exchange information between themselves and external environment. The dense deployment of IoT devices demand high bandwidth, high bit rate, security, low latency, and electromagnetic interference free data transmission. The implementation of 5G technology in IoT, high speed FSO links extends the coverage area and the overall system performance by means of spectral efficiency and robust data transmission capability \cite{dhasarathan2019development}. In other words, the integration of FSO technology with IoT over 5G connectivity can satisfy the requirement to realize global IoT. 

\item  \textit {Satellite communications:} 

FSO technology can be effectively deployed in space applications together with radio astromy and remote sensing. FSO transmission scheme provides power efficient large distance inter-satellite orbital links with better receive sensitivity. Orbital angular momentum (OAM) beams have the intense level of data carrying capability providing an enhanced system capacity and spectral efficiency. Therefore, FSO is a preferred choice for attaining ultra information carrying capacity using OAM beams in deep space communications.

\item \textit {High-capacity backhaul network:}

A backhaul network establish a bridge between access network and core network. A high-speed backhaul connectivity is a compelling part while exchanging a large volume of data information between access network (AN) and core network (CN) in 5G/B5G or IoT perspectives. Densification of cellular networks by means of mass deployment of small cells to increase frequency utilization as well as system capacity. The deployment of additional infrastructure in the access networks are considered in densely populated areas where an intensive level of traffic volume is generated. Furthermore, the shorter user-base station distance ensure the greater level of attainable throughput due to lower multipath fading in 5G/B5G networks. To address the network densification issue, the facility of high-speed FSO connectivity guarantee the backhaul densification. In addition, a high capacity backhaul is essential for long-range applications such as inter-building communications, inter-city connectivity and satellite-to-satellite links. Unless the high capacity backhaul network, the entire communication will be worthless even though the AN or CN supports the Gbps communication equipment, hence, leads to bottleneck in the system. Current backhaul networks are equipped with dedicated fiber optic infrastructure, coaxial cable, copper, microwave and mmWave links, and sometimes satellite paths \cite{jaber20165g, artiga2018shared}. Together with the wired optical fiber networks, FSO system offers an excellent features to solve low-capacity backhauling issue effectively especially in long distance outdoor backhaul link. FSO enabled high speed backhaul networks provides the connectivity in far distance areas for example, underwater, space communication, isolated island, etc. Moreover, the latency is less than one millisecond and achievable data rate of 40 Gbps for FSO point-to-point communication technology. In summary, FSO systems has the potential to handle high traffic volume, allowing massive telecommunications and IoT devices, least power consumption, cost effectiveness, and greater spectral efficiency in the backhaul network platform.

It is widely acceptable that the throughput of optical fiber cable is the maximum among all other related technologies till now. Since both fiber optic technology and FSO systems use the similar type of optical transmitter and receiver, so attaining similar throughput performance will possible in the forthcoming future. Unlike NLOS link, FSO LOS link achieves higher system capacity because of the absence of multipath propagation delays and effective power link budget. On the other hand, a diffused light sources is used for the NLOS links to disperse a light beam for multipath propagation caused by the reflections of different obstacles. Thus, NLOS links are more robust when dealing with objects along communication paths. However, increasing the FSO backhaul capacity is a challenging task with the enormous growth of traffic volume.

\item Device-to-device (D2D), machine-to-machine (M2M), vehicle-to-infrastructure (V2X), multipoint-to-multipoint, end-to-end communication mechanisms can be applied in healthcare, railway stations, shopping malls and industry. Scalability limitations of RF-based networks for instance, fairness and throughput are limited for the large hop number, can be easily overcome with high speed FSO network over existing wireless mesh connections.

\end{itemize}

\subsection{FSO in IoT, 5G and B5G}

\begin{figure}
  \centering
  \includegraphics[width=0.9\columnwidth, height = 110mm]{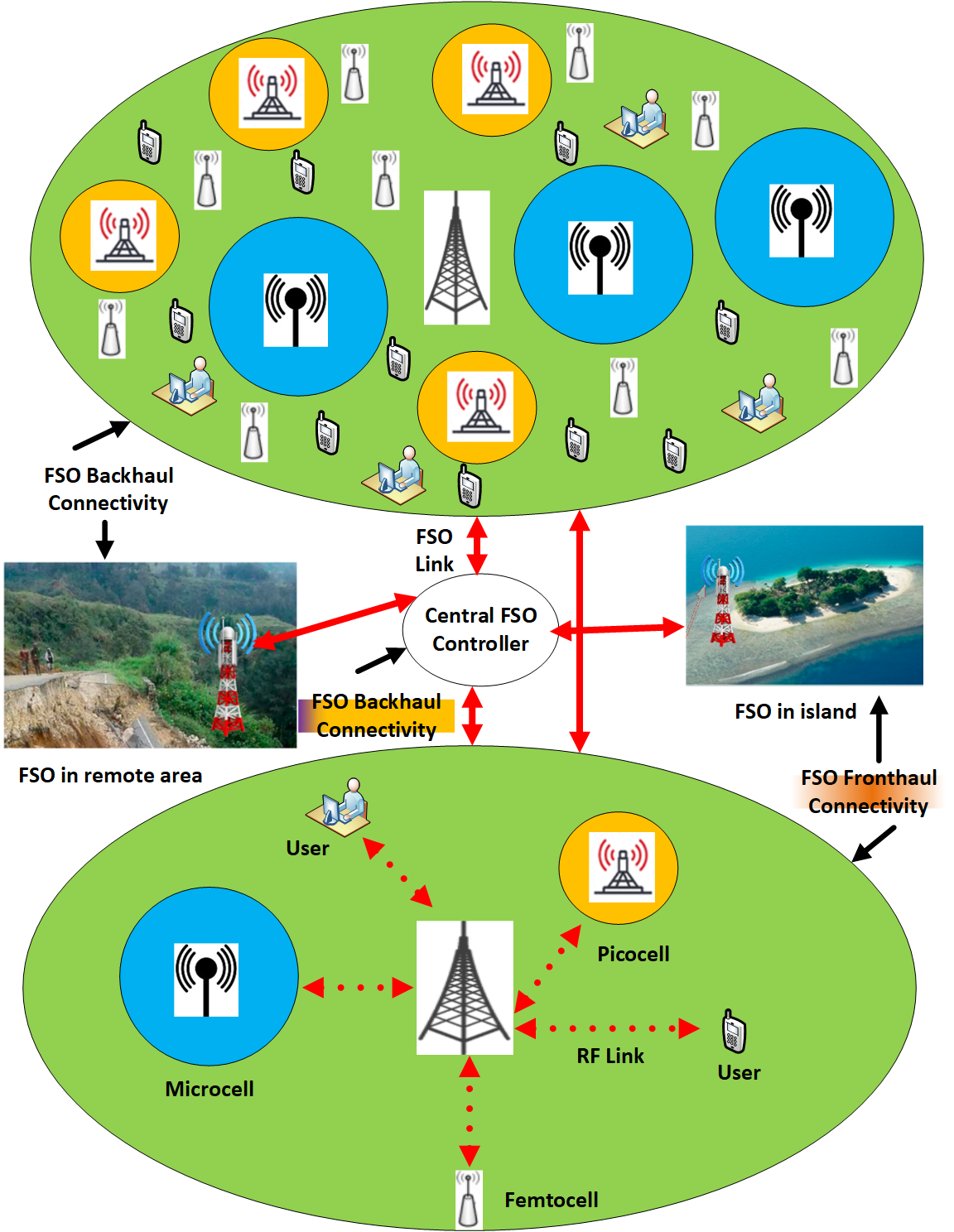}\\
  \caption{FSO Backhaul Connectivity for Cellular Networks.}\label{Fig:2}
\end{figure}

\begin{figure}
  \centering
  \includegraphics[width=0.9\columnwidth, height = 110mm]{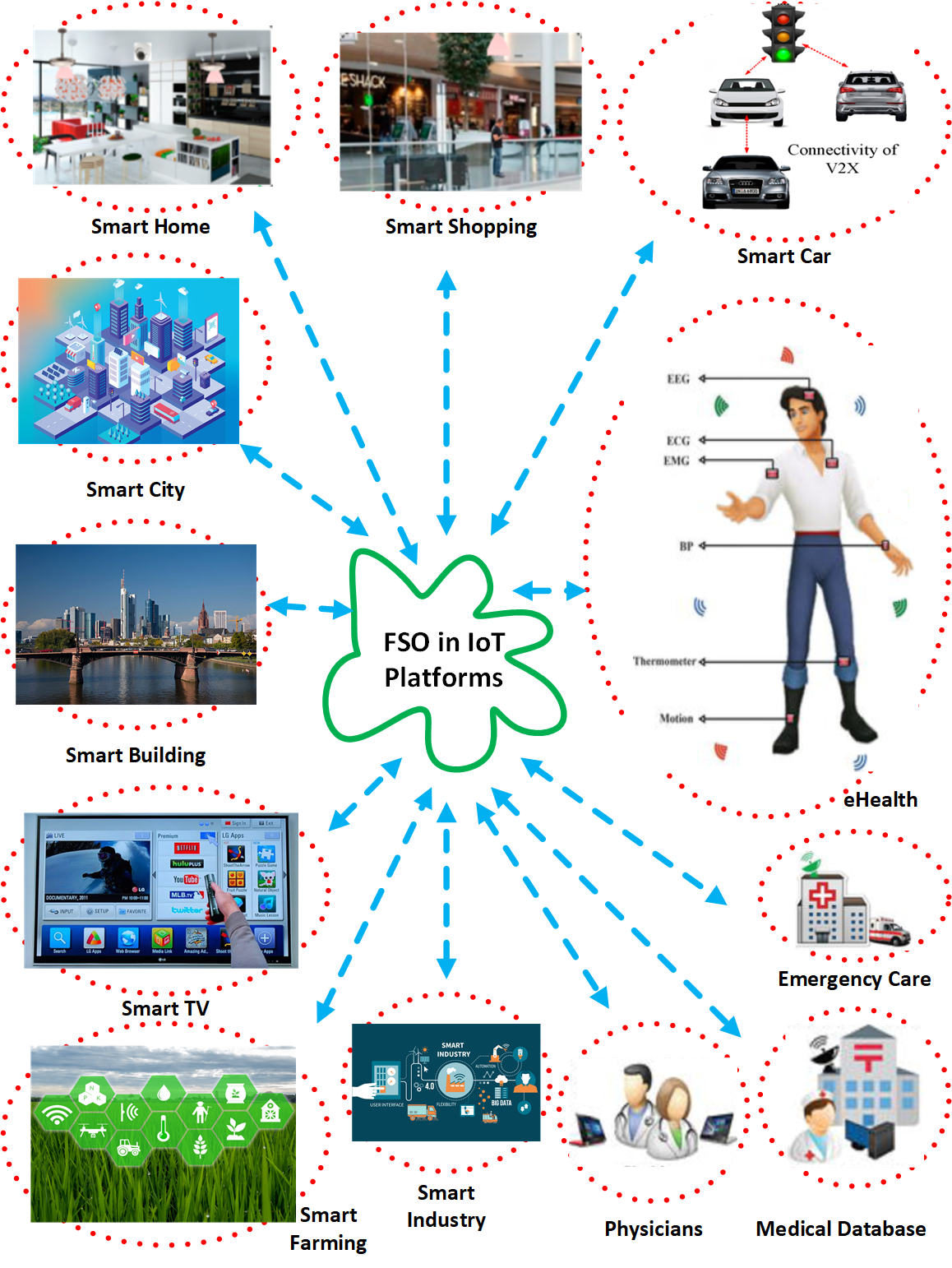}\\
  \caption{FSO in IoT Platform.}\label{Fig:3}
\end{figure}

5G communication offers enormous smart device connectivity, ultra-high throughput, extremely low latency, ultra-high encryption, incredibly low power consumption, and ultra-level quality of service \cite{ghosh20195g, habibi2019comprehensive}. 5G network and beyond are now widely developed including nested small cells in heterogeneous (i.e. a combination of macrocell, microcell, picocell and femtocell) configuration in order to increase network capacity in an effective way with desired quality of experience (QoE) \cite{khan2019survey}. To ensure high QoE under an intensive data demand and wide range of large device connectivity, 5G/B5G networks are anticipated to ultra-dense HetNet deployment. 5G/B5G networks has the capability to support up to 10 Gbps downlink data rate which will be ten to 100 times bigger than 4G. The number of connected electronic devices is hundred times higher than 4G LTE networks results in thousand times volume data \cite{ijaz2016enabling}. 5G technology consume 90\% less power compared to 4G networks and provide extremely low latency in fraction of millisecond. B5G networks are expected to provide four times spectral efficiency over 4G networks and lower network deployment cost. Offloading massive volume of data to indoor small cells from the macrocell is another significant characteristic of 5G/B5G networks \cite{khan2019survey, 5g}. Fig. \ref{Fig:2} demonstrates the implications of FSO transmission in 5G/B5G wireless cellular networks and Fig. \ref{Fig:3} presents the application of FSO in massive IoT platforms. 
 
FSO support high speed bit rate 5G communication together with high-capacity backhauling and the massive connectivity of IoT. Energy consumption is a key concern for designing a large scale 5G cellular and IoT infrastructure in where FSO helps to curtail additional energy expenditure in comparison to tradition RF systems \cite{chettri2019comprehensive}. Some other demands of IoT systems are low implementation cost, high level of security, large energy efficiency, cost effective and support a huge number of smart devices. FSO technology is capable to support massive smart device connection in IoT paradigm through low power LED or LD technologies. A reliable connection by means of strong encryption can be assured using FSO technology \cite{chowdhury2019role}. 

\subsection{FSO in Space Communications}

\begin{figure}
  \centering
  \includegraphics[width=0.9\columnwidth, height = 100mm]{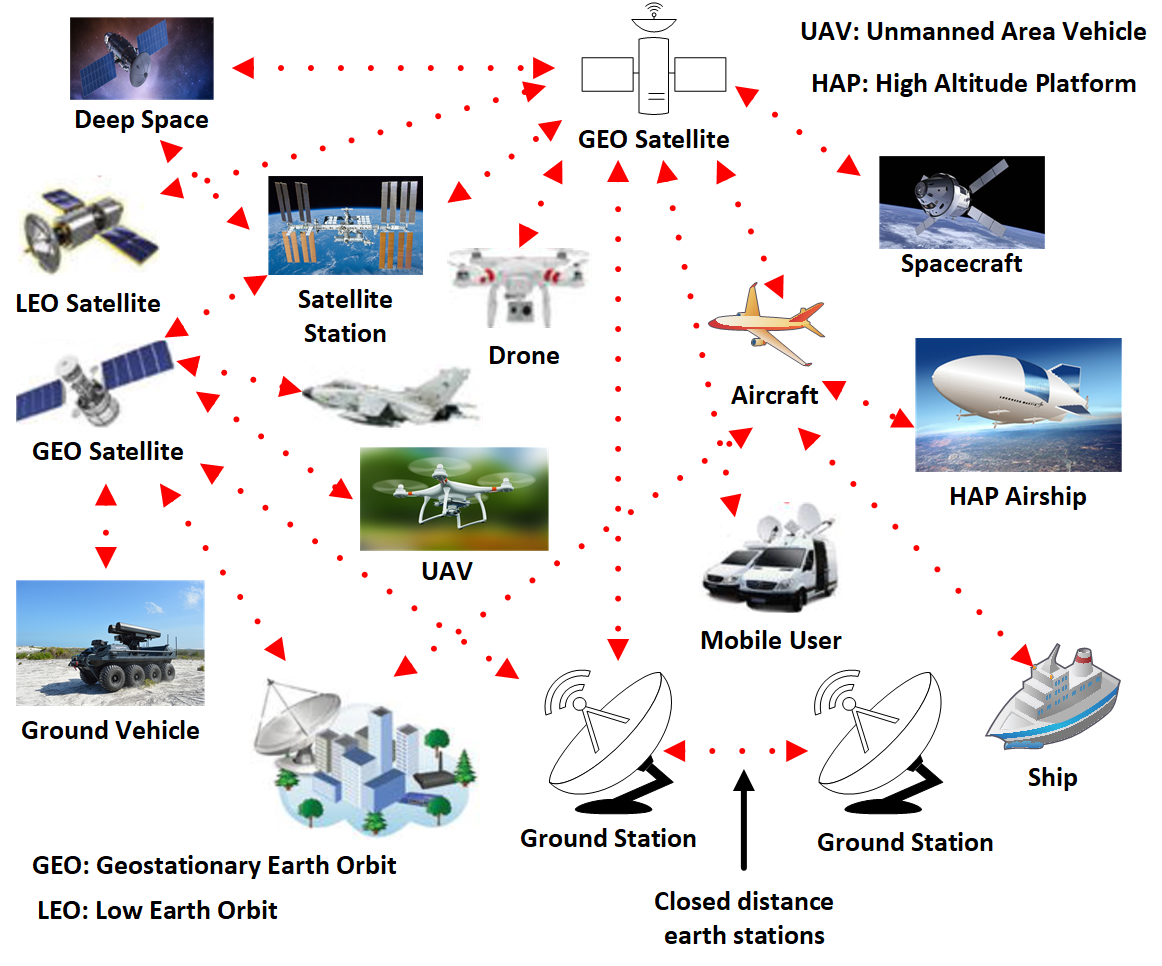}\\
  \caption{FSO in Space Communications.}\label{Fig:4}
\end{figure}

With the advancement of space and FSO technology, satellite communications including uplink (i.e. ground-to-satellite), downlink (i.e. satellite-to-ground), inter-satellite links, and deep space communications incorporating FSO have received considerable attention among researchers and industries \cite{kaushal2016optical, aveta2019free}. The losses incurred though the atmosphere is much higher for FSO uplink compared to the downlink communications as the optical beams are spread and distorted at the ground-based point. The variation of temperature, concentration of aerosol particles, and encountered air pressure with altitude affect the laser beam propagation in the passage through atmosphere. This losses can be defined as absorption (i.e. absorbers may be the water molecules, $CO_{2}$ gas) and scattering loss. Note that absorption loss and scattering loss (dB/km) is a function of wavelength. The scattering loss is more noticeable when wavelength is below 1 um especially ultraviolet or visible range for FSO communications. Thus, the selection of transmission window (e.g, ultra-short IR, short IR) is a crucial FSO design parameter. Rayleigh scattering is produced when the atmospheric particles are comparatively smaller to the optical wavelength and Mie scattering is generated for vice-versa \cite{kaushal2016optical}. Air molecules and haze particles typically causes Rayleigh scattering, in contrary, aerosol, rain, snow, fog particles are the primary contributors for Mie scattering. The application of FSO in space communications are explicitly presented in Fig. \ref{Fig:4}.

Under the presence of dense fog (visibility is below 50m), the signal losses can be higher than 350 dB/km \cite{nadeem2010continental}. In such condition, high power lasers operating at the third transmission window help to uplift the link reliability. It is obvious that the heavy rain droplets leads to higher attenuation loss. According to \cite{suriza2013proposed}, the heavy rain (25 mm/hr) incur attenuation of ten times than the light rain droplets (2.5 mm/hr) ranging from 1 dB/km to 10 dB/km for 1500 nm wavelength. The hybrid RF/FSO systems is a suitable choice to enhance the link availability under this condition. On the other hand, the attenuation due to snow particles are lies between rain droplets and fog particles. Atmospheric turbulence induced particles called eddies with different size and refractive indices generated beam wandering effect when the size of eddies are greater than beam radius. In contrast, smaller eddies size in comparison to the beam dimension also resulting in tempo-spatial irradiance fluctuations of the transmitted signal at the receiver end \cite{kaushal2011experimental}. Turbulent cells also degrade the coherence properties of the optical beam and hence, create interference.  

The beam divergence is an important factor of FSO uplink design in space communications. A beam divergence is incurred before the receiver aperture due to diffraction of light. Some fraction of laser beams are not collected at the receiver hence, causes beam divergence loss or geometrical loss. This loss considerably increases with the transmission distance regardless the diameter of receiver aperture. In such case, a narrow beam laser source is preferable and receive diversity techniques can be employed. It is worth to mention that sky radiance can be happened both daytime and night time which is driven by the altitude of receiver, cloud thickness, atmospheric particles concentration and solar irradiance. In particular, the magnitude of sky radiance is primarily depends on the geometrical position of sun, earth and the FSO receiver. Pointing loss may happen due to small vibration of satellite, slight stress on mechanical or electronic devices, or jitter platform. A little misalignment between the transceivers could cause FSO link failure, thus, it is very essential to control LOS directivity with narrow beam divergence \cite{ma2012influence}. In addition, beam wandering effect may shift transmitted beam from the propagation path that cause pointing loss. In any cases pointing error can remarkable reduce the received optical power as well as SINR resulting in high BER and increase the chances of FSO link failure. Proper care must be taken of vibration free FSO devices installation and maintain adequate bandwidth control to compensate residual jitter. Pointing error is more apparent for lower earth orbital (LEO) satellite than GEO satellite \cite{ma2012influence}. Pointing error considerably degrades the system performance in terms of BER and outage particularly at visible wavelength. Furthermore, noise in the tracking sensors and mechanical vibration of satellite causes tracking error. A compensation technique to address pointing and tracking errors for the uplink FSO space communications is presented in literature \cite{viswanath2015analysis}.

The presence of thick cloud disrupt the signals or occasionally completely block the link regardless uplink or downlink communications. These intermediate disruptions can last from fraction of minutes to few hours based on the geolocation and season. The opaque clouds could lead severe attenuation as big as tens of dB. The tempo-spatial diversity deployment can combat the optical beam loss due to attenuation. Owing to effect of atmospheric turbulence, the optical signals fluctuating at the focal plane of photodetector or leads to spot motion. This effect is known as fluctuation of arrival angle that can be mitigated by implementing fast beam steering mirror or adaptive optics. The aforementioned limiting factors for FSO space communications can be handle by proper choice of system design parameters for instance, operating wavelength, beam divergence, intensity of transmitted power, transceiver field of view (FOV).

Inter-satellite space FSO links are not affected by weather and atmospheric limitations as the satellite obits are far distant of atmosphere layer; however, they are restricted by other key challenges such as acquisition and tracking (AT), stability of satellite station, Doppler shift, and background radiations. Coherent detection method is more appropriate over direct detection for inter-satellite FSO communications since it covers longer link distance. Cooperative frequency tuning between the transmitted signal and local oscillator (LO) is adopted to combat the impact of Doppler shift. Coherent tacking scheme can be implemented to combat AT including background radiations. Coherent technique either homodyne or heterodyne maintain power efficiency of the link and offer good receiver sensitivity with lower level of power penalty \cite{zech2015lct}. The concept of detector arrays contemplating in deep space FSO communications helps to collect the light signals even if misalignment exist between transceivers \cite{bashir2019free}. Bashir \cite{bashir2019free} pointed out that a receiver detector array outperforms a lot better in terms of probability of bit error, pointing error and SINR than a single detector system. 

\subsection{FSO Transceiver}

FSO communication can be broadly categorized into three stages: optical transmission through the atmosphere that follows the Beer-Lambert’s law, channel link where different noises and turbulences are accumulated, and receiver section equipped with electronic devices to process the received signal. 

\subsubsection{FSO Transmitter}

Optical transmitter comprised of a light source, a modulator, optical amplifier (optional) and beam forming mechanism. The incoming information data bits are optionally encoded before modulation. The modulated light beams (optical intensity) are boosted using optical pre-amplifier. Then, the laser beams are collected and refocused via beam forming optics before transmission. The most commonly used optical sources such as light emitting diodes (LEDs) and laser diodes (LDs) offer some distinctive advantages and disadvantages of FSO systems based on their applications. 

\textit{Advantages:}

With the recent development of solid-state lightning, high intensity LEDs are replacing fluorescent and incandescent lamps. Several key advantages provided by LEDs are greater energy efficiency, longer lifecycle, lower heat production, minimal usage of toxic materials for manufacturing, and better color interpretation \cite{pathak2015visible}. In addition, LEDs are capable of faster switching in different light intensities and this functionality makes the LEDs to use as high-speed OWC transmitter and highly efficient optical source simultaneously. On the other hand, laser sources are highly monochromatic and produced a single wavelength output. LDs act as a coherent light sources in where all the emitted light spectrum follow the same direction, same manner and at the same time. The coherent characteristics (i.e. concentrated and intensive unidirectional nature) of LDs travel longer distance and provide a high data rate transmission with less interference compared to LED sources.

\textit{Disadvantages:}

Incoherent nature of low power light emission is the prime limitation of LED light source which can cause significant interference with natural and artificial light sources. Because of narrow aperture of LD, only point-to-point communication take place. In some cases LDs are not a suitable choice for mobile communication due to thermal effect of radiation in human body, cost and color mixing complication \cite{chowdhury2018comparative}.

However, semiconductor laser diode is typically used in the FSO systems due to intense optical power operating, although high power LEDs with beam collimators are commercially available in the market. In addition, LDs provide wide modulation bandwidth over a wide range of temperature. Another important factor for laser transmitter is eye safety issue. On the other hand, LEDs are preferred for indoor applications with low data rate. LEDs are extended sources that cover broad emitters and can be operated safely even at high power. However, VCSEL sources are commonly used for first transmission window operated on 850 nm, and Fabry-Perot and distributed feedback (DFB) lasers ae used for operation at third WDM transmission window of 1550 nm \cite{khalighi2014survey}. In order to ensure eye safety from radiation exposure, LDs are allowed for transmission at 1550 nm that are considered for long-range communications. Another wavelength of 10 µm has been recently considered in FSO systems that have improved transmission characteristic through deep fog \cite{whitepapaer2009}. Apart from this, currently UVs are also considered for FSO communication owing to robust feature against pointing errors and beam blockage. UV spectra exhibits lower sensitivity to sunlight and other background interference \cite{xu2008ultraviolet}. 

\subsubsection{FSO Receiver}

At the optical receiver, the radiated optical signal in terms of photons are collected optically and the photodetector (PD) converts the photons into electrical current by means of trans-impedance circuit. The FSO receiver handles the detected electrical current through electronic devices and recover the original transmitted data signals. In other words, the converted optical to electrical signal is then passed through a low-pass filter (LPF) in order to restrict background noise and thermal noise levels. Positive-intrinsic-negative (PiN) PD and avalanche photodetector (APD) are widely used in FSO systems. PiN PDs are low cost, can operate in low-bias and tolerate wide range temperature fluctuations that are normally used for low data rate and short range FSO links \cite{hamza2018classification}. The performance of PiN PD is limited by the thermal noise. Compared to PIN PDs, APDs have high multiplication gain (thanks to the process of impact ionization) as well as improved SNR as it operating high reverse bias and hence, APDs are preferred choice for high data rate and long distance FSO systems. Although APD shows superior performance, they are expensive, high power consumption, and avalanche gain is temperature sensitive \cite{khalighi2014survey}.  

Solid state devices having good quantum efficiency are widely used in commercial FSO systems operating in commonly available wavelengths such as 850 nm, 1550 nm. Si PDs are suitable for operation at shorter wavelength around 850 nm, whereas InGaAs PDs have maximum sensitivity in terms of very low transit time, rapid response detectors around at 1550 nm. With the recent advancement of graphene materials, plasmonic nanomaterials, two dimensional materials, quantum dots nanoparticles facilitates to develop ultra-speed photodetectors over a wide range of wavelengths \cite{liu2014graphene, ferrante2018raman}.

The deployment of optical amplifiers (OAs) in the long distance multi-hop FSO communication has been proposed to improve link performance \cite{aladeloba2012improved}. Erbium doped fiber amplifier (EDFA) is a preferred choice over semiconductor optical amplifier (SOA) operating at 1550 nm wavelength. However, the optical amplifier generates amplified spontaneous emission (ASE) noise which can deteriorates the receiver performance in terms of achievable SNR. Nevertheless, OAs has the potential to cope with electronic noise and reduce the scintillation effect in the case of weak turbulence \cite{abtahi2006suppression}. 

Different types of associated noises at the receiver comprises of dark current noise, shot noise (also called quantum noise) and thermal noise. The PD dark current noise can be ignore most of the practical cases. Another noise at the receiver causes photocurrent fluctuations called laser relative intensity noise (RIN) results from the instability of LDs intensity. Likewise PD dark current, RIN imposes a negligible impact on the receiver SNR performance \cite{xu2011impact}. Primarily the two main noise sources shot noise and thermal noise affecting the receiver performance if the background illumination is considered negligible. A PIN PD is normally thermal noise limited. In contrast, APD based receiver sensitivity depends either of the noise in accordance with impact ionization gain and load resistance. However, the thermal noise is a function of load resistance that originates from the electronic circuitry, which can be modeled as a zero mean Gaussian random noise. In contrary, shot noise arises due to random fluctuations of electrical currents flowing in PD or background radiations which can be modeled by Poisson process \cite{bhatnagar2016performance}. However, the shot noise can be approximately modeled as a Gaussian process if the number of absorbed photons are comparatively large. In most of the FSO applications, the shot noise distribution can be approximated by a Gaussian process \cite{bayaki2009performance}.  
 
\section{Reliability of FSO communication systems}	\label{5}

A reliability issues is an important characteristic for any sort of communication systems. The FSO systems assure a high level of SINR both for indoor users and long-distance outdoor communications addressing associated shortcomings. Besides, FSO networks opens the opportunity of provisioning additional tier network which inevitably enhances the reliability of a communication systems. Several limiting factors such as scattering, absorption, poor atmospheric visibility variations interrupt the high speed data transmission through the wireless optical link that deteriorate the FSO system performance and reliability as well. Atmospheric turbulence is the key source for link reliability degradation. The installed FSO systems should retain to operate under severe weather conditions i.e., under the influence of beam diversity, multipath fading and received signal fluctuations due to atmospheric turbulence. The tight beam propagation with a confined divergence of few mrad and the field of view at the receiving end temporarily causes poor link connectivity. As a result, narrow beam property cause link loss or misalignment issues and thus pointing, acquisition, tracking (PAT) mechanism is essential. Under link budget design, the received optical power can be easily estimated in a fibre optic communication, whereas, it is unpredictable, complex and inhomogeneous in air because of spatial distribution of FSO channel and atmospheric fluctuations. Since optical propagation through space is severely affected by atmospheric conditions, resulting in signal loss or phase shifting/fluctuations of signal intensity, thereby limiting the SINR performance and link availability as well with increased BER. Fog prominently deteriorates visible and NIR radiations and the greater level of humidity percentage in air also lessen the optical signal wave front quality. Scintillation effect is more apparent for long transmission distance that degrade the overall reliability of FSO system. On the other hand, atmospheric molecular absorption is a function of optical wavelength, attenuates the signal particularly in the higher wavelength region \cite{majumdar2014advanced}. As a consequence, the choice of operating optical wavelength consistent with the on-site atmospheric conditions is the key design issue for successful deployment of FSO devices. However, FSO devices are place in such a position where the probability of disturbance occurrence much lower and the weather conditions are better on average. It is important to taking into account of point-to-point delay, packet loss, jitter and fairness while designing protocols and algorithms in FSO systems. For example, the existing congestion control mechanism in TCP could result inferior throughput performance due to incompatibility in FSO networks and hence more adaptive protocols should be developed.  

To model the long-term reliable FSO data link design, the average visibility information should be surveyed based on the statistical behavior of atmospheric conditions at a given geometrical location. After these analyses, the link budget can be expressed as following \cite{raj2019historical}

\begin{equation}
R_{LM} = P_{TX} - S_{RX} - L_{P}		\label{reliability}
\end{equation}

where $P_{TX}$ represents transmit power, $S_{Rx}$ denotes the receiver sensitivity, and propagation loss is indicated by $L_{P}$. Equation \ref{reliability} gives basic link margin (LM) to express the reliability of optical data link. The LM can also be defined by \cite{katsilieris2017accurate}

\begin{equation}
R_{LM} = \frac{P_{RX}}{P_{RX}^{'}}
\end{equation}

where $P_{RX}$ represents the available receive power and $P_{RX}^{'}$ identifies the received power required to maintain a specified BER at the given transmission rate. $R_{LM}$ defines the loss margin required to compensate the turbulence effects at a given link coverage. For the telecommunication purpose, FSO networks must meet the high data rate requirements as much as possible which is equivalent to five minutes down time per year (link reliability around 99.99\%) \cite{majumdar2014advanced}. For example, if laser transmit power $P_{TX}$ = 30nW, photodetector sensitivity $S_{RX}$ = 25nW, misalignment loss $L_{a}$ = 3dB, optical loss $L_{o}$ = 4dB, then the estimated $R_{LM}$ =54 dB. For a particular BER value, the range of received power ($P_{RX}$) level defined the reliability of link functionality. The FSO system fails to attain desired BER when the minimum SINR do not meet the condition of $P_{RX}$ > $P_{Th,r}$ or $P_{RX}$ < $S_{Rx}$, where $P_{Th,r}$ represents the specified saturation value of $P_{RX}$ level. However, the data link availability ($T_{a}$) can be defined when the achievable throughput exceeds the threshold limit at a specified BER. In practice, $T_{a}$ is a crucial design parameters in the presence multiple turbulences.  

\begin{equation}
T_{a} = \int_{0}^{\alpha_{a}} p(\alpha_{a}) \,d\alpha_{a}
\end{equation}

where $\alpha_{a}$ presents the atmospheric attenuation coefficient and the probability distribution function can be defined as \cite{uysal2016optical}

\begin{equation}
\alpha_{a} = \frac{17.3}{V} (\frac{\lambda}{550})^{-q} , \quad dB/km
\end{equation}

where V is the visibility measurement to estimate $\alpha_{a}$, the value of q is chosen based on the V parameter. 


$T_{a}$ is very useful to assess FSO link performance particularly in FSO assisted access networks at a given location and coverage distance. Note that pointing errors due to jitter is not taken into account to compute $T_{a}$ for the automated FSO systems \cite{das2015radio}. However, the link reliability can be expressed as 

\begin{equation}
R(l) = \int_{l_{th}}^{\infty} p(I) \,dI  = 0.5 - 0.5erfc(\frac{ln(I_{th}/I_{0})}{2\sqrt{2}\sigma})
\end{equation}

where $p(I)$ is cumulative probability irradiance and a threshold intensity ($I_{th}$), $l$ is the transmission link distance in m, $erfc(.)$ is the complementary error function, $I_{0}$ is the average intensity with no turbulence and $\sigma$ is the variance parameter. FSO data link reliability varies in the range between 0 to 1 that depends on the link distance, maintain inverse relationship with $R(l)$. FSO link can be recognized as an established when reliability is greater than threshold limit $I_{th}$.

The broadband functionality of FSO networks is key benefits due to wide frequency range and it offers unlimited ultra-speed data services in compared to RF wireless networks. It has been reported that 1550nm laser operating at 200 THz bandwidth exhibits 200000 times greater data rate over a 2000 MHz microwave link \cite{davis2003flexible}. Notably the FSO systems provide an attractive solution to the last-mile bottleneck and recognized as a pivotal component for the next generation broadband wireless connectivity. However, the total achievable throughput of the FSO link determined by the aggregate traffic carrying capacity to the destination under the given constraints. FSO transceivers require an advanced level of automation allowing self-healing and self-configurations algorithms for topology control \cite{desai2005autonomous}. Authors in \cite{desai2005autonomous} proposed a heuristic algorithm addressing congestion called bottom-up minimum spanning policy according to the transfer matrix. Demir and Yilmaz \cite{demir2020investigation} proposed design of experiment based on Taguchi’s optimization method to investigate the reliability of the FSO systems analyzing different parameters such as visibility, link distance, scatter particle size variations per unit volumes to avoid large computational time.

\section{Mitigation Techniques}		\label{6}

In order to attain high link reliability, some viable approaches have been presented in literature including PAT technique \cite{kaymak2018survey, wu2020design, kaymak2017divergence, yousif2019atmospheric}, diversity mechanisms \cite{aghajanzadeh2012information, miglani2019statistical, sharda2020diversity, al2020performance, jaiswal2018investigation}, hybrid RF/FSO systems \cite{zhou2017distributed, el2017effect, balti2018mixed, salhab2016power, trinh2016mixed, enayati2016deployment, varshney2017cognitive, soleimani2015generalized, ai2020secrecy, bhowal2020relay, hassan2019hybrid} considering atmospheric fluctuations. Optical signal can be blocked fully or partially by the temporary obstacles such as drones, birds, industries smoke, unmanned area vehicles, building sway, tree limbs, etc. in the line of sight connectivity. These sorts of temporary fluctuations can be handle by adopting multi input multi output (MIMO) FSO transmit/receive diversity system \cite{miglani2019statistical}. Under MIMO FSO scheme, an object blocking signal either in Tx or Rx side  does not cause any considerable impact on overall FSO performance. Today’s commercial FSO networks are augmented with beam divergence combination, beam tracking-alignment, clock-recovery phase locked loop (CR-PLL) that have the capability to maintain beam centroid capability. It is widely recognized that high power laser beams alleviate the atmospheric disturbances while maintain desired quality of data rate. Albeit data transmission through FSO links do not create mutual interference but extreme power laser sources beyond threshold limit are harmful for eyes. Therefore, safety regulations is paramount issue for FSO performance and it is important to handle the laser emission enforcement at high intensity in optical wireless communications. The adaptive control mechanism of laser sources via dynamic power adjustment in accordance with the wireless optical channel conditions offer a reliable FSO data link services. It is obvious that the system requires low transmit power for clear weather condition, whereas, a relatively high intensive laser power is needed to ensure the desired level of throughput, BER and QoS.

The temporal diversity of the FSO channels directly affected by the variations of temperature, geographical location, wind velocity, altitude, air pressure, humidity level, etc. \cite{raj2015comparison}. In order to address spatial distortion of the received beam, numerous techniques such as pointing, acquisition, pointing (PAT), aperture averaging and wave front correction. PAT technique is essential to maintain beam centroid stability for long-haul communications like inter-satellite FSO, terrestrial FSO systems. According to \cite{son2017survey}, optical wireless inter-satellite require 1-10 µrad pointing accuracy for 10 Gbps bit rate. Wave front aberrations implies the optical phase variations due to atmospheric disturbances along the wireless propagation path. Generally, a wave front sensor or charged coupled camera device is used to measure the wave-front distortions, whereas, deformable mirror (DM) or electromechanical mirrors are employed for error compensation. All the components are working based on closed-loop control process where a laser power source transmit a Gaussian beam from a distant position and get aberrated as it travels through the free space. The optical lenses (DM) accumulate all the incoming beams and the DM split into two beams: reflected beams incident on the charged coupled camera device (CCCD) and propagating signals falls on the photodetector. The DSP controller drive the DM to rectify the measured distortions through generating appropriate control signals based on CCCD output. Integration of advanced optics in FSO communication improves the system reliability by eliminating phase errors \cite{li2017bp}. However, some promising mitigation techniques are illustrated in the following to address the technical challenges encountered by FSO networks.


\subsection{Aperture sizing}

\begin{figure}
  \centering
  \includegraphics[width=0.9\columnwidth, height = 60mm]{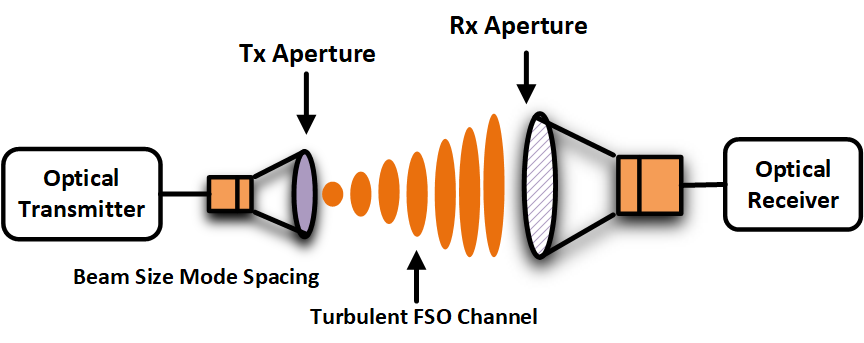}\\
  \caption{Aperture Averaging of FSO transceiver.}\label{Fig:5}
\end{figure}

Averaging of receiver aperture is the popular technique to mitigate received signal fluctuations. The aperture area is determined by the transmission length and strength of aforementioned limiting factors. Increasing the receiver dimension helps to reduce channel fading and eddies induced fast fading. Aperture averaging manifest the amount of fading reduction in physical layer platform. Note that the quantity of background noise increases with the increment of receiver aperture. Thus, the choice of aperture radius should be optimum to uplift the power efficiency of the FSO link.The graphical illustration of aperture averaging of FSO transceiver is shown in Fig. \ref{Fig:5}.

The aperture averaging factor, $F_{a}$ at the receiver side can be expressed as 

\begin{equation}
F_{a} = \frac{\sigma_{I}^{2}(D)}{\sigma_{I}^{2}(D_{0})}
\end{equation}

where $D$ is the receiver aperture, $\sigma_{I}^{2}$ is the signal power, and $\sigma_{I}^{2}(D_{0})$ means the variance of signal intensity at the centre of the receiver. For a plane wave the $F_{a}$ can be approximated by \cite{raj2019historical}

\begin{equation}
F_{a} = [1 + 1.07(\frac{kD^{2}}{4l})^{7/6}]^{-1}
\end{equation}

where $l$ represents the link distance, k is the wave number (2$\pi$/$\lambda$). For the smaller aperture, ($\frac{kD^{2}}{4l}$)<< 1), $F_{a}$ =1.0 and the signal intensity reduces with the increment of aperture sizing i.e. for larger aperture, ($\frac{kD^{2}}{4l}$)>> 1). Therefore, aperture averaging is a significant parameter to address the power variations since larger aperture antenna collects all the signal beams falling into it, thereby, exhibits comparatively lower turbulence induced scintillation. 

\subsection{Adaptive optics (AO)}

\begin{figure}
  \centering
  \includegraphics[width=0.9\columnwidth, height = 3in]{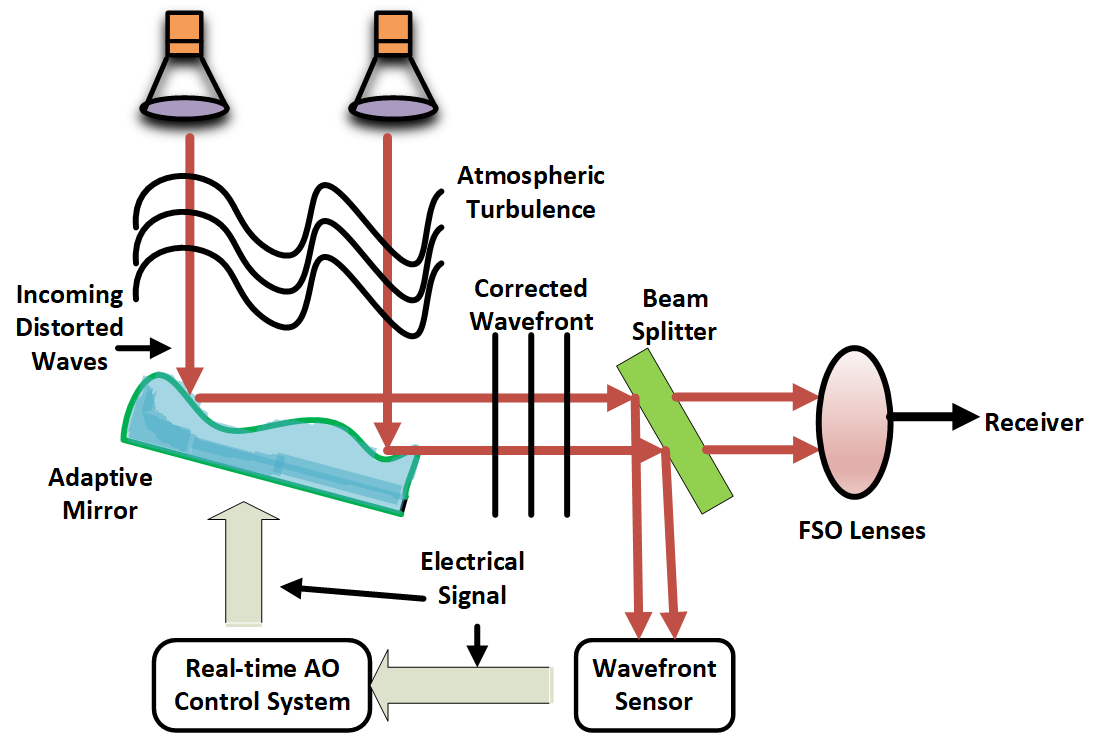}\\
  \caption{Illustration of Adaptive Optics System.}\label{Fig:6}
\end{figure}

Adaptive optics is based on the closed-loop control principle that helps to transmit distortion free laser beam through atmosphere. Precise beam pointing in terms of position and angle is accomplished by implementing adaptive optics with the assistance of steering mirrors in beam propagation path. However, piezoelectricity driven electromechnical devices are commonly used for beam steering, but it suffers due to huge space occupancy and longer processing time. AO system consists of wavefront sensor and corrector, and a set of deformable mirrors at the each transceiver unit to compensate phase floor fluctuations. The study of AO using micro-electromechanical systems (MEMS) for longhaul FSO communication is pointed out in \cite{viswanath2015design}. MEMS based beam steering system enables fast tracking capabilities and precise pointing between two FSO terminals. In addition, MEMS based AO systems control laser beams on a narrow size scale allowing integrated electrical, mechanical and optical systems. The application of AO in downlink FSO satellite communications is investigated in \cite{vedrenne2016adaptive}. Fig. \ref{Fig:6} depict the simplified demonstration of adaptive optics system in FSO perspectives.

\subsection{Relay transmission}

\begin{figure}
  \centering
  \includegraphics[width=0.9\columnwidth, height = 3.2in]{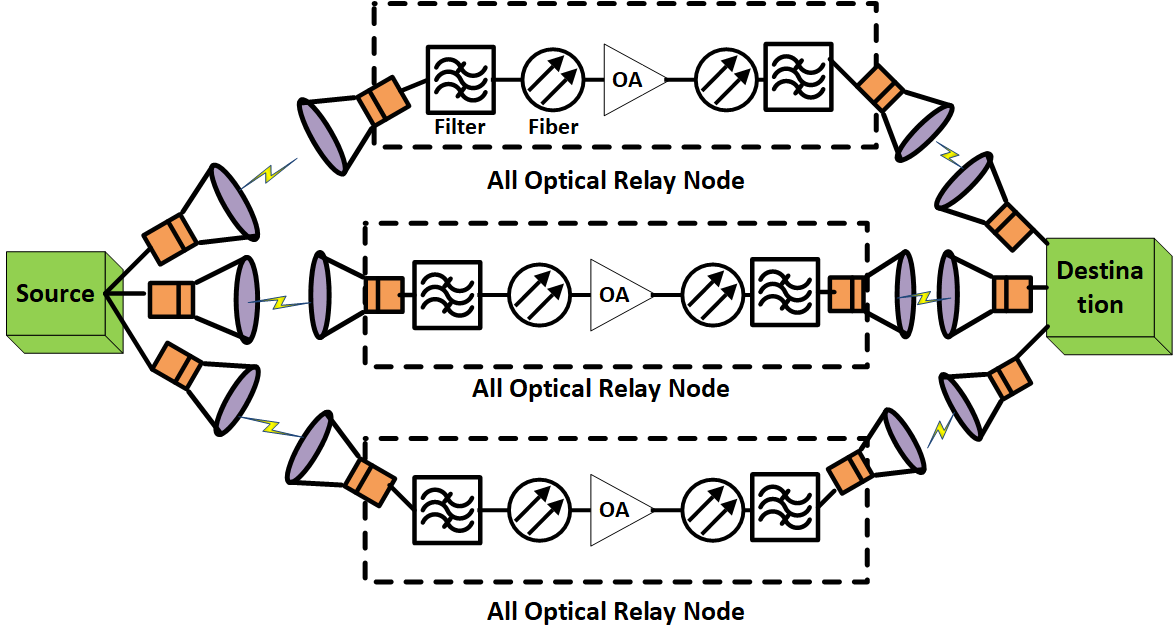}\\
  \caption{Illustration of Relay Aided FSO Transmission.}\label{Fig:7}
\end{figure}

Relay aided transmission is a form spatial diversity where all the antennas are arranged in distributed fashion to share their resources by a cooperative communication. The fundamental objective of introducing relay nodes in the FSO system to extend transmission distance without sacrificing the QoS level. The relay nodes draws intensive attention when the source-destination link suffer poor quality. Relay nodes provide a superior system performance combating the impact of turbulence and offer high SINR manifesting the greater level of received signal intensity over noise. Particularly, relay assisted transmission guaranteed an enhanced level of user experience quality service under severe shadow fading or path loss effects. Regenerative relay transmission mechanism also improve the coverage of mobile wireless systems exploiting additional diversity benefits. Relay based transmission eliminates the need of multiple aperture at the transmitter or received side, a single antenna has the potential to attain significant diversity gain without sacrificing the quality of experience. The choice of placing relay nodes is an important design criteria to achieve huge diversity gain employing the cooperative FSO communications. The study of all optical relay placement and system performance combating the limiting factors is carried out in several literature \cite{kashani2013optimal, boluda2015impact, yang2014performance, trinh2015all}. Fig. \ref{Fig:7} manifests the operations of relay assisted FSO transmission.

\subsection{Modulation}

A modulation technique is chosen based on the type of applications. Various modulation schemes have distinctive features such as spectral efficiency (SE) and bit per energy level or energy efficiency (EE). EE implies the maximum achievable data rate at a target BER for a given transmit power. Although unlicensed optical band consists of huge bandwidth, still SE is key design parameters as it proportionally related to the speed of electronic switch circuitry in FSO systems in practical case.  At the receiver end, it is important to know the instantaneous channel state coefficients for optimal signal detection. A few pilot carriers are used in practice to estimate the channel state information (CSI) with perfect accuracy \cite{yue2020performance}. A symbol-by-symbol maximum likelihood (ML) detection technique is used as an alternative approach based on the partial knowledge of the distribution of channel fading coefficients and ML detection based on the information of joint temporal fading coefficients statistics  \cite{xu2017channel, le2019receiver}. 

The most commonly used a binary level modulation scheme called On-Off Keying (OOK) in FSO systems due to its simplicity. For ultrafast data communication and complex system architecture, OOK is less efficient due to poor spectral and energy efficiency (EE). Numerous intensity modulation schemes have been carried out to suppress some disadvantages. Alternatively, pulse position modulation (PPM) or variable PPM is a preferred choice for addressing EE applications particularly in deep space communications \cite{sevincer2013lightnets, khalighi2014underwater}. Unlike OOK, PPM does not need any dynamic threshold for optimal detection while performing hard signal identification. However, OOK and PPM schemes are single carrier based pulsed modulation and can be unreliable during severe channel impairments for FSO links. Single carrier modulation techniques are inappropriate for high data rate applications due to high level of inter-symbol interference (ISI) \cite{islim2016modulation}. Subsequently, subcarrier intensity modulation (SIM) and orthogonal frequency division multiplexing (OFDM) based multi-carrier SIM (MCSIM) techniques are adopted in several literature \cite{balti2017aggregate, arezumand2017outage, bekkali2010transmission}. A pre-modulated RF signal drive the incoming optical signal for carrying data in SIM technique. In order to maintain all the amplitude of optical pulses are positive, a DC bias is incorporated to the signal before drive an optical source \cite{bekkali2010transmission}. A SIM based approaches have some advantages over single carrier modulation like mitigate channel impairments, simplicity, cost-effective solution, and improved spectral efficiency \cite{hassan2016subcarrier}.

The addition of an external DC bias with RF signal to neglect non-negative amplitude pulses causes reduced power efficiency. With the increment of number of subcarriers in MCSIM scheme, the DC bias signal also need to increase large to prevent non-linear distortion and clipping effects. Thus, OFDM based MCSIM modulation technique severely experienced by peak-to-average power ratio (PAPR) which degrades the power efficiency \cite{hassan2016subcarrier}. The nonlinearity of light signals is also a critical challenge in MCSIM technique which leads to inter-carrier interference and widening the signal spectrum resulting inter-modulation distortion (IMD) \cite{bekkali2010transmission, tsonev2013complete}. As a consequence, small number of subcarriers are need to employ to limit IMD, but this surely reduce the transmission rate. Another solution to reduce IMD effect is to use separate optical source for each carrier resulting complexity and cost as well \cite{yu2014iterative}. In compared to PPM, multi-pulse PMM (MPPM) reduces PAPR and increases the spectral efficiency as well but having higher demodulation complexity \cite{nazrul2019effect}. MPPM exhibits a superior performance over PPM under peak transmit power and PPM outperforms MPPM for an average transmit power \cite{xu2009coded}. Zhang \textit{et. al.} \cite{zhang2018improved} proposed an improved linear non-symmetrical transform to reduce PAPR in OFDM based wireless optical communications. A PAPR reduction technique make the signal less vulnerable in terms of nonlinear distortion, hence, improve the performance of MSIM techniques \cite{zhang2014time}. A precoding matrix techniques have been proposed to improve the PAPR, SNR, BER performances of the DC-biased optical OFDM (DCO-OFDM) system in perspective of OWC systems \cite{jiang2018investigation}. In order to achieve the same bandwidth efficiency and enhanced spectral efficiency with the traditional (DCO-OFDM), authors proposed \cite{chow2019performances} a modified asymmetrically clipped optical OFDM method.   
   
Another well-known modulation technique is pulse width modulation (PWM) has an improved spectral efficiency, demand lower peak transmit power, and better performance to inter-symbol interference (ISI) in comparison with PPM \cite{ebrahimi2018investigation}. However, an extra guard slot is required to add to suppress sending consecutive positive pulses over symbol period. Both PPM and PWM are known as synchronous modulation since they require slot and symbol synchronization. On the other hand, digital pulse interval modulation (DPIM) is a type of asynchronous modulation scheme which is more spectrally efficient than PPM, PWM and does not require symbol and slot synchronization. The possible error propagation during demodulation at the receiver is the potential limitation of DPIM \cite{abdullah2014adaptive, guo2018ordered}. Following to the concept of MPPM, overlapping PPM (OPPM) \cite{noshad2013application, chizari2016designing}, differential PPM \cite{abdullah2014adaptive} and digital pulse interval and width modulation (DPIWM) \cite{mu2014analyzing} are proposed for improving the spectral efficiency. In OPPM, the optical pulses occupy adjacent slots and symbols end with a pulse can be exploited for synchronization in DPPM, whereas, the binary sequence is encoded into pulse width of alternating amplitude in DPIWM. The fundamental limitations of all these modulations are the poor energy efficiency, high receiving complexity, and the probability of error propagation while detecting encoded bit sequence. 

Being inspired from the aforementioned concerns, multi-level modulation schemes could be employed in FSO systems to attain better SE in comparison with binary modulations. For example, the optical pulse is modulated on M levels in M ary-PAM \cite{tsai2020500}, M ary-PSK \cite{shakir2017performance}, and M ary-QAM \cite{lei2020performance}, which requires a laser source with different emission intensity.  All of these M-level modulation methods provide higher SE and EE with respect to binary level modulation schemes. Apart from this, other multilevel intensity modulation and direct detection (IM/DD) schemes such as differential phase shift keying (DPSK) \cite{amirabadi2019performance}, carrier-less amplitude and phase (CAP) \cite{khalighi2017pam} modulation have been considered in FSO systems.

\subsection{FSO channel modeling}

FSO channels are subject to atmospheric turbulence and fading, so the definitions of channel capacity by means of ergodic capacity and outage probability should be considered because of the random nature of channel fading coefficients. Ergodic capacity is referred as the expectations of the instantaneous channel capacity which can be computed via mutual information expression in comparison with different fading coefficients. If the mutual information fails to reach the information rate in FSO channels, an outage event is specified which is known as probability of fade or outage probability. The performance of FSO link significantly hampered by means of signal fade under the severe atmospheric turbulence factors. In particular, FSO channel is highly variable, unpredictable and vulnerable due to inhomogeneity in pressure and temperature of the atmospheric layer. The variation of refractive index and multiple scattering along the propagation path results in phase and intensity fluctuations of the received optical signal. The effect of scintillation induced by temperature and spatial fluctuations of light intensity lessen the FSO link performance even under clear weather. As a consequences, bit error rate increases considerably especially for the long distance FSO communication even for the small effect of aforementioned issues. The link margin (LM) approach is a statistical parameter that is used to quantify the performance of FSO transmission system. LM can be expressed in decibel (dB) which can be defined as the ratio of received optical power to the power required for the target bit rate and error probability. The term LM accounts all the limiting factors associated in the FSO link including transceiver station parameters and link budget models support the designers for optimum implementation under all weather conditions at a given location \cite{shen2014temporal, uysal2016optical, majumdar2014advanced, ghassemlooy2019optical}. Some other factors such as beam divergence loss, sky radiance, background noise, angle of beam arrival variations, free space seeing, and cloud blockage are need to be considered during FSO link implementation. 

A numerous research works have been conducted to investigate the channel capacity of turbulent FSO channels. The study of ergodic capacity of FSO channels is carried out in \cite{salhab2016power, el2016security, ansari2013impact, usman2014practical, trinh2016mixed, bag2018performance, al2018two, palliyembil2018capacity, balti2018mixed} for the cases of Gamma-Gamma, Rayleigh, Nakagami-m, Malaga, Rician and log-normal fading taken into consideration of AWGN model for the receiver noise. On the other hand, the outage probability \cite{rakia2015power, makki2016performance, salhab2016power, el2016security, ansari2013impact, zhang2015unified, usman2014practical, balti2017aggregate, trinh2016mixed, el2017effect, bag2018performance, shakir2017performance, amirabadi2019performance, petkovic2015partial, soleimani2015generalized, wu2017dynamic, chen2016multiuser, al2018two, arezumand2017outage, touati2016effects, varshney2017cognitive, palliyembil2018capacity, odeyemi2020secrecy, varshney2018cognitive, lei2020performance, balti2018mixed}, bit error rate analysis \cite{islam2019effect, nazrul2019effect, islam2015analytical, ansari2013impact, zhang2015unified, usman2014practical, balti2017aggregate, trinh2016mixed, yang2015performance, bag2018performance, shakir2017performance, amirabadi2019performance, petkovic2015partial, soleimani2015generalized, chen2016multiuser, al2018two, arezumand2017outage, touati2016effects, palliyembil2018capacity, varshney2018cognitive, lei2020secure, lei2020performance, li201982, xing2019joint, kong2015performance, balti2018mixed}, and throughput performance \cite{islam2019effect, nazrul2019effect, islam2015analytical, tang2014link, rakia2015power, makki2016performance, najafi2017optimal, salhab2016power, jamali2016link, trinh2016mixed, yang2015performance, odeyemi2020secrecy, lei2020secure, lei2020performance, tsai2020500, li201982} are studied considering the mentioned fading models with AWGN. 

The unguided beam fluctuations on the free space due to the atmospheric turbulence induced channel fading leads to a remarkable FSO system performance degradation in the form of receiving error bits. Channel coding is one of the promising fading mitigation technique under weak and strong turbulence conditions \cite{kaushal2016optical, safi2019adaptive}. To handle the atmospheric turbulence, several fading reduction techniques such as tempo-spatial diversity \cite{jaiswal2018investigation, garrido2014spatially, jurado2016hybrid, boluda2015miso}, sequence data detection algorithms \cite{song2013robust, dabiri2017glrt, dabiri2018receiver}, and relay enabled communications \cite{huang2019game , kong2015performance, amirabadi2019performance, safari2008relay, nor2016experimental} have carried out in the context of FSO communications. However, majority of the proposed methods are complex and difficult to implementation while processing high data rate communications.

The channel fading of FSO systems is similar to quasi-static property that is slow varying and the CSI can be estimated with better accuracy and fed back to the transmitter. The FSO transmitter is then adjust its parameter for example, transmit power, modulation level and code rate according to the current CSI. Note that adaptive transmission is the most suitable approach for full duplex operations of feedback purpose in FSO link \cite{safi2019adaptive}. Therefore, adaptive channel coding and power control scheme is very important for commercial FSO systems under practical constraints to satisfy the target BER and desired ergodic capacity with minimum outage. Reference \cite{safi2019adaptive} proposed a jointly adaptive transmission and power control schemes assuming Gamma-Gamma turbulence channels aiming to achieve target BER and outage probability. The rate compatible punctured convolutional (RCPC) codes are used to perform adaptive transmission for controlling the channel coding rate. Viterbi algorithm and RCPC codes based maximum likelihood decoder is used to reduce the complexity of the FSO transceivers. Odeyemi \textit{et. al.} \cite{odeyemi2017optical} studied the performance of FSO spatial modulation (FSO-SM) systems considering the influence of Gamma-Gamma distribution and pointing errors. The symbol error rate (SER) performance is investigated employing convolution coding. Authors in \cite{malik2015performance} examined the bit error rate and pair-wise error probability performance of FSO systems based on bit interleaved coded modulation (BICM)-enabled subcarrier intensity modulation (SIM) employing convolution coding. Authors proposed probabilistic shaping schemes in FSO systems of 16-ary quadrature amplitude modulation (16-QAM) and convolution coding technique under the influence of Gamma-Gamma disturbance in \cite{wang2019qc, wang2017fast}. The bit error rate performance for IM/DD FSO systems with binary phase shift modulation and convolution coding over Gamma-Gamma turbulence channels is investigated in \cite{yang2013performance}. The expressions of BER and pairwise error probability (PEP) for convolution codes and Q-ary PPM signaling schemes are thoroughly examined in \cite{liao2013coded}. In \cite{ajewole2019error}, the performance of BPSK OFDM-FSO over the Gamma-Gamma turbulence channel is investigated integrating forward error correction (FEC) coding to improve the error probability. Authors designed error control protocol and analyzed throughput performance employing RCPC code over atmospheric turbulence channels for FSO burst transmission in satellite communications in \cite{le2019throughput}. The operation of novel space-time trellis codes (STTCs) for FSO communication using IM/DD under atmospheric turbulence and misalignment of fading channels is investigated in \cite{garcia2015novel}.

A simple rate adaptive transmission approach based on OOK modulation formats with memory is demonstrated in \cite{jurado2010efficient}. Authors in \cite{hassan2018delay, hassan2017statistical} analyzed QoS aware delay constraint power allocation for a coherent FSO systems considering the atmospheric turbulence fading channel. An optimal power allocation problem under eye safety constraints is studied for adaptive WDM-FSO systems in \cite{zhou2015optical}. Authors proposed an adaptive uncoded transmission policy for a channel state dependent transmit power and modulation level in FSO communications in \cite{karimi2012novel}. Reference \cite{anguita2010rateless} proposed a coding rate adjustment policy based on the temporal conditions of fading channels under fixed transmit power. To improve the FSO link performance in the presence of strong turbulence, an adaptive low-density-parity-check (LDPC) coded modulation is considered in \cite{djordjevic2010adaptive}. In \cite{ali2019performance}, authors presented a dual-hop relay-assisted hybrid RF/FSO communications using Alamouti coding in the presence of pointing error over the Gamma-Gamma atmospheric turbulence channels. Authors investigated indoor FSO link reliability in terms of packet error rate using rateless code (RaptorQ code) under the locally generated turbulence effect in \cite{pernice2016moderate}. A summary of optical signal modulation schemes, wireless link models, and channel coding schemes are presented in Table \ref{Table:6}.

\begin{table}
\centering
\caption{A comparison of FSO cahnnel model, channel coding and modulation scheme}\label{Table:6}
{\tabulinesep=1.3mm
\begin{tabular}{|p{15mm}|p{36mm}|p{45mm}|p{27mm}|} \hline
\textbf{References} & \textbf{Channel model} & \textbf{Channel coding}  & \textbf{Modulation}  \\ \hline
\cite{varshney2017cognitive} & Nakagami-m & OSTBC & IM/DD OOK  \\ \hline
\cite{balti2017aggregate}	 & Malaga & Repetition coding  & SIK  \\ \hline
\cite{safi2019adaptive, jurado2016hybrid} & Gamma–Gamma & FEC RCPC & OOK  \\ \hline
\cite{jaiswal2018investigation} & Generalized gamma & OSTBC & PAM, OSSK   \\ \hline
\cite{song2013robust, dabiri2017glrt}  & Gamma–Gamma  & Convolution  & OOK, PPM    \\ \hline
 \cite{odeyemi2017optical} & Gamma–Gamma  & Convolution  &  Spatial modulation   \\ \hline
\cite{malik2015performance}  & Gamma–Gamma  & BICM, Convolution  &  $M^{2}$-QAM   \\ \hline
\cite{wang2019qc}  & Gamma–Gamma  & Convolution  &  16-QAM   \\ \hline
\cite{wang2017fast, mudge2016development}  & Log-normal, K-distribution  & LDPC  & OOK    \\ \hline
\cite{yang2013performance}  & Gamma–Gamma  & Convolution  &  BPSK   \\ \hline
\cite{liao2013coded}  & Log-normal  & Convolution  &  Q-ary PPM   \\ \hline
\cite{ajewole2019error, le2019throughput}  & Gamma–Gamma  & FEC RCPC  &  BPSK   \\ \hline
\cite{garcia2015novel}  & Gamma–Gamma  & STTC  &  IM/DD OOK   \\ \hline
\cite{jurado2010efficient}  & Non-fading  & RC  &  OOK   \\ \hline
\cite{ali2019performance}  & Gamma-Gamma  & Alamouti  & DPSK    \\ \hline
\cite{zhu2017block}  & Gamma-Gamma  & BMST  & PPM    \\ \hline
\cite{priyadarshani2017effect, boluda2019outage, boluda2017effect}  & Gamma-Gamma  & RC  &  OOK   \\ \hline
\cite{khodadadi2017analysis}  & Gamma-Gamma  & BCHB  &  PPM   \\ \hline
\cite{jaiswal2017performance}  & Gamma-Gamma  & RC  & PAM    \\ \hline
\cite{ji2018performance}  & Gamma-Gamma  & OOC  &  PPM   \\ \hline
\cite{bhowal2019outage}  & Gamma-Gamma  & PLNC  & BPSK   \\ \hline
\cite{shang2018performance}  & Gamma-Gamma  & IMC  & BPSK, 8-PPM    \\ \hline
\cite{arya2019amplify}  & Gamma-Gamma  & RC  & BPSK    \\ \hline
\cite{amhoud2019unified}  & Generalized gamma  & Alamouti, space-time   &   4-PAM  \\ \hline
\cite{varshney2018cognitive}  &  Gamma-Gamma &  OSTBC &  BPSK   \\ \hline
\cite{liu2017aber}  &  Exponentiated Weibull  & LDPC  &  16-PSK, 16-QAM   \\ \hline
\cite{li2017novel}  & Gamma-Gamma  & OCDMA  &  BPSK   \\ \hline
\cite{khallaf2017performance}  & Gamma-Gamma  & Reed-Solomon   &  OOK, QAM-MPPM   \\ \hline
\cite{fang2018polar}  & Gamma-Gamma  & Polar  &  OOK   \\ \hline
\cite{tapse2011hybrid}  & Gamma-Gamma  & Turbo  &  OOK   \\ \hline
\cite{nayak2014performance} & Gamma-Gamma  &  LT, BCHB &   BPSK  \\ \hline
\cite{liverman2018wifo} & Non-fading  & Reed-Solomon   & OOK    \\ \hline
\cite{huang2019design} &  Non-fading & Reed-Solomon   &  CSK   \\ \hline
\cite{lu2016achieving} & Non-fading  & Miller, Concatenation Convolution   &  OOK   \\ \hline
\cite{ndjiongue2018hybrid} & Non-fading  & Trellis  &  MPSK-CSK   \\ \hline
\cite{mejia2019coding} & Non-fading  & Trellis  &  CSK   \\ \hline
\end{tabular}}
\end{table}

\subsection{Background noise reduction}

The background noise is created due to solar irradiation in the daytime. The background radiation noise is a function of wavelength, higher background noise at lower operating wavelength. The introduction of spatial filters with an adaptive modulation technique having high PAPR can able to mitigate this noise. In such case, some crucial design parameters like Doppler shift, angle of signal arrival, and spectral width of laser source are need to be considered to build spatial filters. M-ary PPM is the best choice for FSO links as it is more energy efficient and sharply reduce the background radiation noise. Mapping narrow FOV with the help of AO including an array of actuators in the receiver side, and corresponding filter selection is another approach to combat background noise \cite{hashmi2014analysis}.

\subsection{Diversity}  

\begin{figure}
  \centering
  \includegraphics[width=0.9\columnwidth, height = 70mm]{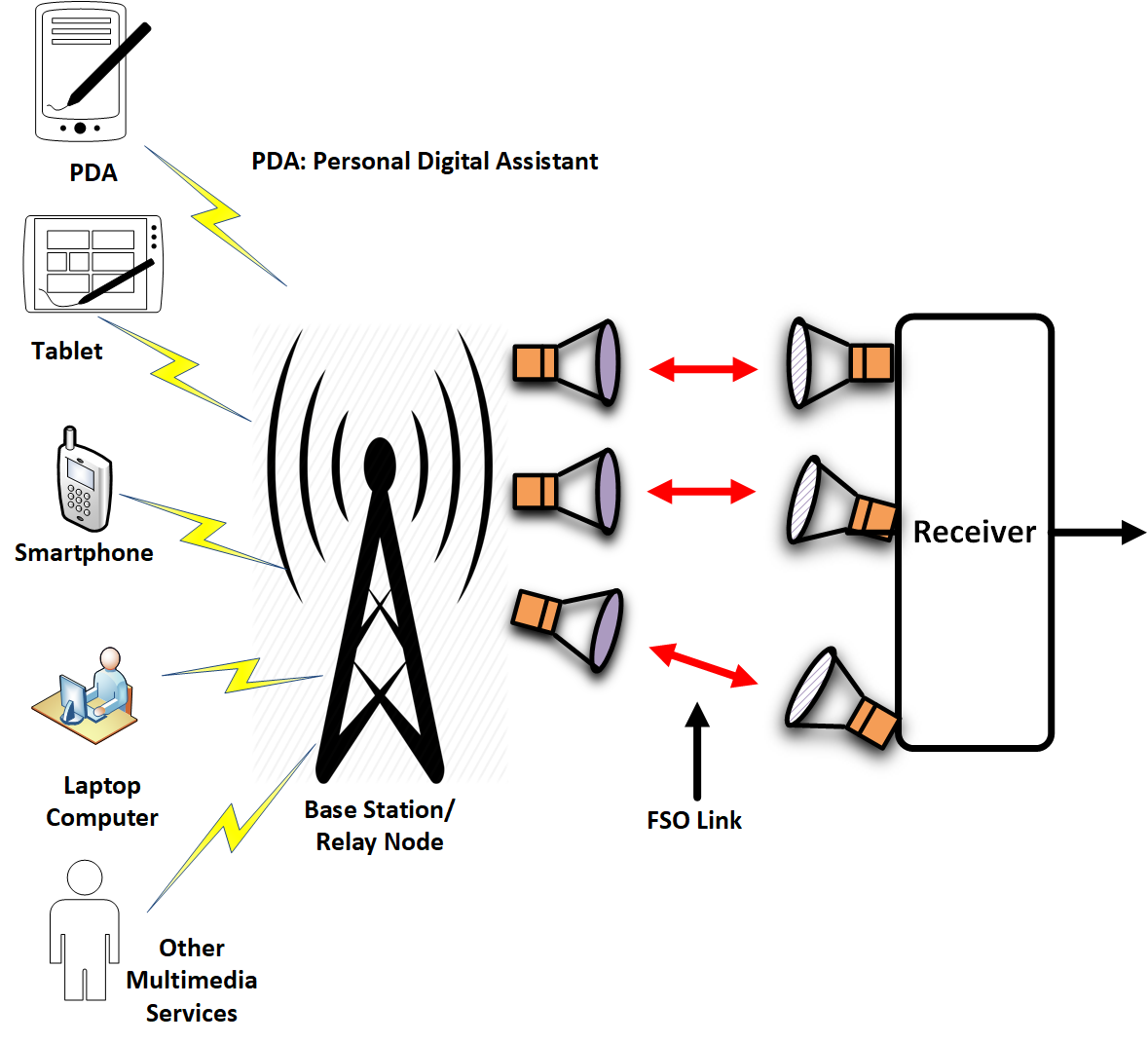}\\
  \caption{Spatial Diversity Mechanism in FSO Networks.}\label{Fig:8}
\end{figure}

Diversity techniques can operate in three different domains such as time, frequency and space to diminish atmospheric disturbances. As contrast with single large aperture, an array of multiple transmitters and receivers are used to generate/receive  multiple copies of mutually correlated signals either in three above domains. It is widely recognized that diversity techniques restricts the need of active tracking during misalignment that has the capability to improve BER performance with desired QoS. To achieve the full benefits of spatial diversity, the separation between antennas should be equal or greater than the coherence length at either transmitter or receiving end. However, spatial correlation is a function of beam width, beam divergence and variance of beam wander. For SIMO receive diversity, the diversity gain can be derived by averaging the multiple independent laser beams. However, three receive diversity techniques namely selective combining (SC), maximal ratio combining (MRC) and equal gain combing (EGC) can be used at receiving end. The diversity gain for MRC scheme is comparatively higher that proving maximum level of SINR. On the other hand, ECG is a preferred choice over two other methods because of simplicity and low cost installation \cite{ma2015performance}. FSO MIMO is quite similar to RF MIMO that increases the channel capacity linearly scaled with the number of transmitting antennas. Time diversity is normally deployed for time selective fading channels where repetitive symbols are transmitted in different coherence time. However, time diversity can be implemented either incorporating coding or bit interleaving in accordance with the difference between the length of data frame and channel coherence time. The generalized spatial diversity technique in FSO networks is depicted in Fig. \ref{Fig:8}.     

\subsection{Coherent FSO}

The fluctuations of amplitude and phase of the received signal results in decreasing SINR and degrade the BER performance. Based on the detection techniques, FSO communication can be classified into: direct detection (DD) FSO and coherent detection (CD) FSO. In DD-FSO communication scheme, the intensity of source light carry the information and photodetector directly detect the changes of all photons entering into it without any external mechanism at the receiver side. In contrary to IM/DD systems, amplitude, frequency or phase can be used for information encoding in coherent FSO systems allowing to enhance spectral efficiency. The receiver unit equipped with local oscillator (LO) beam where the received signal mixed with LO generated optical field before photo detection. Thereafter, the signal is amplified and passed through the filter in order to rejection of background noise and interferences \cite{al2020amplitude}. It has been recognized that coherent FSO scheme provides the greater level of the receiver sensitivity over IM/DD systems. Numerous modulation techniques such as M-ary quadrature amplitude modulation (M-QAM), multilevel quadrature phase shift keying (M-QPSK), or multilevel polarization phase shift keying (M-PolSK) are commonly used in CD-FSO systems \cite{liu2017aber, khallaf2017performance}. 

\begin{figure}
  \centering
  \includegraphics[width=0.9\columnwidth, height = 80mm]{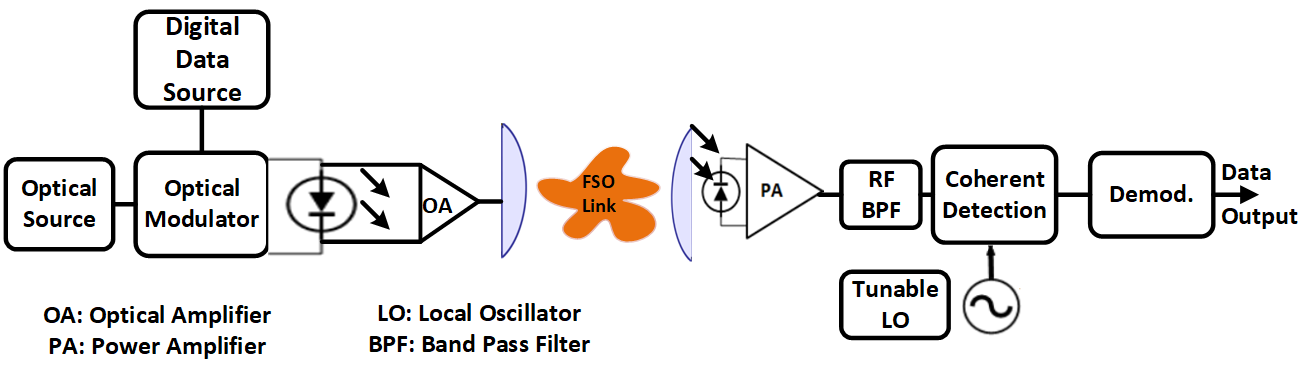}\\
  \caption{Coherent Detection in FSO Systems.}\label{Fig:9}
\end{figure}

Despite the potential benefits of coherent FSO by means of noise rejection, handling turbulence induced multipath fading and improved level of receiver sensitivity, DD-FSO systems are mostly used because of simplicity and lower cost. With the advancement of ultra-speed digital processing integrated units, coherent FSO schemes are implemented increasingly \cite{li2018free}. Atmospheric disturbances distorts the transmitted signal and resulting imperfection of wave front with the LO at the receiving end. The relevant phase distortions severely curbing the system performance particularly when the receiver aperture is higher than coherence length of the incoming signal wavefront. Phase compensation technique through tracking the optical beam offers a promising solution of turbulence-induced distortions. Two different approaches named homodyne and heterodyne detection are followed in coherent receivers. Homodyne reception technique requires a precise level of phase matching i.e. proper phase locked loop that makes the device very expensive. Because of this constraint, heterodyne receiver is widely adopted although the prior technique provides a better detection sensitivity \cite{barua2020fdm}. Recently, several research works have been done to improve spectral efficiency, link distance and system capacity employing advanced modulation formats in the context of coherent FSO communications \cite{kakati2020640, wang2016320, wang2018single, li2018independent, zhu2018terabit}. Reference \cite{toyoda2012marked} demonstrated a 160 Gbps of 256 QAM OFDM over 160 km single mode fibre (SMF) for optical backbone and 4 km FSO links. Authors in \cite{wang2016320} presented a 200 Gbps, 20 Gbaud dual polarization (DP) 1024-QAM transceiver scheme of spectral efficiency (SE) 14 bit/sec/Hz, whereas, Kakati and Arya \cite{kakati2020640} demonstrated maximum 640 Gbps data rate with 40 Gbps symbol rate, SE of 15.3 bit/sec/Hz for both in optical core infrastructure and free space optics over the same transmission distance (i.e. 160 km SMF and 4 km FSO). However, receive diversity under different turbulence nature has attracted special attention in coherent systems providing much benefits of mitigating fading, noise and interfering signals. Fig. \ref{Fig:9} presents the graphical illustration of coherent receiving technique in FSO systems.

An optical WiMAX transmission system enables longer distance free-space communication, massive information carrying capacity, and proximity repeatability combing the merits of optical and wireless technologies. In the FSO-WiMAX scheme, OFDM divided the channel into N number of narrowband slots assigning appropriate subcarriers. OFDM technique improve the spectral efficiency and dispersion rate. A tightly coherent optical beam adopts the spatial light modulator to a broadband optical WiMAX transport system over the long distance FSO data channel. 

\subsection{Sub-carrier Multiplexing (SCM)}

Employing multiple wavelength using sub-carrier multiplexing (SCM) technique can achieve 1 Tb/s bit rate over 3 Tb/s-km bandwidth-length product and curtails the overall cost per bit \cite{raj2019historical}. As a result, the integration of SCM based modulation with FSO systems offers an emerging solution to the last-mile applications particularly for the large temporal and spatial bandwidth of light signal. According to the definition, N subcarriers employing multiplexing are transmitted and Mach-Zehnder modulator (MZM) is used for modulating the optical carrier signal. Thereafter, the modulated signal is passed through FSO channels; photodiode demultiplexed the incoming information to their destinations. Due to the coherent electrical detection at the receiving end, the full spectrum can be efficiently utilized in SCM-FSO systems.

\subsection{Hybrid FSO}

Each individual technologies of RF based network and optical wireless have some advantages and limitations. The fundamental limitations of FSO systems including blocking of communications by obstacles can be effectively overcome using the concept of the coexistence of RF and FSO networks taking their individual advantages. The presence of hybrid RF/FSO systems improves the link reliability and facilitates load balancing under different scenarios. The convergence of heterogeneous architecture including RF and FSO networks incorporating diverse frequency bands plays a pivotal role to provide higher degree of QoS and work well under all adverse weather conditions. The simultaneous operation of hybrid approach endeavor seamless system capacity and end users avail the inherent benefits from the RF/FSO technology. In other words, RF systems address the blockage hole providing wider coverage area at the user end and FSO ensure high data rate that can operate in the same environment without causing interference among each other. Hybrid technology can play a significant role in link-reliability enhancement, energy-efficient operations, seamless wireless connectivity in remote distances, interference minimization, and security issues. The goal of combined wireless systems to attain better performance through eliminating the limitations of individual technologies. Some technical hurdles like shadow effects, inter-symbol interference (ISI), phase induced noise, multi-access interference (MAI) and multipath fading reduces the intensity of the received signal. The RF networks performs as a backup path whenever blockage occurs, NLOS conditions, antenna misalignment, or the severe degradation along FSO link. The transmission link switch back to optical link when the signal is recovered. 

\begin{figure}
  \centering
  \includegraphics[width=0.9\columnwidth, height = 3.2in]{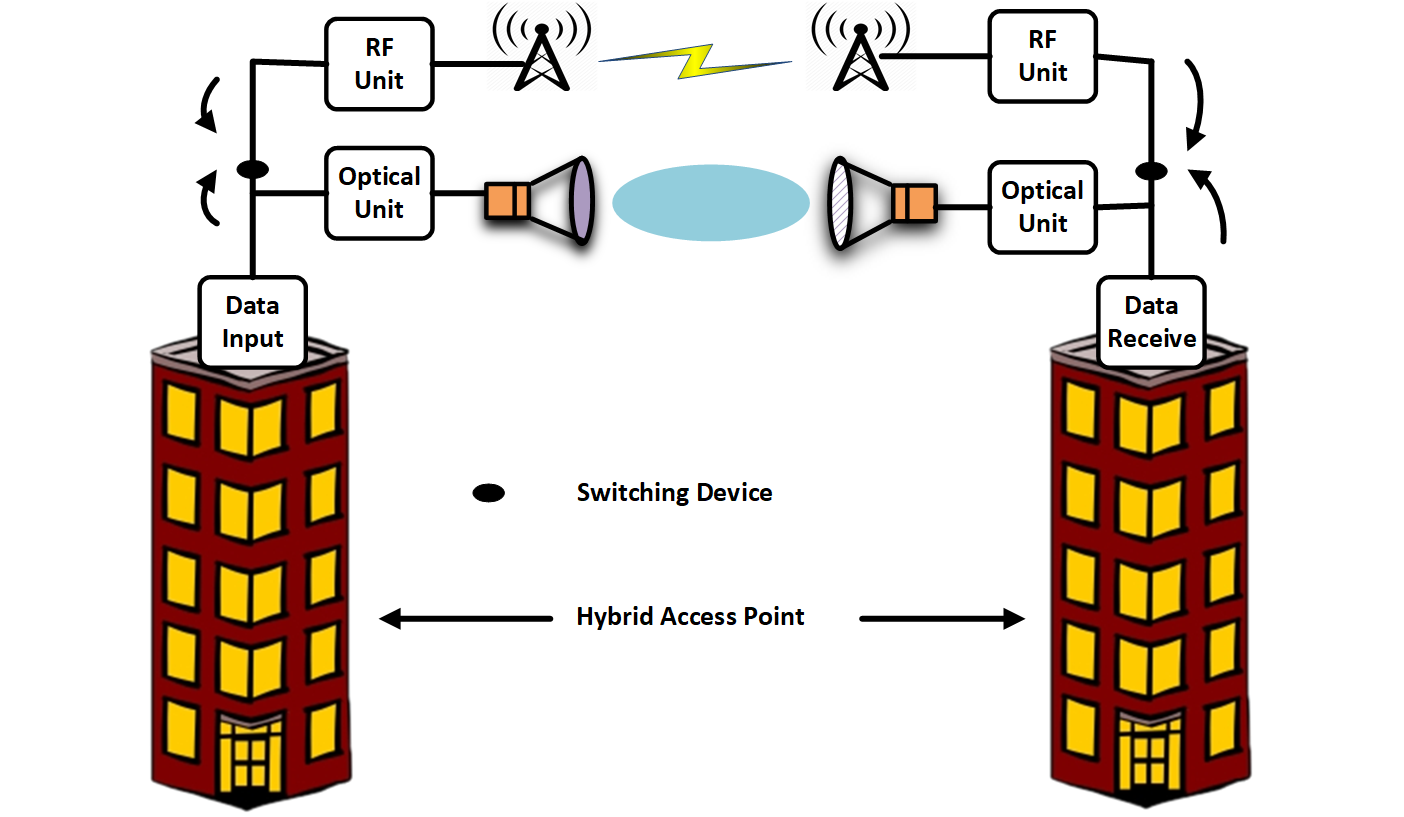}\\
  \caption{Demonstration of hybrid RF/FSO mechanism.}\label{Fig:10}
\end{figure}

In the response to escalating FSO link reliability, RF link serving as a backup path in parallel with the FSO channel propagation path offers a potential solution during FSO link outage. Thermal expansion, heavy wind loads, and little earthquake causes the vibration, thereby break the tight pointing (i.e. misalignment) of the FSO transmitter toward the photodetector position \cite{majumdar2010free, al2017outage}. The combined use of RF systems like microwave and mmWave with FSO can suppress these restrictions. Although RF system exhibits inferior data rate performance in compared to FSO link, it can guaranteed the connectivity while FSO channel fail. However, RF channels is less affected by pointing errors, fog and atmospheric turbulence. Notably RF link is typically configured in the millimeter (mm) wave bands ranging around 60 GHz that facilitate high bandwidth incorporating with FSO link. Moreover, RF-based communications enable improved performance in NLOS conditions and support better mobility, but RF technology is electromagnetic interference sensitive. For the long-span combined RF/FSO systems, RF link can be used for beam pointing and acquisition for ensuring better reliability \cite{khalighi2014survey}. These special features of RF technology mitigate the limitations of FSO deployment. The energy efficiency (bps/Watt) of FSO system is comparatively higher than an RF system. Shifting the traffic load to optical networks from RF minimize the overall energy consumption. Provisioning the RF/FSO systems is the best choice for spectrum utilization particularly for the indoor user’s performance. Mobility support with extremely high speed is paramount benefits of the RF/FSO hybrid systems. Hybrid FSO scheme offers excellent communication distance for vehicular systems (V2X) handling dense traffic densities over the wider geographical region \cite{cuailean2017current}. An FSO networks plays as a primary link and RF system used as a secondary backup path to establish high reliability. The wireless connectivity through relay nodes is termed as mixed RF/FSO network.

The problem encountered in hybrid FSO system, a re-transmission approach is adopted in literature \cite{pratama2018bandwidth} where the lost packets are retransmitted via RF link. The proposed technique based on throughput-optimal scheduler that has the capability of handling handover between optical network and RF architecture efficiently. The function of hybrid infrastructure gateway is scheduling, re-transmitting and handling all the incoming data from data packet scheduler. However, the key challenge of vertical handover is to maintain expected QoS. Markov decision process \cite{hou2006vertical} and fuzzy-logic approach \cite{purwita2018handover} is proposed to manage vertical handover issue. Markov decision scheme minimize the pin-pong effects i.e. the redundant switching is avoided when blocking occurs in LOS link. The optical link interruption is primarily depends on the user distribution based on space, time and duration. In accordance to the user mobility predictions, the packet scheduling of hybrid FSO networks can be determined using Markov chain process. On the other hand, fuzzy-logic process address the uncertainty of decision metrics. Reference \cite{saber2019physical, tubail2020physical, abd2017effect} demonstrates the physical layer security of hybrid RF/FSO systems formulating the minimum power consumption satisfying the users secrecy rate. Fig. \ref{Fig:10} demonstrates the fundamental hybrid RF/FSO communication system.

\paragraph*{WiFo (WiFi-FSO) Communication}

WiFo is a hybrid (i.e. a combination of both RF and FSO technologies) high capacity indoor wireless communication systems based on femtocell architecture to increase spectral efficiency and reduce interference \cite{liverman2016hybrid, nguyen2017embedded}. WiFo can be used to handle multi user FSO communication problems where end users received multiple FSO signals simultaneously in overlapped areas. With the added benefits of high modulation bandwidth of VCSEL, the WiFo typically adopt embedded coding PAM technique for short distance indoor communications resulting low energy consumption and PiN photodiode \cite{nguyen2017embedded}. A set of FSO transmitters are comprised in WiFo directly under the roof. All FSO transmitters generate invisible light that are connected to high speed 100 Gbps Ethernet network controlled by the access point (AP). Whereas, AP manages the simultaneous data communications of FSO signals and the existing WiFi links. WiFo permits denser deployment of light sources where a single receiver can capture all FSO transmitted signals like MISO systems. These crowded installations enable end users in overlapped areas to achieve higher data rate.  

Underwater optical wireless communications (UOWC) by means of the aggregate technology of RF, acoustics and optics has drawn deep attention for multiple potential applications for instance, oil pipe monitoring, underwater environment and offshore investigation. UOWC based of RF and acoustic technology cannot support long distance communication and high speed data rate as well. Acoustic waves enables maximum 20 km underwater link distance with bit rate ranging from tens Hz to hundred of kHz \cite{alimi2017challenges}. On the other hand, RF systems also exhibits extremely poor performance that are limited to short link distance. The excellent technical benefits of hybrid RF/acoustic/optical technology can overcome limitations such as link delay, low coverage, and strong signal attenuation.   

Recently, optical/optical hybrid wireless technologies such as FSO/VLC, FSO/OCC showing different characteristics has also received much attraction among researchers. FSO/OCC hybrid solution provides long distance V2X communications while maintain stable link performance eliminating their individual limitations. OCC support comparatively short distance communications between vehicles with low data rate. In contrast, FSO networks provides communication between cars far apart. Therefore, the hybrid system satisfy user demand and increase reliability as well. However, the performance is considerable suffered by the atmospheric conditions in outdoor particularly precise control of the transceiver \cite{vats2019outage}. To address precise indoor and outdoor localization, pointing, navigation, interference effect, handling of diverse data rates, bit rate requirement, coverage area, a hybrid FSO/VLC heterogeneous interconnection is proposed in literature \cite{pesek2018demonstration, huang2017hybrid}. The implementation and deployment of FSO/VLC hybrid network coordination together with user mobility support, localization, routing algorithm, traffic management, and handover issues are clearly discussed in \cite{huang2017hybrid}.  

In order to measure the effectiveness of any system, performance analysis is a vital issue. The performance metrics assessment can be carried out in various ways such as analytical evaluation, numerical method, simulation environments, and experimental verification. Different works used different performance metrics such as throughput analysis, BER, spectral efficiency, outage probability, secrecy rate analysis, symbol error rate, ergodic capacity, SINR, and power efficiency to investigate received signal quality, system capacity and reliability. Some applications of hybrid FSO systems with brief description are presented in Table \ref{Table:7}. Table \ref{Table:8} and Table \ref{Table:9} summarizes the key objectives and detailed discussion on different aspects on recent research trends of hybrid FSO.


\begin{table}
\centering
\caption{Summary of Hybrid FSO Applications}\label{Table:7}
{\tabulinesep=1mm
\begin{tabular}{|m{20mm}|m{20mm}|m{85mm}|m{25mm}|c|}
\hline
 	Hybrid type & Objective & Contribution & References \\\hline
 	Microwave/FSO and mmWave/FSO & Backhaul connectivity & Seamless microwave photonic link composed of RoFSO, RoF and RF, traffic shifted from RF to FSO network. Effect of channel impairments like chromatic dispersion, atmospheric turbulence, and multi-path induced fading are considered. Complementary properties of the FSO and mmWave channels, diversity selection combining technique improves the link reliability in terms BER of under strong turbulence & \cite{nguyen2019m, alavi2016towards, bohata201824, zhang2018ultra, dat2018seamless, esmail20195g,  rakia2017cross, zhou2017distributed, shakir2017performance, lei2017secrecy,  kumar2014quantize, khan2017adaptive, rakia2015outage} \\\hline
 	RF/FSO  & Backhaul connectivity & Hybrid RF/FSO improve the link reliability for long distance communication. Extensive simulation is conducted to evaluate SINR, BER taking into account of limiting factors under different network scenarios & \cite{kumar2014quantize, jamali2016link, kong2015performance, salhab2016power, trinh2016mixed, enayati2016deployment, varshney2017cognitive, soleimani2015generalized, tang2014link, palliyembil2018capacity, varshney2018cognitive, hassan2019hybrid, odeyemi2020secrecy, lei2020secure} \\\hline	
 	Acoustics UWC/FSO & High bit rate, link reliability enhancement & Based on the link distance,  oceanic turbulence induced scintillation and misalignment in underwater links, and pointing, acquisition, and tracking (PAT) establishment, LOS/NLOS condition, a UWC/FSO system is established for reliable communication & \cite{kaushal2016underwater, zeng2016survey, lei2020performance, yadav2017optimal , tsai2020500, sun2020review} \\\hline
\end{tabular}}
\end{table}

\begin{table}
\centering
\caption{Summary of Hybrid FSO Research Trends}\label{Table:8}
{\tabulinesep=1.3mm
\begin{tabular}{|p{12mm}|p{21mm}|p{32mm}|p{65mm}|p{13mm}|}
 	\hline
 	Reference & Metrics & Objectives & Description & Remarks\\\hline	
 	\cite{zhou2017distributed} & Throughput, Outage probability & Data rate maximization of vehicular ad-hoc network (VANET) & Developed a joint channel allocation and rate control to enhance throughput and solve the cross layer design problem incorporating carrier sense multiple access (CSMA). Capacity and outage probability are also satisfied through alternating detection method & System capacity \\\hline
 	\cite{enayati2016deployment} & Throughput  & Orthogonal frequency  division multiple access (OFDMA) based scheme ensure the backhaul performance of FSO link & Proposed a relay assisted OFDMA based throughput maximization of hybrid RF/FSO link considering multiuser resource allocation scheme including power and subcarrier allocation & System capacity, signal quality \\\hline
 	\cite{tang2014link} & Throughput & Network control approach for throughput optimization based on various channel conditions & Studied the achievable throughput enhancement using mixed linear integer problem. Also, proposed a heuristic scheduling and traffic demand routing adopting physical channel interference model & System capacity, reliability \\\hline
 	\cite{rakia2015power} & Outage probability, SINR & Power adaption technique driven by truncated channel inversion for ensuring fixed SINR at receiver & Determined the improvement of outage performance  assuming two distinctive power adoption strategies with analytical derivations & System capacity, signal quality \\\hline
 	\cite{makki2016performance} & Throughput, outage probability, power efficiency & Performance evaluation assuming precise channel state information at receiver & Derived the closed form expressions for the signal decoding. Also, analyzed the throughput and outage probability in consideration of adaptive power allocation and state of channel conditions & System  capacity  \\\hline
 	\cite{najafi2017optimal} & Throughput, power efficiency & Throughput maximization using parallel relay assisted networks & Focused on the optimal system design according to sub-optimal buffer aided mechanism such as the optimal relay selection strategies and optimal time allocation & System capacity, reliability \\\hline
 	\cite{salhab2016power} & Ergodic capacity, outage probability, blocking probability, SINR & System performance optimization with opportunistic scheduling & Closed-from expressions are derived to evaluate the system performance analyzed varying different atmospheric turbulence conditions. Monte-Carlo simulations are conducted to justify the exact and asymptotic results in the presence of turbulence fading and pointing error & System capacity, signal quality, reliability \\\hline
 	\cite{el2016security} & Ergodic capacity, power efficiency, outage probability & Investigate the performance of multiuser SIMO hybrid relay networks with opportunistic user scheduling & Outage probability, symbol error probability and ergodic channel capacity are assessed based on closed-from expressions using maximal ratio combining and selective combing in the presence of different fading model  & System capacity, signal quality \\\hline
 	\cite{ansari2013impact} & Ergodic capacity, outage probability, BER & Performance evaluation of asymmetric link of dual-hop relay transmission system with pointing error & Develop the closed-from expressions of CDF, PDF, and moments of SINR validated with the help of different metrics & System capacity \\\hline
 	\cite{usman2014practical} & Ergodic capacity, outage probability, BER & Performance investigation of switching-based transmission scheme & Investigate average BER, ergodic capacity and outage probability considering gamma-gamma fading and Nakagami fading & system capacity \\\hline
 	\cite{balti2017aggregate} & Ergodic capacity, SINR, SER, outage probability & Hardware impairments for heterodyne IM/DD with fixed gain relay scheme & Impact of hardware impairments for different SNR is illustrated. Analytical expressions for different performance parameters are derived and compared the numerical results through the numerical integration method & system capacity, signal quality \\\hline
 	\cite{jamali2016link} & Throughput, delay tolerant performance & Cascading of small cell networks to hybrid RF/FSO backhaul link & Developed optimal fixed and adaptive link allocation algorithms based on channel state information to investigate delay limited and delay tolerant transmission schemes & system capacity, signal quality \\\hline
 	\cite{trinh2016mixed} & Throughput, BER, outage probability, ergodic capacity, SNR & Cost reduction and scalability improvement of mmWave RF/FSO  & Examined the performance metrics considering the effect of pointing errors owing to the misalignment between the transmitter and receiver. Note that the cascaded channels are modeled by the Rician and Malaga distributions & system capacity, reliability \\\hline
 	\cite{yang2015performance} & Throughput, BER, outage probability, SINR & Mitigation technique of multipath fading and atmospheric turbulence for a dual hop hybrid system & Proposed a transmit diversity at the sending end and selection combining technique at the receiver end to investigate the effect of pointing errors on the FSO link. Also, examine the the system performance and provide a light of way to mitigate limiting factors & System capacity, reliability, signal quality \\\hline
 	\cite{el2017effect} & Outage probability, power efficiency & Determine the optimal transmit power and co-channel interference & Exact closed-form expression is derived to find an optimal power allocation. The secrecy performance and physical layer security performance is studied under different constraints & Reliability \\\hline
\end{tabular}}
\end{table}	
 	
\begin{table}
\centering
\caption{Summary of Hybrid FSO Research Trends (Continued)}\label{Table:9}
{\tabulinesep=1.3mm
\begin{tabular}{|p{12mm}|p{23mm}|p{32mm}|p{65mm}|p{13mm}|}
 	\hline
 	Reference & Metrics & Objectives & Description & Remarks\\\hline
 	\cite{bag2018performance} & Ergodic capacity, outage probability, BER & Estimation higher link reliability under extreme atmospheric constraints  & To avoid needless switching, a single FSO link is considered to active whereas an extra mmWave RF/FSO link is used as a backup. The primary link is modeled by gamma-gamma distribution assuming strong turbulence & Reliability, system capacity \\\hline
 	\cite{shakir2017performance} & SNR, BER, Outage probability  & Switching strategies to minimize the bandwidth wastage maintain the same data rate & Performance investigation using selection combining diversity based on closed-form expression of BER and outage probability. Ensure the same data with the same data rate over the FSO link without having CSI & Reliability, system capacity \\\hline
 	\cite{petkovic2015partial} & BER, Outage probability & Impact of outdated CSI based relay selection on the system performance & Considered partial amplify-and-forward relay selection. RF link is subject to Rayleigh fading and FSO link is modeled by gamma-gamma distribution. Mathematical expressions of various performance metrics are derived & Reliability \\\hline 	
 	\cite{wu2017dynamic} & Outage probability, power efficiency & Cost minimization due to power consumption while ensuring packet success probability & Performance investigation by joint consideration of power allocation and dynamic link selection for guaranteeing the long-term reliability requirements. The closed form of power allocation policy is derived for link selection and Lyapunov optimization algorithm is used to solve the problem & Reliability \\\hline
 	\cite{chen2016multiuser} & BER, Outage probability & Effective link quality scheduling using multiuser diversity & Analyzed a hybrid RF/FSO point-to-multipoint system assuming multiple FSO users, RF users and access points. The asymptotic closed form expressions for average BER, outage probability are derived considering multiuser diversity gain & Reliability, signal quality \\\hline
 	\cite{al2018two} & SINR, SER, ergodic capacity, Outage probability & Two way multiuser relay networks with opportunistic scheduling and non-uniform channel fading is investigated & An efficient optimal power transmission algorithm is developed for a heterodyne detection scheme. The system performance is evaluated for two way relaying network and single way relaying network in the presence of weather turbulence and pointing error effect & System capacity, reliability, signal quality \\\hline
 	\cite{arezumand2017outage} & SINR, BER, Outage probability & Dual hop hybrid system based on cognitive amplify and forward relay networks & Fixed gain and channel assisted relaying policy is derived according to the asymptotic expressions. Heterodyne IM/DD detection receiver with double gamma fading channel is considered. In addition, the diversity order and diversity multiplexing trade-off is carried out to investigate the overall system performance  & System capacity, reliability \\\hline
 	\cite{touati2016effects} & BER, Outage probability & Controlling the FSO link fragility and reduction of outage probability  & Analytical expressions of outage probability and BER are derived varying different modulation scheme with its different channel impairments & System capacity, reliability \\\hline
 	\cite{lei2020performance} & SINR, BER, Outage probability & Performance analysis of dual hop RF/UWOC hybrid systems under fixed and variable gain relaying schemes  & Evaluate SINR, BER, and outage probability under the effect of water bubble levels, temperature gradient, water conditions and detection techniques of underwater wireless optical communication & Reliability \\\hline
 	\cite{tsai2020500} & Throughput, BER & Boosting of total channel capacity with FSO backbone UWOC & Total achievable throughput of FSO/UWOC link is 500 Gbps integrating PAM4 modulation with five (5 R/G/B LD transmitters of two stages) wavelength-polarization multiplexing scheme  & System capacity, signal quality \\\hline 
\end{tabular}}
\end{table}

\section{Radio over FSO (RoFSO) System}

\begin{figure}
  \centering
  \includegraphics[width=0.9\columnwidth, height = 1.6in]{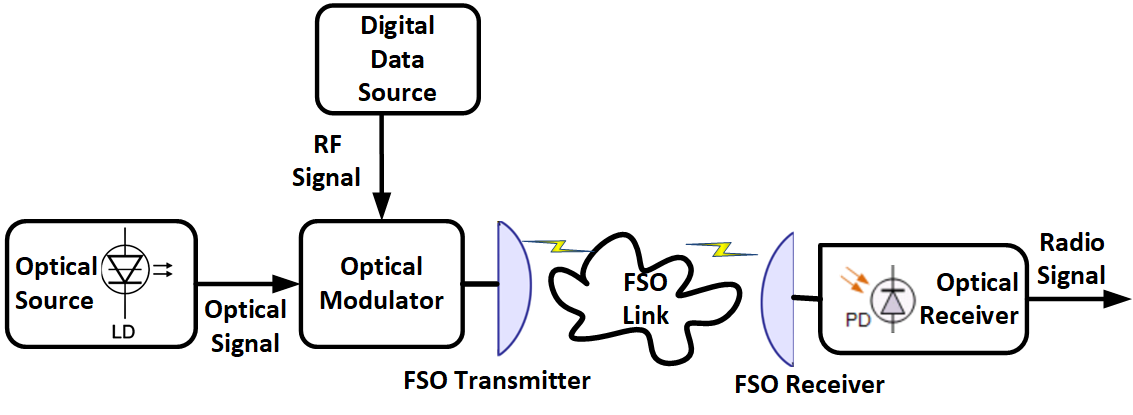}\\
  \caption{Simplified Block Diagram of RoFSO System.}\label{Fig:11}
\end{figure}

Radio over free space optics (RoFSO) has emerged as a prominent technologies in future communications due high data transmission and low energy consumption. The concept of RoFSO system is composed of radio over fiber (RoF) and FSO system where the optical signal is modulated with the RF signal at a transmitter end before transferring to the receiver. In other words, RoFSO is technique that allows simultaneous data transmission of broadband RF signals in bidirectional path over FSO link. A technique of modulating RF subcarriers onto an optical carrier for transmission over an optical fiber link is known as RoF. However, RoFSO system can overcome the last mile problems for broadband services with desired reliability of communication link in area where the installation of fiber cable is restricted \cite{bohata201824}. To meet the unprecedented growth of user demands in 5G and B5G communications, a telecom operators need to install more number of base stations (BSs) in order to avoid spectral congestion in lower RF frequency bands which significantly increases the net present cost and system complexity. Whereas RoFSO provides the greater level of reliability, flexibility in deployment, low signal loss, immunity to electromagnetic interference with the advantage of high optical bandwidth in a cost effective way \cite{chaudhary2014role}. 
The principle of RoFSO is similar the operation of WDM technique. The generalized block diagram of radio over FSO (RoFSO) system is clearly presented in Fig. \ref{Fig:11}. Under the RoFSO scheme, the central base stations distribute optical signals through free space to multiple remote stations. In general, Mach–Zehnder modulator (MZM) is used to perform the modulation of an optical beam as a function of a RF signal by combining phase-delayed optical beams. In MZM, a beam-splitter separated the incident laser light waves into two paths to illustrate a linear electro-optic (EO) effect that produces birefringence in an optical medium when the electric field is applied. By tuning the electric field i.e., RF input signals the phase delay of upper and lower path can be controlled and then the beams are combined at the output either constructively of destructively. The EO mechanism changes the refractive index of the optical medium and thus, the MZM can be used for amplitude, phase, frequency modulations \cite{das2015radio}. 

\section{Multi-user FSO Communication}

\begin{figure}
  \centering
  \includegraphics[width=0.9\columnwidth, height = 1.6in]{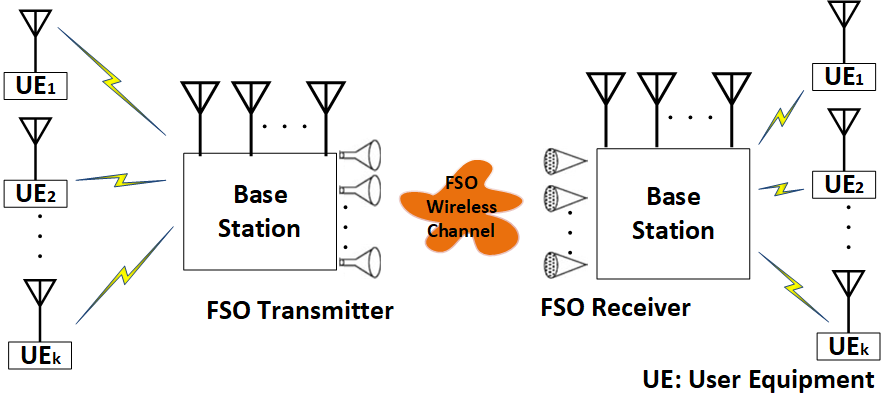}\\
  \caption{Generalized Multiuser MIMO FSO System.}\label{Fig:12}
\end{figure}

Multiple users transmit different signals, which is mixed in the wireless propagation medium. Blind source separation (BSS) technique is used to extract the original information from the mixed signals. This problems exist in numerous applications such as biomedical data analysis, speech signals identification, machine learning based communications, etc. \cite{aveta2017multi}. Since users are randomly distributed over space, different channels will experience independent fading. Exploiting multiple antenna diversity technique can considerably improve the FSO wireless transmission performance. Multiuser diversity offers some inherent benefits over antenna diversity as it explores independent fading channels. It also support a simple receiver structure where a single antenna is equipped per receiver. Introducing a number of access points through long-range FSO link that amplifies the received signal via short distance RF link and forward to next hop increase the transmission distance and reliability of data services as well. Authors \cite{amirabadi2019performance} reported a multi user multi hop combined RF/FSO communication systems that offers lower energy consumption, low delay processing and minimum complexity. Amplify and forward relaying is used for fixed gain amplification when channel state information (CSI) is unknown, in contrast, adaptive gain is a suitable choice when CSI is known. Chen \textit{et. al.} \cite{chen2016multiuser} studied the multi-user diversity RF/FSO point-to-multipoint MISO communication systems consisting hybrid access points aiming to minimum error probability and BER. A multiuser mixed RF/FSO two way relaying (TWR) method considering asymmetric channel gain is presented in \cite{al2018performance} to examine the outage probability and energy-efficient power allocation. However, Multiuser MIMO transmission is a better choice over spatial multiplexing as it can considerably extend the capacity of wireless communication systems \cite{yang2020advanced}. Fig. \ref{Fig:12} shown the basic multiuser MIMO communications in FSO perspectives.

It is essential to sense the optical spectrum periodically in order to avoid interference among multiple users particularly to detect the channel occupancy. Spectrum sensing plays a significant role in multi-user FSO communications that allow user to monitor the available optical frequency spectrum in their operating range \cite{kong2020multiuser}. A generalized blind spectrum sensing [i.e. applicable to all optical wavelengths] policy based on SNR estimation for an unknown optical signal over strong atmospheric turbulence conditions is reported in \cite{arya2019generic}. An extensive investigation of energy detection method based optical free spectrum sensing for FSO communication over exponentially distributed channel is conducted in \cite{ho2019spectrum2}.

\section{MIMO FSO Systems}

\begin{figure}
  \centering
  \includegraphics[width=0.9\columnwidth, height = 1.6in]{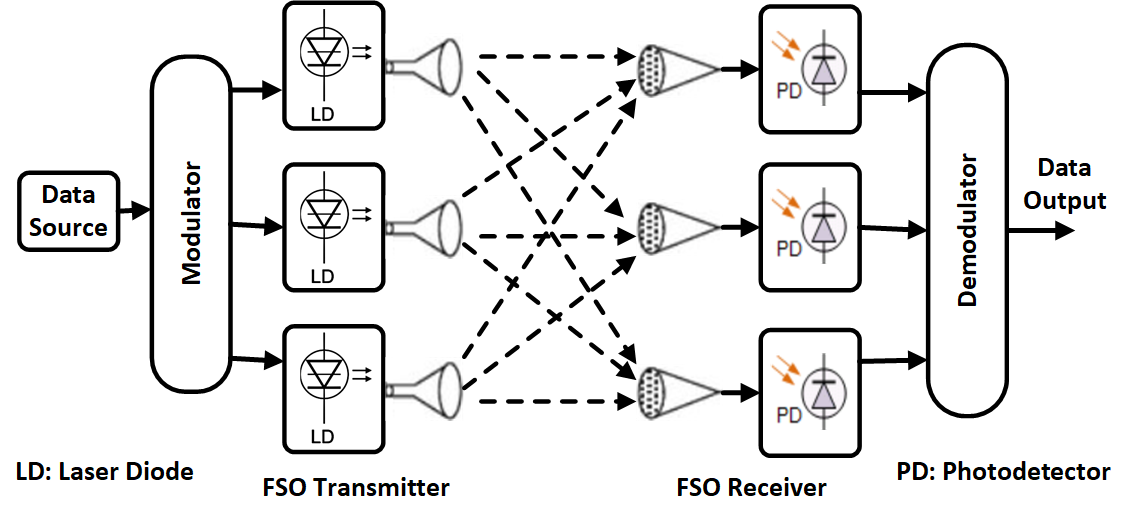}\\
  \caption{Simplified Illustration of MIMO FSO System.}\label{Fig:13}
\end{figure}

Diversity techniques e.g. spatial diversity realized multiple beams at either the transmitter or multiple apertures at the receiver end can considerably improve the wireless FSO transmission over fading induced channels. The spatial diversity technique helps to enhance the FSO transmission reliability introducing inherent signal redundancy \cite{ren2013high, sharda2020diversity}. It is worthy to mention that turbulence-induced fading considerably deteriorates single input single output (SISO)-FSO link \cite{alimi2017effects}. As an indication of the context, additional power margin is employed to achieve desire QoS, this power penalty can be alleviated by increasing the transmission power. However, this approach is impractical due to eye safety regulations and relatively lower reliability. Typically, a conventional Gaussian light signal can experience beam wandering effect, causes power fluctuations and wavefornt distortions due to unwanted atmospheric turbulence. In addition, misalignment between transceivers causes significant power loss at the receiver end as explained beforehand. There have been several approaches to handle possible signal degradations. Aperture averaging is a potential technique of improving system performance by means of incorporating wider lens at the receiver side to mitigate fading effect via averaging intensity fluctuations. This technique is efficient under moderate-to-strong atmospheric turbulence when the receiver aperture is greater than fading correlation length, $\sqrt{\lambda}l$, with $l$ denote the transmission distance and $\lambda$ represents the wavelength \cite{khalighi2014survey}. Using multiple small apertures instead of large aperture at the receiver is a suitable way to reduce fading with taking the advantage of aperture averaging. Employing several apertures is more advantageous under strong turbulence regime but involves implementation complexity. For single input multi output (SIMO) spatial diversity systems, equal gain combining (EGC) method is used that offer less complexity but shows comparatively inferior performance to optimal maximal ratio combining (MRC) \cite{yang2020advanced}. On the other hand, multi input single output (MISO) FSO scheme at the transmitter end is mostly used to transmit the same signal on the separate beams, which is referred as a repetition coding (RC). Authors in \cite{garcia2015novel} pointed out that MISO FSO based spatial diversity technique combat the fading effects implementing the multiple laser sources at the receiving end. Moreover, generalized selective combining (GSC) method recognized as a low complexity diversity scheme to select the best possible path from subset and combining them using MRC fashion. However, the weak channel estimation limits the merits of MRC scheme that may degrade the system performance considerably. A simplified illustration of MIMO FSO system is displayed in Fig. \ref{Fig:13}.

Likewise RF systems, MIMO technique is very popular to combat turbulence induced multipath fading increasing the bit rate and quality of signal transmission through spatial multiplexing \cite{yang2020advanced}. Under this scenario the FSO transceiver serve multiple end users simultaneously using the same frequency band where multiple co-located antennas provide the larger multiplexing gain. In order to control inter-user interference, several precoding techniques are studied for multiuser MIMO transmission. Typically, intensity modulation/direct detection (IM/DD) method is implemented in MIMO-FSO systems, where IM is used for signal transmission and DD is employed at the receiver. According to \cite{yang2020advanced}, nonlinear precoding schemes attain better performance over linear precoding method, which consider the total rate capacity of all users but involves increased implementation complexity. For instance, the dirty paper coding (DPC) cancel out inter-user interference prior to data transmission with full knowledge of CSI \cite{jindal2005dirty}. On the other hand, zero-forcing beamforming (ZFBF) \cite{yoo2006optimality} and random unitary beamforming (RUB) \cite{sharif2005capacity} are popular linear precoding schemes have significantly lower implementation complexity. The precoding matrix is modeled pseudo inverse matrix of selected users in ZFBF, whereas, the precoding matrix in RUB is designed based on the channel quality of randomly selected users on distinctive beamforming directions. Reference \cite{safari2008we} studied the orthogonal space time block code (OSTBC) and repetition code (RC) under log-normal fading channels conditions. However, the log-normal atmospheric channel model is not sufficient when the huge number of heterogeneous devices are connected in 5G \cite{alimi2017toward}. Recently, numerous research works conducted focusing on MIMO-OSTBC FSO applications due to higher diversity gain and wide range of FSO applications \cite{wang2017space, wang2018high, amhoud2019oam, biagi2020analysis}. Besides the OSTBC, the RC scheme outperforms than OSTBC in terms of diversity gain but the RC applications with MIMO is challenging because of timing asynchronous alignment between transceivers and thereby, resulting the time difference of signal propagation \cite{alimi2020simple}. The use of repetition coding in MIMO FSO systems endeavoring to combat turbulence effects by introducing redundancy is studied in \cite{djordjevic2018advanced}. However, several laser beams having identical frequency are interfere among each other when simultaneously received at the same aperture. 

To address timing misalignment of asynchronous MIMO FSO channels, OSTBC based MIMO scheme offer better performance and robust to inter-symbol interference (ISI) maintain the orthogonality conditions. However, the system performance gap between RC and OSTBC schemes is more significant for higher order of MIMO FSO systems. Saidi and Hamdi \cite{saidi2020end} presented BER performance for multihop hybrid RF/FSO systems using M-ary pulse positon modulation (MPPM) and MIMO technique under gamma-gamma turbulence channels. The enhancement of average secrecy capacity and improvement of secrecy outage probability for MIMO-FSO link with equal gain combing reception technique over gamma-gamma atmospheric turbulence channels is demonstrated first time in literature \cite{boluda2020enhancing}. Ref \cite{song2020demonstration} investigated satisfactory BER, power penalty reduction, and outage probability performance for 10 Gb/s QPSK FSO links using aperture averaging receive diversity and MIMO spatial technique. Eghbal and Abouei \cite{eghbal2019performance} studied received signal intensity of MIMO FSO systems accounting destructive interference with small variations of path length between optical transmitter and photodetector. A disaster management method for MIMO-OFDM FSO links employing polarization division multiplexing (PDM) with Malaga channel model is presented in \cite{jeyaseelan2020disaster}. With PDM, the two high speed OFDM signals are mutually orthogonal polarizations is an excellent way to uplift spectral efficiency. Unlike fibre-based communications, PDM is a suitable choice for FSO systems as it is immune from cross polarization modulation, polarization mode dispersion and associated losses. Under the proposed scheme, two different OFDM subcarriers i.e., eve and odd subcarriers are multiplexed using PDM that enable to transfer two independent data signals on a single wavelength. Das \textit {et. al.} \cite{das2020free} employed Almouti STBC transmit diversity and switch-and-examine combining (SEC) scheme at the receiver side for MIMO-FSO communication systems to investigate ergodic capacity, BER and outage probability over the Malaga turbulence wireless channels.  

\section{FSO Transmission in TCP Layer}

Numerous research works are conducted focusing on signal re-transmission, re-routing, cross layer design, delay tolerant networking in the perspectives of transmission control protocol layer in conjunction with the physical layer \cite{nguyen2016tcp, nguyen2018cross, zedini2014performance}. 

\begin{itemize}

\item	\textit{Re-transmission:}
An automatic repeat request (ARQ) is a re-transmission protocol that is commonly used in reliable data communication. Under this scheme, data transmission is carried out in the set of certain packet frame lengths. The packet is re-transmitted when the receiver does not acknowledge after certain period of time. This process is continuing until the positive acknowledgement is received at the transmitter end. As a result, this go-back-stop-wait N-ARQ re-transmission scheme leads to high bandwidth penalty, big delay and large power consumption as well. Another technique of selective repeat ARQ (SR-ARQ) where the packets are repeatedly send from the transmitter without any feedback response from the receiver end either at data link level or at transport layer. Here, the receiver unit will continue to accept data packets and acknowledge to the transmitter. If the acknowledgement is not received after certain time, it presumed to be data lost and transmit again. On the hand, hybrid ARQ (H-ARQ) combining FEC coding and ARQ error control mechanism in FSO communications is studied in \cite{zedini2014performance}. It is found that superior performance in terms of outage probability and BER is obtained under strong turbulence but it suffers high delay latency and bandwidth penalty. In contrary, cooperative diversity ARQ (C-ARQ) have shown remarkable performance combating turbulence induced fading in FSO links. In addition, modified C-ARQ (MC-ARQ) outperforms a lot better by means of lower transmission delay and higher power efficiency \cite{aghajanzadeh2012information}. Rateless Round Robin algorithm is used for error controlling in practical FSO deployment under strong atmospheric turbulence \cite{hammons2011diversity}.

\item	\textit{Re-routing:}
Data path re-routing in between optical link or low bit rate RF link increase the FSO link reliability during adverse weather conditions. An autonomous dynamic path reconfiguration including topology control mechanisms and beams re-routing algorithms in the context of hybrid RF/FSO link is demonstrated in \cite{yang2015performance}. The proposed method considerably improve the system reliability but increase the processing delay and net present cost. Therefore, the routing algorithm should be designed in such a way to minimize re-routing delays or minimum number of hops with least cost. A common routing protocol called proactive routing where all the possible routes are calculated in prior and stored in the routing table. However, this scheme is not appropriate for large-scale networks as it exploit heavy overhead that makes it less bandwidth efficient. Another routing protocol named reactive routing produce less overhead to the networks in comparison with former proactive routing. This scheme compute new paths based on real-time demands during the malfunctions of existing route but it causes prolonged latency of data transmission. Therefore, hybrid routing scheme combing this two protocols can be employed into clusters of networks for example proactive algorithms should apply within each cluster and reactive routing policy apply among different clusters \cite{yang2015performance}. 

\item	\textit{QoS control:}
QoS can be defined in terms of throughput, jitter delay, latency, signal loss, spectral efficiency and energy efficiency. In order to achieve reliable FSO communication between two nodes, the designed system should meet the specified requirements satisfying QoS. The primary challenges to maintain desired QoS in the FSO systems are the fluctuations of delay, data packet rejection ratio, and overhead. The idea of buffer introducing a IP packet investigation and scheduling processing according to the QoS specifications and link availability. Reference \cite{peach2010performance} proposed 10 Gbps buffer encountered QoS to minimize the packet rejection rate under strong atmospheric conditions in hybrid RF/FSO networks. According to \cite{peach2010performance}, QoS based buffer scheme provides 8 dB less fading margin. An efficient scheduling and buffer algorithm, tight error controlling and proper channel access policies improves the QoS level in medium access control (MAC) layer.

\end{itemize}

\section{Next Generation FSO Communication Networks}

\begin{figure}
  \centering
  \includegraphics[width=0.9\columnwidth, height = 90mm]{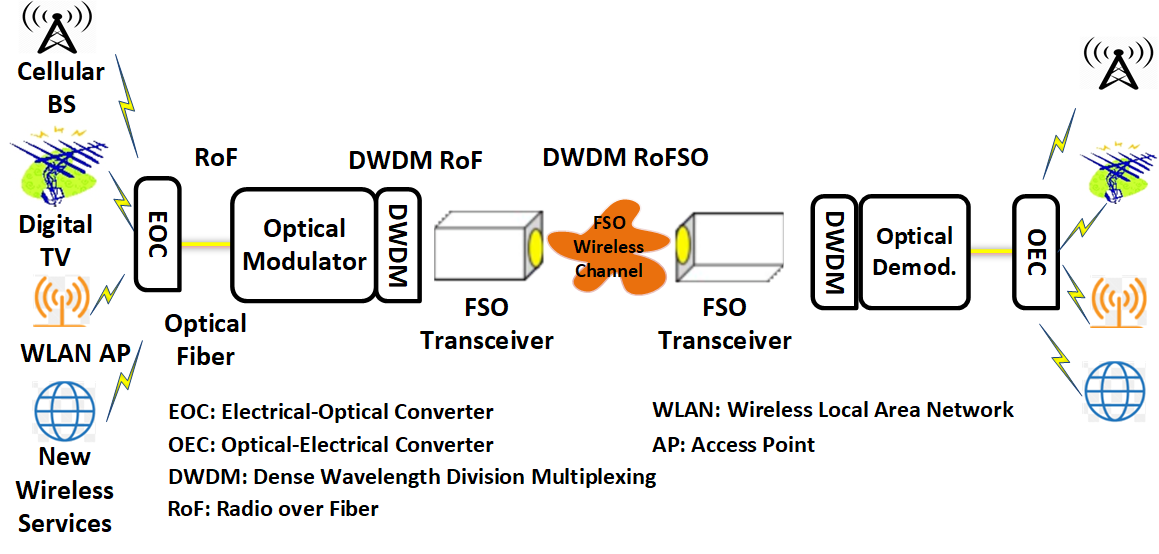}\\
  \caption{Next Generation FSO Networks.}\label{Fig:14}
\end{figure}

In recent years, FSO systems as an alternative wireless communication technology has gained incredible attention to cope up high traffic demand in a cost effective way. In addition, FSO technology address the interference issues and reliable connectivity via high level of security over the existing RF infrastructure. The FSO inter-channel interference can be limited by the controlling the optical beam width at the transmitter station. Exploiting the coexistence of RoFSO, OWC/FSO or RF/FSO compensate the weakness of each channels for the application of intelligent and secured data transmission, battlefield environment monitoring. FSO established a heterogeneous structure (RF, Coax, FSO, optical fibre) to provide extreme capacity of wireless services both in ground and space \cite{raj2019historical}. The optoelectronic devices for example 100 Gb/s optical source, mach-zehnder modulator (MZM), wavelength division multiplexing (WDM), photodetector etc. that are being used for high speed fiber optic infrastructure can also be used in FSO systems. The design of integrating the conventional high data rate signals into FSO systems allows to transfer terrestrial digital HD TV signals, WLAN, perform 5G communication protocols, and IoT/IoE wireless services. A performance evaluation of FSO link in terms of BER, SNR, power-in-bucket using rectangular coherent flat-topped beam in consideration of atmospheric turbulence has been investigated in \cite{majumdar2018optical}. To handle the effect of atmospheric scattering and absorption, all optical relaying and intermediate repeater approach is demonstrated. Reliability analysis contemplating polarization shift keying modulated FSO system and optimization of FSO enabled cellular backhauling for the next generation mobile networks are also investigated in \cite{kaushal2016optical}. A simplified illustration of next generation FSO communication systems is depicted in Fig. \ref{Fig:14}.

The perfect FSO channel estimation is a challenging task as it involves high cost and power consumption. It is not an appropriate choice in cellular communications system since pilot transmission deteriorates the data rate. Inspired from this encountered problems, authors \cite{amirabadi2020deep} proposed deep learning based low cost and low complexity channel estimation for FSO system. Although blind channel estimation does not require pilot transmission, in addition, it offers lower complexity but this approach is not realistic because of fixed channel conditions \cite{zhu2015fast}. Recently, machine learning \cite{thrane2016machine} has attracted to investigate optical performance monitoring, which is very useful when the system model is complex. On the other hand, deep learning \cite{goodfellow2016deep} could solve the complex problems when the input is not enough for computation and provides dynamic statistical signal processing tools to generate perfect probabilistic channel model for received data impairments. Deep neural network (DNN) \cite{wang2017intelligent, li2018joint} algorithms can be used in high-level simulation to analyze input data, offering fast response. DNN algorithms is employed for best modulation scheme identification \cite{khan2017joint}, controlling of optical amplifiers \cite{huang2017dynamic}, routing paradigms \cite{rottondi2018machine}, and quality evaluation of transmitted data signals \cite{zhong2019routing}. 

From the aforementioned discussion, machine learning (ML) for future generation FSO networks becomes a hot topic. Few researches are conducted in ML based FSO communication such as detection purpose \cite{zheng2015svm}, error correction for advanced optics \cite{li2017bp}, and perfect demodulator design for angular optical beams \cite{tian2018turbo, li2017adaptive}. 

\section{Summary}		\label{12}

With the proliferation of diversified multimedia applications, RF systems is not capable to fulfil the unprecedented demand by IoT, 5G, and beyond communication systems. Various wireless access technologies have variety range of appealing features with some limitations. It is widely known that RF and optical wireless technologies (e.g. FSO) perform reciprocal characteristics. The key benefits of the RF technology are mobility support, transmission performed under LOS/NLOS conditions; provide wider coverage area, whereas it significantly suffered due to bandwidth shortage, electromagnetic interference, and smaller throughput performance. On the other hand, FSO systems among all other OWC technologies offers the potential advantages of longer coverage area and enormous bandwidth. Unlike RF signal, the optical beam cannot be detected by spectrum analyzers and laser beams cannot penetrate walls. Moreover, FSO transceiver are easy to deploy, lightweight, compact, easily expandable. The narrow optical beams between FSO transceiver has the potential to establish electromagnetic interference free longhaul ultra-speed data communication guaranteeing a greater level of energy efficiency. However, the reliability issue of FSO communications is degraded due to some limiting factors such as fog, heavy rain, dense snowfall, thick cloud, pointing errors, scintillation effect, and atmospheric molecular absorption.    
An extensive comparison of different optical wireless technologies is carried out at the beginning of this article to pointing out the key difference of FSO technology compared to other OWC schemes. Thereafter, the wide range of FSO applications in different perspectives such as space communications, cyber security, solution of last mile problems, and core networks are clearly described. The recent advancement of optical light source e.g. solid-state based monochromatic LASER transmitter and the performance of various photodetectors as optical receivers are explicitly presented pointing on merits and demerits. The coherence property of laser diodes (LDs) enables fast data transmission with less interference against LEDs. At the receiver end, PiN photodiode (PD) is a preferable choice for short distance FSO links; in contrast, Avalanche PD (APD) exhibits satisfactory system performance because of improved level of receiver sensitivity for longer link distance. 
The goals of FSO systems are different, and numerous approaches have been proposed to address reliability issues. Pointing, Acquisition, and Tracking (PAT) technique to maintain LOS directivity between two endpoints \cite{yousif2019atmospheric, aghajanzadeh2012information, miglani2019statistical}, adaptive modulation schemes \cite{bekkali2010transmission, hassan2016subcarrier, tsonev2013complete, yu2014iterative, xu2009coded, zhang2018improved, zhang2014time, jiang2018investigation, chow2019performances, ebrahimi2018investigation}, appropriate channel coding based on different optical wireless link models \cite{jurado2010efficient, hassan2018delay, hassan2017statistical, zhou2015optical, karimi2012novel, anguita2010rateless, djordjevic2010adaptive, ali2019performance, pernice2016moderate}, background noise reduction technique \cite{hashmi2014analysis}, spatial diversity mechanisms \cite{kaymak2017divergence, yousif2019atmospheric, aghajanzadeh2012information, miglani2019statistical, sharda2020diversity}, coherent FSO systems \cite{mudge2016development, zhu2017block, priyadarshani2017effect, boluda2019outage, boluda2017effect}, aperture averaging for eddies induced fading \cite{bayaki2009performance}, adaptive optics \cite{viswanath2015design, vedrenne2016adaptive} for wavefront fluctuations, relay nodes \cite{kashani2013optimal, boluda2015impact, yang2014performance} to attain high channel gain, and hybrid FSO systems \cite{pratama2018bandwidth, hou2006vertical, purwita2018handover, saber2019physical, tubail2020physical, abd2017effect, liverman2016hybrid} to overcome the limitations of a single technology. A number of research works have been performed to address several technical issues, however, some issues remain. For the successful deployment of FSO systems, several issues have been counted as paramount research concerns, including interference mitigation \cite{chen2015fractional, kazemi2016downlink, liu2017interference}, automatic error correction \cite{nawaz2019quantum}, throughput enhancement \cite{shakir2017performance, lei2017secrecy}, energy efficiency \cite{jahid2017pv, shams2017comp, jahid2019toward} for 5G, B5G and IoT applications. In addition, the efficient FSO system design issues for quantum and THz communications \cite{kshatriya2016estimation, majumdar2019principles}, multi-hop, multi-user and cooperative communications based on relay transmission \cite{peppas2013capacity, boluda2018amplify, abou2010cooperative, bhatnagar2011performance, bandele2016saturation, dabiri2018performance, huang2018performance}, WDM FSO communications \cite{arimoto2008320, mbah2016outage, lin2014high, tan201812, zhao2018200, choi2011hybrid}, adaptive switching mechanism in underwater optical wireless communications \cite{saeed2019underwater}, horizontal and vertical handover \cite{lee2016internet} issues in heterogeneous networks, adaptive user association policies for optimal load balancing for hybrid FSO systems \cite{stefan2014hybrid, wang2016fuzzy}, physical layer security \cite{skorin2016physical, boluda2019secure, lei2016secrecy, wang2014secure, van2018physical, lopez2015physical, wang2018secrecy, lei2018secrecy1, lei2018secrecy2, pan2019secrecy, saber2017secure}, software defined FSO networking \cite{kobo2017survey, cox2017advancing, haque2016wireless, lin2017indoor, kashani2015performance, hammouda2018link} issues are also discussed. Thereafter, radio over FSO (RoFSO) technology is presented in 5G and other communications perspectives \cite{bohata201824}. The comprehensive study of MIMO FSO spatial diversity is studied in \cite{wang2017space, wang2018high, amhoud2019oam, biagi2020analysis, alimi2020simple, djordjevic2018advanced, saidi2020end, boluda2020enhancing, song2020demonstration, eghbal2019performance} targeting to achieve robust system performance. In light of existing research works, we focused the prospect of the future generation FSO networks incorporating machine learning, deep neural networks, efficient routing algorithms, QoS controlling methods.

Table \ref{Table:10} and Table \ref{Table:11} discusses the summary of different FSO systems in terms of distinctive performance metrics. 

\begin{table}
\centering
\caption{Summary of FSO channel capacity and link design}\label{Table:10}
{\tabulinesep=1.3mm
\begin{tabular}{|p{12mm}|p{25mm}|p{40mm}|p{13mm}|p{15mm}|p{13mm}|p{15mm}|} \hline
Reference & Modulation & Link model & Ergodic capacity & Outage Probability & BER/SER & Throughput \\ \hline
\cite{islam2019effect} & OCDMA-WDM, PPM & Non fading & No & No & Yes & Yes \\\hline
\cite{nazrul2019effect} & OCDMA-WDM, PPM & Gamma-Gamma & No & No & Yes & Yes \\\hline
\cite{islam2015analytical} & OOK&Non fading & No&No & Yes& Yes\\\hline
\cite{tang2014link} & OFDMA OOK &Rayleigh & No&No &No& Yes \\\hline
\cite{rakia2015power} & M-QAM& Gamma-Gamma, Nakagami-m& No&Yes & No&Yes \\\hline
\cite{makki2016performance} &IM/DD BPSK& Gamma-Gamma, log-normal&No &Yes & No&Yes \\\hline
\cite{najafi2017optimal} &OOK &Gamma-Gamma &No & No& No& Yes\\\hline
\cite{salhab2016power} & BPSK&Gamma-Gamma, Rayleigh &Yes &Yes &No &Yes \\\hline
\cite{el2016security} & BPSK&Gamma-Gamma, Nakagami-m &Yes & Yes& No& No\\\hline
\cite{ansari2013impact} &DBPSK, CBPSK, CBFSK &Gamma-Gamma, Rayleigh &Yes &Yes &Yes &No \\\hline
\cite{zhang2015unified} &OOK &Rayleigh, Nakagami-m &No &Yes &Yes &No \\\hline
\cite{usman2014practical} &QAM &Log-normal & Yes& Yes& Yes&No \\\hline
\cite{balti2017aggregate} &SIM-BPSK &Malaga &Yes &Yes &Yes & No\\\hline
\cite{jamali2016link} & OOK&Rician & No&No & No& Yes\\\hline
\cite{trinh2016mixed} & M-QAM, M-PSK&Rician, Malaga &Yes &Yes &Yes &Yes \\\hline
\cite{yang2015performance} & BPSK& Rayleigh&No & Yes& Yes&Yes \\\hline
\cite{el2017effect} &IM/DD,BPSK &Gamma-Gamma, Nakagami-m &No &Yes & No& No\\\hline
\cite{bag2018performance} &OOK &Gamma-Gamma, Rayleigh & Yes&Yes &Yes &No \\\hline
\cite{shakir2017performance} &BPSK, QPSK, 8-PSK & Gamma-Gamma&No &Yes &Yes & No\\\hline
\cite{amirabadi2019performance} &DPSK &Gamma-Gamma &No &Yes &Yes &No \\\hline
\cite{petkovic2015partial} &DBPSK, SIM BPSK &Rayleigh &No &Yes &Yes &No \\\hline
\cite{soleimani2015generalized} &DBPSK, CBPSK, CBFSK &Gamma-Gamma, Nakagami-m & No &Yes &Yes &No  \\\hline
\cite{wu2017dynamic} &OOK & Nakagami-m&No & Yes&No &No \\\hline
\cite{chen2016multiuser} &BPSK &Gamma-Gamma &No &Yes &Yes &No \\\hline
\cite{al2018two} & BPSK& Gamma-Gamma&Yes &Yes & Yes&No \\\hline
\cite{arezumand2017outage} & SIM (CBFSK, CBPSK, DBPSK)&Nakagami-m &No &Yes &Yes &No \\\hline
\cite{touati2016effects} &OOK &Gamma-Gamma &No &Yes & Yes&No \\\hline
\cite{varshney2017cognitive} &OSTBC &Nakagami-m &No &Yes &No &No \\\hline
\cite{palliyembil2018capacity} &IM/DD BPSK &Nakagami-m, Malaga &Yes & Yes& Yes&No \\\hline
\cite{odeyemi2020secrecy} &IM/DD BPSK &Gamma-Gamma, Rayleigh &No &Yes &No &Yes \\\hline
\cite{varshney2018cognitive} &BPSK, DBPSK, BFSK &Gamma-Gamma &No &Yes &Yes &No \\\hline
\cite{lei2020secure} &IM/DD BPSK &Nakagami-m, Malaga & No& No&Yes &Yes \\\hline
\cite{lei2020performance} &OOK, BPSK, QPSK, M-QAM & Generalized Gamma, Nakagami-m &No &Yes &Yes &Yes \\\hline
\cite{tsai2020500} &four-level PAM &Non fading &No &No &Yes &Yes \\\hline
\cite{li201982} & PAM4& Non fading&No &No &Yes &Yes \\\hline
\cite{xing2019joint} &M-PSK &Non fading &No &No &Yes &No \\\hline
\cite{kong2015performance} &M-PSK, DPSK & Gamma-Gamma&No &No &Yes &No \\\hline
\cite{balti2018mixed} & CBFSK, CBPSK, DBPSK &Generalized Gamma, Nakagami-m &Yes & Yes&Yes &No  \\\hline
\end{tabular}}
\end{table}

\begin{table}
\centering
\caption{Comparison of throughput and latency \cite{chowdhury2018comparative, chowdhury2019role}}\label{Table:11}
{\tabulinesep=1.3mm
\begin{tabular}{|p{17mm}|p{23mm}|p{18mm}|p{27mm}|} \hline
Technology & Data rate (Gbps) & Latency (ms) & Communication link  \\ \hline
Optical fiber & 100 & $<$1 ms & E2E  \\ \hline
FSO	 & 40 & $<$1 ms & E2E  \\ \hline
mmWave & 10 & $<$1 ms & E2E  \\ \hline
Microwave & 1 & $<$1 ms/hop & P2P/P2mP  \\ \hline
\end{tabular}}
\end{table}

\section{Research Challenges, Directions and Open Issues}		\label{13}

Diverse type of multimedia applications and high data rate digital wireless services offered by 5G and B5G, the existing RF system is incapable to fulfill the intended data transmission rate. As a result, FSO scheme offering an ultra-band frequency range of license free spectrum is an attractive viable solution. FSO communication is typically take place with a narrow directional beam using line-of-sight (LOS) propagation technique over a long-haul link distance. A 40 GB/s data rate through FSO communication has already been implemented. Recently, FSO system can cover link distance ranging from nm (inter-integrated circuit design) to 10000km (ground base to satellite FSO link) \cite{raj2019historical}. Despite the potential benefits of FSO, to handle ever-increasing growth of smartphones, cloud big-data, electronic devices, IoT/IoE devices, artificial intelligence, etc. developing a robust infrastructure is become challenging. We discussed different application scenarios ranging from residence to space for the FSO links and systems, still research on fully harvesting the inherent potential are immature. A number of important challenging problems are need to be figured out for the efficient deployment of FSO systems. In this section, we shed the light on research challenges, future research directions and open problems related to FSO systems and application.   

\subsection{5G/B5G and IoT/IoE solutions}

It is obvious that FSO is the most emerging technology to support next generation cellular networks and gigantic connectivity of IoT devices. The substantial deployment of LDs for the FSO technology creates high inter-cell interference in the 5G/B5G and IoT/IoE networks which become a challenging issue. In addition, different modulation techniques of LDs may cause flickering (i.e. fluctuation of modulated signals) \cite{khan2017visible}, thereby avoiding this issue also challenging. A coordinated multipoint (CoMP) enabled cellular network remarkably reduce the inter-cell interference assuring throughput maximization \cite{jahid2017pv, shams2017comp, jahid2019toward}. It is commonly accepted that with the augmentation of deployed OWC nodes lead to increase the inter-cell interference (ICI). ICI coordination and mitigation techniques in OWC domain have been studied utilizing the recognized approached used in the RF domain \cite{chen2015fractional, kazemi2016downlink, liu2017interference}. It is important to realize how to manage interference in optical wireless link. A last mile bottleneck problem may arise if the backhaul system for next generation mobile networks fails to handle an enormous volume of data traffic to support high speed services at the end users. Hence, the enhancement the FSO backhaul capacity for data rate improvement to address last mile problem is become a challenging issue. Furthermore, a massive amount of throughput demand by the access network is generated owing to support huge connectivity of high data rate devices to the network. Therefore, an extensive backhaul capacity is required to solve aforementioned issues. Notably user traffic backhauling via distinctive backhaul networks needs precise synchronization. Some of the future FSO applications include both at cellular backhaul and front haul levels at the transport and access network levels. Therefore, FSO system optimization for mobile network backhauling by means of cost and reliability metrics is an important issue.

Finally, machine-learning/deep-learning based networking is one of the prime key in future generation cellular networks. Nowadays the demand of artificial controlling and remote decision making increasing which can be tackled by supervised learning such as smart healthcare and smart home load controlling \cite{hasan2019real}. Integrating reinforcement learning and machine learning in 6G networks enhance data rate, manage network traffic, enables intelligent network re-assignment, automatic error correction and efficient decision making among surrounding networks \cite{nawaz2019quantum}. 

The seamless deployment of IoT devices trigger a demand to incorporate and inter-operate a variety of hybrid connectivity technologies. For example, numerous applications need the integration of technologies like RFID, wireless sensor networks (WSN), cloud network using Bluetooth, WiFi, and ZigBee connectivity on to a single or hybrid network. As a consequence, conventional RF spectrum is more congested to support variety of IoT applications. One promising key technology is hybrid FSO under OWC technology since it do not interfere with RF spectrum and using the visible light spectrum more than thousand times the modulation bandwidth of RF signal \cite{chatzidiamantis2011distribution, bruzzi2012dual}. To realize the IoT/IoE paradigm, OWC technologies must evolve to accommodate the huge data volume and expected transmission speed.  Because of the aforementioned benefits exploited by an unregulated OWC technology, FSO will play a key role in the future IoT connectivity. The co-deployment of FSO and VLC ensure the capability of physical systems of 5G/B5G network, smart cities (connectivity of heterogeneous wireless services and IoT devices to urban infrastructure) with a number of thins in synergy to connect and reach \cite{yoshida2013assisted, li2015vehicular}. Moreover, connecting smart devices to IoT and beyond IoT network, FSO plays a core backbone network in a convenient way adopting OFDM modulation scheme particularly in an indoor environment \cite{tian2016lighting, hussein2018lightweight}. Hence, an efficient and practical solution of FSO schemes for the smart approach of internet of everything (IoE) is a major design factor.

\subsection{THz and Quantum Communications}

FSO communication provide high data rates over a long coverage distance of LOS connections especially in outdoor applications. FSO support some indoor applications like connect devices inside offices, residences, hospitals, shopping malls etc. The THz spectrum can provide the concrete solution for the data transmissions in non-LOS (NLOS) applications where the transmitted signal reelected off by various objects. However, THz wireless system can be applied for both indoor and outdoor applications with very high data rate including LOS and NLOS links. Therefore, the efficient design of NLOS FSO in consideration of shadow fading and non-linear channel response is a critical design issue. On the other hand, the aspect of quantum computing applications achieve extensive secure are reliable communication in atmosphere and space. So, FSO in optical wireless communications integrating quantum communication will enter a new era in inter-space links \cite{kshatriya2016estimation, majumdar2019principles}.

\subsection{Relay-enabled FSO networks}

A terrestrial FSO link performance can be severely degraded by beam misalignment, attenuation and atmospheric turbulence \cite{wu2014visible}. In this regard, relay assisted FSO networks has the capability to cope with atmospheric turbulence through allowing the transmitted data to an intermediate relay node in order to avoid direct link \cite{jamali2016link}. A high transmission power can be forced to overcome the degradation of the end-to-end performance due to such impairments. Nonetheless, such additional transmitted power may affect the level of secrecy. So, optimal power allocation in RF/FSO hybrid networks is a key concern for the relay assisted optical wireless systems. The relay assisted buffer based heterogeneous networks for unmanned area vehicles (UAVs) with fixed and moving relay stations are presented \cite{fawaz2018uav}. Multi hop transmission (i.e., serial) and cooperative diversity (i.e., parallel) relaying are this two types of relay configuration reported in \cite{safari2008relay, hamza2018classification}. Multi hop transmission relaying technique extend the coverage range where the signal transmitted in between relay node to destination node in a serial fashion. Under parallel relaying approach, the same information is transmitted to the both receiving node and relay node from the sending end and thereafter relay node retransmit the data again to the destination node. Data forwarding protocols using the relay nodes particularly amplify and forward \cite{peppas2013capacity, boluda2018amplify} and decode and forward are discussed in \cite{abou2010cooperative, bhatnagar2011performance}. With the introduction of all optical relaying has the potential to avoid the need of optical-electrical-optical (OEO) conversion at relay each node \cite{bandele2016saturation, dabiri2018performance, huang2018performance}. The cascaded use of saturation gain semiconductor amplifier (SOA) in FSO systenms are capable of eliminating scintillation, atmospheric turbulence even without having the knowledge of channel state information (CSI) \cite{bandele2016saturation, boucouvalas2013first, yiannopoulos2013fade}. However, the idea of relay assisted networking is established for RF technology, still the scope of relaying is immature in optical wireless communication literature.

\subsection{WDM FSO Links}

The integration of wavelength division multiplexing (WDM) techniques with FSO communication pushes toward a new dimension by expanding the capacity of FSO links. With the introduction of dense WDM (DWDM), FSO systems capacity can be greatly enhanced \cite{arimoto2008320, mbah2016outage, lin2014high, tan201812, zhao2018200}. A realization of 320 Gbps LOS/Long WDM FSO links were developed and experimented using OOK modulation \cite{arimoto2008320} and a FSO link of 200 Gb/s bit rate is demonstrated in \cite{zhao2018200}. The outage probability of WDM FSO links in the presence of inter-channel crosstalk and turbulence is analyzed \cite{mbah2016outage}. On the other hand, optical CDMA technique received more attention for future FSO networks to support ever-increasing user demand especially fiber to the home (FTTH) service via optical access network. A few research works \cite{choi2011hybrid, islam2015analytical} have been devoted on OCDMA-WDM optical ring networks in the context of FSO communication considering pointing error effect using received diversity which provides augmented capacity and overall system performance as well. Multi wavelength (MW) OCDMA offers an attractive solution to enhance channel capacity by means of reducing multi access interference (MAI) and crosstalk. A combined solution of WDM-OCDMA support massive number of simultaneous users over conventional WDM scheme and also enable dynamic add/drop function for the next generation access network \cite{choi2011hybrid}. The study of hybrid DWDM-MIMO FSO communications \cite{hamzahenhancement} is conducted aiming to increase transmission coverage with minimum BER and outage probability taking into account severe weather conditions. Despite the recent advances, more research is needed to contemplate low cost and high speed integrated FSO WDM links under different scenarios such as terrestrial, point coverage, cellular coverage, LOS, transmission range, fixed or mobile relay structure.

\subsection{Channel Characterization}

Over the past few decades, various channel models namely log-normal, lognormal-Rician and Gamma-Gamma distributions have been proposed to quantify the impact fading properties and turbulence effect of FSO links \cite{son2017survey}. The performance of FSO links is heavily affected by atmospheric turbulence such as scattering, scintillation, air absorption, free space loss and refraction. Besides, storm, fog, heavy rain and dust may severely degrade the successful communication link between transmitter and receiver in the outdoor environment. Hence, the mitigation technique to address atmospheric turbulence is difficult issue. Statistical modeling of combining the effect of atmospheric turbulence, mobility of obstacles, and pointing error of intensity fluctuation of receive optical signal has been a challenging research issue. The power allocation and the link performance is strongly depends on atmospheric loss \cite{malik2015free} which may lead to scope to research for the counterbalancing of the atmospheric loss. A high speed NLOS data communication in UV band is a significant option \cite{xu2008ultraviolet}. A methodological assessment in consideration of transceiver geometrical configurations, atmospheric conditions and effective channel modeling is still the issue of future research. However, modeling of optical scattering communication (OSC) channel is more difficult than traditional LOS FSO links. This is due to the longer link range pushes the complexity of combined modeling attenuation and multiple scattering \cite{ghassemlooy2015emerging}. Therefore, designing the channel and system models that capture different limiting factors of OSC is become a great interest particularly for connecting distributed objects in IoT devices. The application of MIMO with optical wireless technology is quite challenging due to the characteristics of IM/DD channel \cite{ghassemlooy2016optical}. The application of MIMO could lead increase the hardware complexity and may limit its application. However, the deployment of MIMO with accurate channel modeling is a good research issue.

\subsection{Pointing, Acquisition, and Tracking (PAT)}

In general, FSO transmitter have a broadband and point-to-point directional link in a beeline between the sending node and destination. PAT scheme are typically incurred in static and mobile FSO communication systems when transceiver shoot out narrow beams and the divergence of the beams is smaller than a few µrad (micro-radian) \cite{kaymak2018survey}. This ultra-narrow beam property results high speed data rate, long reach span, less interference, and more energy efficient along the FSO link. The properties of narrow spectral beam make the FSO link more difficult to build between two endpoints. A precise LOS direction of transmission link is required to maintain connectivity between endpoints. However, PAT problems have not completely addresses despite its promising significance. For the successful deployment of PAT mechanism, an integrated and flexible hardware architecture is need to be developed for dynamic PAT. 

Under the pointing mechanism, FSO node starts to coordinate the potential nodes existing in the free space prior connection procedure in mmWave networks \cite{son2014design, he2015link}. Therefore, synchronizing between coordinating nodes is an inherent network design issue related to the pointing mechanism. The acquisition mechanism associated with the modulation-demodulation techniques. FSO receiver aperture is designed such a way that it can accepts multiple optical beams, then the receiver decides the desired signed for decoding. As an indication of physical architecture, the dimension of receiver aperture required to be adjusted based on the divergence angle emitted from the laser beam the distance \cite{farid2007outage}. Such technique substantially reduce outage probability, high date transfer rate and enhance the efficiency of power budget. The tracking mechanism is related to the problem raised by narrow spectral beam property. The link performance is heavily depends on the geometric alignment of transmitter and receiver and beam shape as well. Tracking mechanism under mobile transceivers, a misalignment of optical beams results to increase the outage probability and reduce the available capacity \cite{cap2008optical}. On the other hand, very high pointing accuracy is to be maintained between stationary transceivers. 

The recent growing interest of an intelligent transportation system such as V2V, V2I, or Vehicle-to-everything (V2X) require to keep LOS connectivity with vehicle \cite{cheng2015routing, lee2016internet}. The optical wireless technology can be applied for traffic management \cite{yamazato2017uplink}. Such scenario cause a challenge for signal acquisition and tracking in the presence of dense building in an urban areas. The high speed moving vehicle for instance, cars, electric trains, or unmanned area vehicles may be the challenging task for FSO system in the area of V2X communications. An agile PAT mechanism can keep track with the high speed vehicles to handle the optical link between V2X \cite{urabe2012high}. In addition, the development of PAT scheme in smaller size and low complexity required by the mobile systems (e.g., battery powered drone) is a challenging issue. The design of a suitable PAT model to provide multi-directional coverage and has the ability to mitigate vibration is another interested area. Moreover, a directive PAT mechanism is an important feature where the laser beam focused towards the moving direction of the tracked vehicle is another scope to research. This directive mechanism for mobile FSO communication reduce reduces the handover time with the advance knowledge of trajectory object. On disadvantage of this mechanism the error incurred due to mismatch of the actual and expected speed of vehicle, therefore, it needs GPS or other localization technology support. A time synchronization between two transceivers in PAT process is another crucial issue. As a solution, transceivers allow to exchange control information, channel state information such as location, mobility, and time for synchronized alignment.  

\subsection{Underwater communication (UWC)}

Nowadays UWC received much attention for oil pipe investigation, offshore monitoring, and object detection. Long coverage distance and high speed optical links is the much priority in many UWC applications. The adaptive modulation and coding (AMC) technique in underwater environments is an important issue \cite{zeng2016survey}. Moreover, the adaptive switching mechanism between an acoustic and optical mode for hybrid optical/acoustic networks in underwater is needed for various applications \cite{saeed2019underwater}.  

\subsection{FSO Networking}

5G/B5G technology is comprised with ultra-dense heterogeneous networks (a combination of mixed networks such as macrocell-pico cell, macrocell-femto cell, macrocell-microcell) to increase the data transportation capacity. A small cell dense heterogeneous networks (HetNet) will generate frequent handover between optical and RF networks which could lead many needless handover \cite{du2018context}. Thus, controlling an unwanted handover and the ping-pong effect is a crucial issue. Supporting of two different type of receivers (heterogeneous) is important for hybrid FSO networks. The properties of RF-based transceiver and the optical transceiver are different, the both networks must be active simultaneously for the hybrid framework. Thus, combining the two different architectures in the same platform and the same data transmission over the distinctive systems simultaneously is a vital issue. On the other hand, the distinctive nature of physical layer properties and data link layer for the hybrid optical and RF wireless networks arise a major challenge for the mobility support. A suitable handover algorithm may overcome the continuity user mobility applications. It is important to enable horizontal handover \cite{haas2015lifi} and vertical handover \cite{sarigiannidis2014architectures} mechanisms to allow seamless user mobility. For example, a user can seamlessly move around LiFi cells (horizontal handover) and among LiFi and WiFi networks (vertical handover) under LiFi—WiFi hybrid systems \cite{wang2015dynamic}. Meanwhile, the effect of user mobility based on channel estimation and handover is also challenging. The time taken for handover process should be short enough to meet the specification of 5G/B5G. The time taken for exchanging the information between user equipment and central station depends on the algorithms \cite{li2015cooperative, jin2016resource}. Note that a smaller coverage area incurred large handover and wireless optical channel is vulnerable to channel barrier due to handover. Also, a precise handover is required for switching among the backhauling networks. The challenging of backhauling include the continuous change of pointing of backhaul link for the moving objects. 

\subsection{Load Balancing}

The effective user association policy among the different available access network is a research concern. An optimal user association mechanism solve the joint resource allocation and user allocating problem \cite{stefan2014hybrid}. User equipment (UE) may need to be transferred to different access point for better performance under load balancing technique \cite{wang2016fuzzy}. Therefore, optimal load balancing can be considered as a potential technical issue.

\subsection{Physical Layer Security}

According to \cite{skorin2016physical}, optical wireless communications e.g. FSO networks are vulnerable to cyber-attacks at the physical layer. In recent years, the security and privacy issues of FSO communications received growing interest in the presence of external eavesdropper. Unlike the conventional encryption in top layer, no secret code is needed in physical layer security for long FSO communications. Authors in \cite{boluda2019secure} analyzed secrecy outage probability (SOP) of FSO schemes taking into account of an external eavesdropper with generic orientation at the physical layer. A new framework focusing on misalignment error and corresponding approximate and asymptotic solutions for the non-orthogonal light beams at the eavesdropper receiver is proposed. An external eavesdropping cause the larger beam waist compared to the receiver size due to the pulse spreading of optical beam through wireless link. An extensive analysis of FSO physical layer security for multiple input single output (MISO), MIMO systems, and cooperative systems are reported in \cite{lei2016secrecy, wang2014secure, van2018physical} respectively. The state-of-the-art of physical layer security is still immature for FSO communications. The SOP analysis over the Malaga turbulence channels without accounting pointing errors is pointed out in \cite{lopez2015physical, wang2018secrecy}. The impact of channel impairments on physical layer security for mixed RF/FSO relay architectures is clearly presented in \cite{lei2018secrecy1}. The average secrecy capacity (ASC) analysis under active eavesdropping for mixed RF/FSO relay networks based on decode-and-forward (DF) is studied in \cite{lei2018secrecy2, pan2019secrecy}. The performance of ASC is examined over Malaga turbulence for zero boresight misalignment errors without considering eavesdropper location in \cite{saber2017secure}. It is important to investigate nonzero boresight pointing errors distributions in the FSO channels including eavesdropper’s location and orientation in order to realize realistic scenarios on secrecy performance.  

\subsection{Software Defined Networking (SDN) Control}

SDN can effectively control and manage the hybrid optical networks through SDN controller under dense deployment \cite{kobo2017survey}. However, several OWC applications based on network demand such as traffic re-routing, network flaw management and security issue can be handled by the first application layer in SDN system \cite{cox2017advancing}. On the other hand, updating the flow control, network selection mechanism along with other essential control can be performed in the control layer \cite{haque2016wireless}. SDN technology can be incorporated into FSO system for the purpose of minimization of energy consumption through data traffic controlling. The approach of real time applications and virtualizing of network function in SDN system can be the emerging research issue.

Extending the optical spectrum beyond UV band allows the advantages from high power and lower cost light sources \cite{ghassemlooy2015emerging} which is another research issue. The RF-enabled indoor positioning systems are greatly affected by multipath fading and penetration loss \cite{lin2017indoor, kashani2015performance}. Both indoor and outdoor positioning with greater accuracy using free space optics can be one of the potential research issue. 
In order to guarantee data loss reduction, selection of optimal transmitter among multiple nodes, and delay minimization, the seamless steering of transferred information become an eminent challenging problem. The optical wireless system is usually designed for downlink communications purpose because of some challenges in uplink direction such as energy constraint of portable devices and the limitations of narrow beam controlling for mobile devices. Hence, the uplink transmission is open research issue that need to be addressed. A few research works have conducted of RF/optical hybrid wireless from the perspective of physical layer whereas barely focus on data-link level metrics \cite{hammouda2018link}. The QoS analysis tool of cross layer design between physical and data link layer is an important research zone.

\section{Conclusions}		\label{14}

The design of pervasive and efficient wireless systems for a wide range of transmission links is essential to meet ever-increasing higher data rate demands in the next generation communication networks. Owing to spectrum scarcity of radio-frequency (RF) counterparts, free space optical (FSO) communication has been recognized as a promising option for next generation optical networking that can support tremendous traffic demand initiated from the internet of things/everything (IoT/IoE) devices and  next-generation cellular communication systems. The realization of unlicensed extremely wide optical band can fulfill the exponential growth of traffic demand and complementarily overcome the RF spectrum shortage in a cost effective way. The aggregate deployment of hybrid wireless system offering different features can suppress the limitations of a single network either standalone RF or optical wireless and make a prominent alternative over RF based communication systems. In this review paper, we have discussed the brief overview of promising OWC technologies from various viewpoints. An extensive research on the deployment issues of different FSO wireless networks addressing the related challenges are underway. Principle, advantages, limitations, design issues, mitigation techniques, significance, worldwide achievements, state-of-the-art of recent developments and revolutions free space optical communications are presented. Advancements of hybrid FSO, coherent FSO, FSO link budget, research challenges of FSO infrastructure deployment of different application scenarios such as acoustic underwater communication, 5G and B5G communication, FSO-IoT/IoE interface are demonstrated. Provisioning of quality of experience (QoE), massive connectivity, high capacity, tight security, low power consumption are also focused in perspective of the coexistence FSO and RF wireless technology. The opportunities triggered by FSO schemes as well as network architecture, application scenarios, the key research directions, the solutions of major challenges for a successful deployment of the considered framework in the context of wireless communication and IoT paradigms are also briefly pointed out. It can be concluded that FSO is increasingly becoming a prominent technology for future generation communication system and extending the realm of FSO technology in indoor, healthcare, industry, offices, shopping malls, offices, stadiums, residences, transportation, space and terrestrial environments. This survey paper provides valuable resource for clear understanding the recent research contribution optical free space communication in different implications and is anticipated to persuade further effort for the eminent deployment of other OWC systems in the forthcoming years.

\bibliographystyle{IEEEtran}
\bibliography{ref_FSO}
\nocite{*}

\end{document}